\documentclass[10pt,journal,compsoc]{IEEEtran}

\usepackage[nocompress]{cite}

\ifCLASSINFOpdf

\else

\fi

\usepackage{makecell}
\usepackage[colorlinks,linkcolor=red,anchorcolor=blue,citecolor=green]{hyperref}

\usepackage{amsfonts}
\usepackage{graphicx}
\usepackage{array}
\usepackage{booktabs} 
\usepackage{bm}
\usepackage{float}
\usepackage{times,amsmath,epsfig}
\usepackage{graphicx}
\usepackage{amssymb}
\usepackage{amsthm}
\usepackage[dvipsnames]{xcolor}

\usepackage{enumitem}
\usepackage{algorithm}
\usepackage{algorithmic}
\usepackage{multirow} 
\usepackage[font={small}]{caption}
\usepackage[caption=false, font=footnotesize]{subfig}
\usepackage{array} 
\usepackage{booktabs} 
\usepackage{longtable}
\usepackage{bm}
\usepackage{verbatim} 
\usepackage{float}
\newcommand{\etal}{\textit{et al.}}
\newcommand{\tabincell}[2]{\begin{tabular}{@{}#1@{}}#2\end{tabular}}
\usepackage{color}

\newcolumntype{H}{>{\setbox0=\hbox\bgroup}c<{\egroup}@{}}
\usepackage{colortbl}
\usepackage{CJKutf8}
\usepackage{pifont}
\usepackage{textcomp}
\newcommand{\cmark}{\ding{51}}%
\newcommand{\xmark}{\ding{55}}%

\newcommand{\greenuparrow} {\textbf{\textcolor{ForestGreen}{ $\uparrow$}}}
\newcommand{\greendownarrow} {\textbf{\textcolor{ForestGreen}{ $\downarrow$}}}
\newcommand{\boldhline} {\Xhline{2\arrayrulewidth}}

\usepackage{arydshln}
\newcommand{\revisetext}[1]{\textcolor{black}{#1}}
\begin{document}
	%
	
	\title{
		A Coding Framework and Benchmark towards Low-Bitrate Video Understanding
	}
	\author{Yuan~Tian,
		Guo~Lu,
		Yichao~Yan,
		Guangtao~Zhai,
		Li~Chen,
		Zhiyong~Gao
		\IEEEcompsocitemizethanks{
			\IEEEcompsocthanksitem Corresponding authors: Guo Lu, Yichao Yan, and Guangtao Zhai.
			\IEEEcompsocthanksitem Yuan~Tian is with the Institute of
			Image Communication and Network Engineering, Shanghai Jiao Tong University, China, and also with the Shanghai Artificial Intelligence Laboratory, Shanghai, China.
			E-mail: ee\_tianyuan@sjtu.edu.cn.
			\IEEEcompsocthanksitem
			Guo~Lu, Li~Chen, and Zhiyong~Gao are with the Institute of
			Image Communication and Network Engineering, Shanghai Jiao
			Tong University, China.
			E-mail: \{luguo2014, hilichen, zhiyong.gao\}@sjtu.edu.cn.
			\IEEEcompsocthanksitem Yichao~Yan is with the MoE Key Lab of Artificial Intelligence, AI Institute, Shanghai Jiao Tong University, China. E-mail: yanyichao@sjtu.edu.cn.
			\IEEEcompsocthanksitem Guangtao~Zhai is with the Institute of Image Communication and Network Engineering, Shanghai Jiao Tong University, China, and also with the MoE Key Lab of Artificial Intelligence, AI Institute, Shanghai Jiao Tong University, China.
			E-mail: zhaiguangtao@sjtu.edu.cn.
			 
	}
	}
	\markboth{Journal of \LaTeX\ Class Files,~Vol.~14, No.~8, August~2015}%
	{Shell \MakeLowercase{\textit{et al.}}: Bare Demo of IEEEtran.cls for Computer Society Journals}

	\IEEEtitleabstractindextext{%
		\begin{abstract}
		Video compression is indispensable to most video analysis systems. Despite saving the transportation bandwidth, it also
		deteriorates downstream video understanding tasks, especially at low-bitrate settings. To systematically investigate this problem, we first thoroughly review the previous methods, revealing that three principles, \textit{i.e.}, task-decoupled, label-free, and \revisetext{data-emerged semantic prior}, are critical to a machine-friendly coding framework but are not fully satisfied so far.
		In this paper, we propose a traditional-neural mixed coding framework that simultaneously fulfills all these principles, by taking advantage of both traditional codecs and neural networks (NNs).
		On one hand, the traditional codecs can efficiently encode the pixel signal of videos but may distort the semantic information. On the other hand, highly non-linear NNs are proficient in condensing video semantics into a compact representation.
		 The framework is optimized by ensuring that a transportation-efficient semantic representation of the video is preserved \textit{w.r.t.} the coding procedure, which is spontaneously learned from unlabeled data in a self-supervised manner.
		 \revisetext{The videos collaboratively decoded from two streams
		 (codec and NN) are of rich semantics, as well as visually photo-realistic, empirically boosting several mainstream downstream video analysis task performances without any post-adaptation procedure}.
		 Furthermore, by introducing the attention mechanism and adaptive modeling scheme, the video semantic modeling ability of our approach is further enhanced. Finally, we build a low-bitrate video understanding benchmark with three downstream tasks on eight datasets, demonstrating the notable superiority of our approach. All codes, data, and models will be available at  \url{https://github.com/tianyuan168326/VCS-Pytorch}.
		\end{abstract}
		
}

	\maketitle

	\IEEEdisplaynontitleabstractindextext

	%
	\IEEEpeerreviewmaketitle

	\IEEEraisesectionheading{\section{Introduction}\label{sec:introduction}}
	\IEEEPARstart{V}{ideo} understanding has gained increasing attention in the computer vision community over the past few years.
  	With continuous efforts, impressive results~\cite{he2016deep}\cite{gao2019res2net}\cite{tran2015learning}\cite{wang2016temporal}\cite{feichtenhofer2019slowfast}\cite{dosovitskiy2020image} have been  achieved on benchmark datasets.
	Nevertheless, few works~\cite{pourreza2019recognizing}\cite{yi2021benchmarking}\cite{tanaka2022does} have systematically investigated how these models perform on compressed videos, especially at low bitrate settings, which are common in real-world video analysis systems to save transmission bandwidth or fit weak network connections.

\begin{figure}[!ttp]
		\centering
		\includegraphics[width=8.8cm]{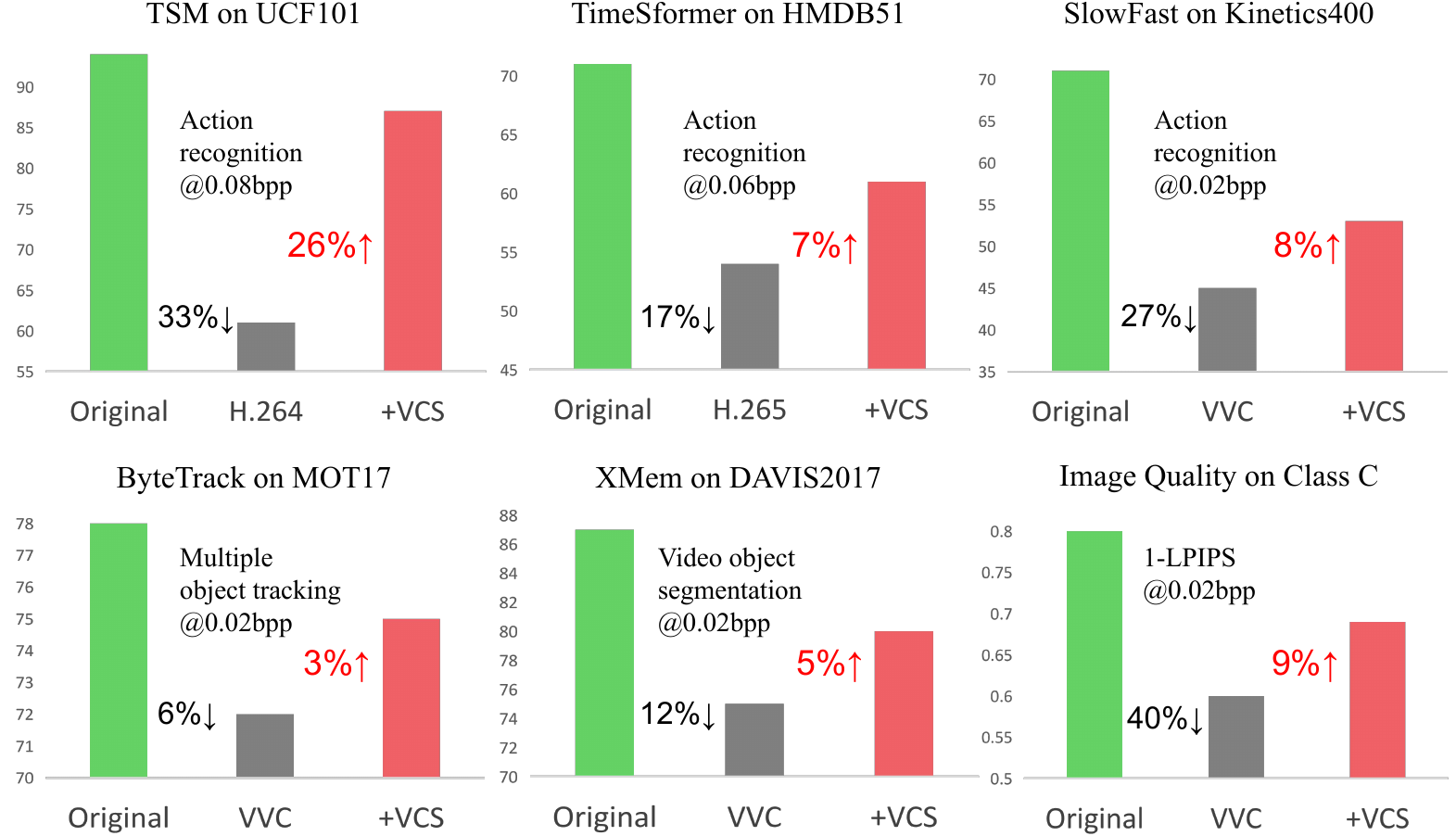}
		\vspace{-2mm}
		\caption {
			Performance improvements on various tasks by applying the proposed VCS framework to different traditional video codecs.
			\revisetext{The bitcost of the proposed VCS framework is calculated by incorporating the additional semantic stream for a fair comparison.}
		}
		\vspace{-7mm}
		\label{fig:head_comp}
	\end{figure}

	As shown in Fig.~\ref{fig:head_comp}, the well-trained video understanding models are degraded severely by the high video compression ratio.
	For example, the recognition accuracy of the SlowFast model~\cite{feichtenhofer2019slowfast} on Kinetics dataset~\cite{carreira2017quo} is dramatically decreased by 27\% at 0.02bpp, even when the advanced VVC codec~\cite{bross2021overview} is adopted.
	Despite the unsatisfactory results when faced with semantics-related tasks,
	probably due to the design goal of preserving low-level pixel signals,
	the traditional codecs have been widely deployed in the industry and enjoy hardware-efficient implementations.
	On the other hand, neural networks (NNs) are well known to be proficient in extracting semantic information as a compact representation.
	Therefore, a natural question is whether we can take advantage of both traditional codecs and NN-based ones to build a unified video coding framework, that preserves general semantic contents required by machine intelligent tasks during the compression.

	To investigate this question, we first review the previous works on Video Coding for Machine (VCM), as shown in Tab.~\ref{tab:summary_VCM}.
	Early works \cite{zhang2016joint}\cite{chao2015keypoint}\cite{baroffio2015hybrid} additionally transport the local descriptors such as SIFT~\cite{lowe1999object} and SURF~\cite{bay2006surf} to better serve the image indexing/retrieval tasks.
	Meanwhile, these also emerge the MPEG standards (CDVS/CDVA)~\cite{duan2018compact}\cite{duan2015overview}\cite{duan2013compact} for visual searching and analysis.
	The transported information in above methods is specified for the target downstream task.
	There are also some methods~\cite{makar2012gradient}\cite{galteri2018video}\cite{choi2018high}\cite{huang2021visual} \cite{choi2020task} \cite{cai2021novel}\cite{zhang2021just} focus on improving the rate-distortion optimization (RDO) target of the traditional codec to better support machine intelligence tasks, namely, the {task-guided RDO} scheme.
	But, they rely on heuristically~\cite{makar2012gradient}\cite{galteri2018video}\cite{cai2021novel} or empirically~\cite{huang2021visual}\cite{zhang2021just} designed task-specific distortion metrics.
	Besides, many other studies are conducted towards {feature compression}. In~\cite{choi2018near}\cite{chen2019lossy}\cite{chen2019toward}\cite{singh2020end}\cite{feng2022image}, the intermediate feature maps of the AI models, instead of the images, are compressed and transported.
	Nonetheless, these methods require carefully deciding which layer features should be transported for a certain task.
	Due to the inner designs are coupled with the specific tasks, the above methods are non-trivial to be deployed to tasks beyond their design.
	Therefore, \textit{task-decoupled} designing principle, \revisetext{\textit{i.e.}, the compression procedure and the inner design should be not tied to specific tasks, is essential for a versatile coding system readily supporting diverse tasks.
	Besides, \textit{task-decoupled} principle is more important for video applications, since the modeling schemes of different video tasks are heterogeneous ~\cite{zhao2023streaming}.}

	Furthermore, scalable coding-based methods~\cite{yang2021towards}\cite{liu2021semantics}\cite{yan2021sssic}\cite{choi2021latent}\cite{sun2020semantic} recently gained much attention,
	which mostly leverage learnable codecs to encode the input image and exploit the cross-task dependencies in the feature space.
	Nevertheless, the architecture of their encoding networks is partially shared with the task models, which brings difficulty to shifting between different models. So, they still cannot completely fulfill the requirement \textit{task-decoupled}.
	More seriously, the annotation of the data labels used to train these methods is laborious.
	Therefore, \textit{label-free} learning scheme, \textit{i.e.}, self-supervised learning from unlabeled video data, is also required for a fast-to-deploy coding system.
	Bedsides, recent methods~\cite{le2021image}\cite{bai2022towards} jointly optimizing the learnable codecs with the downstream tasks are manifestly not {label-free}.
	
	\revisetext{Moreover, effective semantic priors are also crucial for an AI-task oriented video coding system.
	The endeavors can be broadly categorized into hand-crafted, model-driven and data-emerged approaches.
  	For hand-crafted priors, traditional descriptor-based methods~\cite{zhang2016joint}\cite{chao2015keypoint}\cite{baroffio2015hybrid} and recent techniques utilizing manually-defined structural representations~\cite{hu2020towards}\cite{duan2022jpd} or region-of-interest(ROI) regions~\cite{cai2021novel}\cite{galteri2018video} fall into this category. However, these methods are deficient in extracting robust semantic representation from intricate video data.
	As for model-driven priors, most methods rely on pre-trained classification models~\cite{yang2020discernible} or targeted downstream models right for deployment~\cite{choi2022scalable11}, which struggle with unforeseen downstream models.
	We argue that \textit{data-emerged semantic prior}, \textit{i.e.}, the semantic representation spontaneously emerged from the image/video datasets with minimal dependence on human experience or specific models, holds promise for enabling coding system deployment across diverse downstream tasks/models, while adeptly modeling the complex video data.}

	\begin{table}[!tbp]
		\centering
		\renewcommand\arraystretch{1.1}
		\tabcolsep=0.8mm
		\scalebox{0.94}{
			\begin{tabular}{cccccHHc}
				\Xhline{2\arrayrulewidth}
				\textbf{\makecell[c]{Scheme}}  & \textbf{Method} & \textbf{Domain}  & \textbf{\makecell{Task\\decoupled}} & \textbf{\makecell{Label\\free}}  & \textbf{\makecell{Human\\vision}} & \textbf{\makecell{Standard \\reusing}} &\textbf{\makecell{ Semantic prior}} \\
				\hline
				
					\multirow{2}{*}{\makecell{Traditional\\descriptor}}
				
				&\makecell{
					\cite{zhang2016joint}\cite{chao2015keypoint}\cite{baroffio2015hybrid} 
				}& Image& \xmark & \cmark& \cmark & \cmark & Hand-crafted \\
			
				&\cite{duan2018compact}\cite{duan2015overview}\cite{duan2013compact}
				& Video & \xmark & \cmark & \cmark & \cmark &Hand-crafted  \\
				
				%
					\arrayrulecolor{gray}\cdashline{1-8}[5pt/3pt]

				\multirow{1}{*}{Task-guided RDO} &\makecell{\cite{makar2012gradient}\cite{choi2018high}\cite{huang2021visual} \\ \cite{choi2020task}\cite{cai2021novel}\cite{zhang2021just}}
				& Image & \xmark & \cmark & \cmark & \cmark & Hand-crafted \\
				\arrayrulecolor{gray}\cdashline{1-8}[5pt/3pt]
				\multirow{3}{*}{\makecell{Feature\\compression}}
				& \cite{choi2018near}\cite{chen2019lossy}\cite{chen2019toward}
				& Image &\xmark & \cmark & \xmark & \xmark & Hand-crafted  \\
				& \cite{singh2020end} & Image& \xmark&\xmark &\xmark &\xmark &Hand-crafted\\
				& \cite{feng2022image} & Image & \xmark & \xmark & \cmark &\xmark & Data-emerged \\
				
				\arrayrulecolor{gray}\cdashline{1-8}[5pt/3pt]
				
				\multirow{2}{*}{Scalable coding}   &\makecell{   \cite{hu2020towards}\cite{yang2021towards}\cite{liu2021semantics} \\ \cite{choi2021latent}\cite{choi2022scalable11}\cite{yang2021video}
				}
				&Image &\xmark &\xmark & \cmark  &\xmark  & Model-driven\\
				
				&\makecell{   
					\cite{duan2020video}
				}
				&Video &\xmark &\xmark & \cmark  &\xmark  &  Hand-crafted\\
				\arrayrulecolor{gray}\cdashline{1-8}[5pt/3pt]
				\multirow{2}{*}{\makecell{Learnable codec+\\Task-friendly loss}} 
				& \cite{yang2020discernible} & Image & \cmark &\cmark & \cmark &\xmark & Model-driven \\
				& \cite{le2021image}\cite{bai2022towards} 		 & Image & \cmark & \xmark & \cmark &\xmark & Model-driven\\
				
				\arrayrulecolor{gray}\cdashline{1-8}[5pt/3pt]
			
				&\textbf{Ours} & Video & \cmark &\cmark & \cmark &\cmark &Data-emerged \\
				\Xhline{2\arrayrulewidth}
			\end{tabular}
		}
	\vspace{-2mm}
		\caption{
			Comparison of recent VCM methods.
		}
		\vspace{-6mm}
		\label{tab:summary_VCM}
	\end{table} 

In this paper, we propose the Video Coding for Semantics framework (VCS), which is a traditional-neural mixed video coding approach that fulfills all three requirements mentioned above, \textit{i.e.}, \textit{task-decoupled}, \textit{label-free} and \textit{data-emerged semantic prior}.
VCS combines the superior content-coding capability of traditional codecs and the remarkable semantic coding capability of the neural codecs,
aiming to well support various machine intelligent tasks under low-bitrate settings.
The neural codec is learned from purely unlabeled data by leveraging a contrastive learning objective\cite{he2020momentum}, \textit{i.e.}, discriminating each video instance from other ones,
which is inspired by previous methods of unsupervised representation learning~\cite{he2020momentum}\cite{chen2020simple}.
 Moreover, a one-bit map is adopted as the early bottleneck of the video representation, which enforces the semantic information compactly encoded in a sparse low-dimensional space and thus prompts a transportation-efficient system.
 It is noteworthy that the broad idea of combining traditional codec and an auxiliary bitstream has been used in previous works~\cite{baroffio2015hybrid}\cite{zhang2016joint}\cite{veselov2021hybrid}, but few focused on learning task-agnostic semantics from unlabeled video, which is an important step toward a general semantic coding system.

	Furthermore, an attention-based cross-bitstream feature fusion scheme is introduced to reduce the redundancy between the video (from the traditional codec) and the semantic (from the neural codec) streams.
	Additionally, the input-adaptive feature learning scheme is incorporated to enable extracting more accurate semantic information.
	All these designs substantially improve the compression efficiency of our framework.

	We evaluate the framework on three popular video understanding tasks with eight large-scale video datasets,
	\textit{i.e.}, \textit{action recognition} with Kinetics-400, Something-to-Something, UCF101, HMDB51 and Diving48 datasets,
	\textit{multiple object tracking} with MOT17 and SFU-HW-Tracks datasets,
	\textit{video object segmentation} with DAVIS2017 dataset.
	
	To summarize, our {main contributions} are:
		
		\textbf{1.} We propose VCS, Video Coding for Semantics framework, which introduces a neural stream with data-emerged semantic prior beyond the traditional codecs, enjoying their advantages.
		 {VCS can be readily deployed for various downstream video analysis tasks without any task/data-specific adaptation procedure.}
	
		
		

		\textbf{2.} VCS is optimized in a self-supervised manner with a bottleneck-based contrastive learning objective, which facilitates preserving the video semantics and encourages discarding trivial semantic-less information.
		 
		
		
		\textbf{3.} 
		\revisetext{Network architecture of VCS is sophistatedly designed with adaptive and dynamic schemes, aiming to enhance its semantic modeling capability.}
		
		
		\textbf{4.} Our approach demonstrates strong performances on a wide range of tasks/datasets.
			\revisetext{To facilitate future research, we build a systematical coding benchmark for three popular video tasks on eight large-scale datasets by re-implementing and evaluating three traditional codecs, two learnable codecs, and four VCM methods}.

		This paper is organized as follows. In the second section, we conduct a comprehensive review of the previously related works.
		In the third section, we present the proposed framework.
		The experimental results are provided in the fourth and fifth sections, and the conclusions are drawn in the sixth section.


	\section{Related Works}
		\textbf{Video Coding for Machine (VCM).}
	Previous video codecs, including traditional ones~\cite{wiegand2003overview}\cite{sullivan2012overview}\cite{bross2021overview}, the neural-enhanced traditional methods~\cite{tian2021self}\cite{tian2023clsa}, and recent emerging learnable ones~\cite{lu2020end}\cite{hu2021fvc} are designed to achieve better pixel-wise signal quality metrics, \textit{e.g.}, PSNR and MS-SSIM~\cite{wang2003multiscale}, which mainly serve the human visual experience.
	The discrepancy between these metrics and the downstream tasks causes the codecs above to be not optimal for machine intelligence, especially at low bitrate levels.
	Therefore, recent works on VCM~\cite{duan2018compact}\cite{duan2020video}\cite{yang2021video} have emerged to build an efficient joint compression and machine analytics framework.

	Early standards such as CDVA~\cite{duan2013compact} and CDVS~\cite{duan2015overview}\cite{duan2018compact} propose to analyze the image keypoints (\textit{e.g.}, SIFT and SURF) in the edge device and transport this information to the server side, supporting image indexing or retrieval tasks.
	Later works are devoted to compressing the intermediate feature maps by using hand-crafted compression tools such as HEVC~\cite{choi2018near}\cite{chen2019lossy}\cite{chen2019toward} or the learnable entropy model~\cite{singh2020end}  \cite{feng2022image}.
	Considering the videos are usually transported as well, some works~\cite{zhang2016joint}\cite{chao2015keypoint}\cite{baroffio2015hybrid} explore the correlation between the feature stream and the video stream to improve the compression efficiency of the whole system.
	Besides explicitly encoding the features, some other works~\cite{makar2012gradient}\cite{galteri2018video}\cite{choi2018high}\cite{huang2021visual} \cite{choi2020task} \cite{cai2021novel}\cite{zhang2021just}\cite{ge2024task} propose to improve the bit-allocation strategy of traditional codec by using downstream task-guided rate-distortion optimization (RDO) strategy.
	For example, Choi~\etal\cite{choi2018high} proposes to estimate the object importance map from the features of the YOLO9000 network, achieving a better bit-allocation strategy for the object detection task.
	There are also some methods~\cite{le2021image}\cite{bai2022towards}\cite{yang2020discernible} optimizing the learnable codecs by directly incorporating the downstream task during the training procedure.
	However, they are usually designed for one task at one time.
	More recently, a amount of works\cite{hu2020towards}\cite{yang2021towards}\cite{liu2021semantics}\cite{yan2021sssic}\cite{choi2021latent}\cite{choi2020task}
	\cite{wang2019scalable}\cite{duan2020video}\cite{yang2021video}\cite{liu2021semantics}\cite{choi2022scalable11} are built on the scalable coding paradigm, aiming to support multiple AI tasks.
	As one of the representative works, Hu~\etal\cite{hu2020towards} proposes to separately transport the edge structure (base layer) and some sampled reference pixels (enhanced layer) of the face image. Liu~\etal\cite{liu2021semantics} present scalable image
	compression that supports coarse-to-fine classification tasks, where input reconstruction is adopted as the enhancement task.
	Choi~\etal\cite{choi2022scalable11} proposes a unified image coding framework for both humans and machines, which takes advantage of latent-space scalability and achieves obviously superior compression efficiency to previous methods.
	Nevertheless, most of them rely on supervised learning, where the data label annotation is laborious.
	There are also very few works that extend the traditional codec to better support downstream tasks by introducing a feature encoding stream.
	For example, \cite{veselov2021hybrid} uses a task-specific feature extractor (\textit{e.g.}, backbone of a face recognition network) to calculate the features of the original video and the compressed video, and then transports the residual part as the feature stream.
	This hand-crafted feature residual strategy may be unable to achieve the optimal rate-performance trade-off.
	
	Recently, some methods also exploit vectorized edges~\cite{hu2020towards} and semantics segmentation maps~\cite{duan2022jpd} as the highly condensed intermediaries for video semantics, which is intuitively derived from the human prior that these representations are usually recognizable and sparse (easy to be compressed).
	The hand-crafted designs still hinder the system from achieving optimal compression efficiency.
	Very recently, Feng~\etal~\cite{feng2022image} propose to constrain the intermediate-layer features to be semantics-complete with the self-supervised learning scheme.
	But, the downstream task models are still required to be fine-tuned on labeled image data for adapting to the above-learned features.

	Therefore, a task-decoupled framework for VCM that does not rely on task-specific labels or human-intuitive semantic priors is still left blank.
	Besides, the above methods are mainly evaluated on image tasks, but their effectiveness on video tasks, especially on recent large-scale video datasets, remains unknown.
	
		\begin{figure*}[!bthbp]
		\centering
		\centerline{\includegraphics[width=16.8cm]{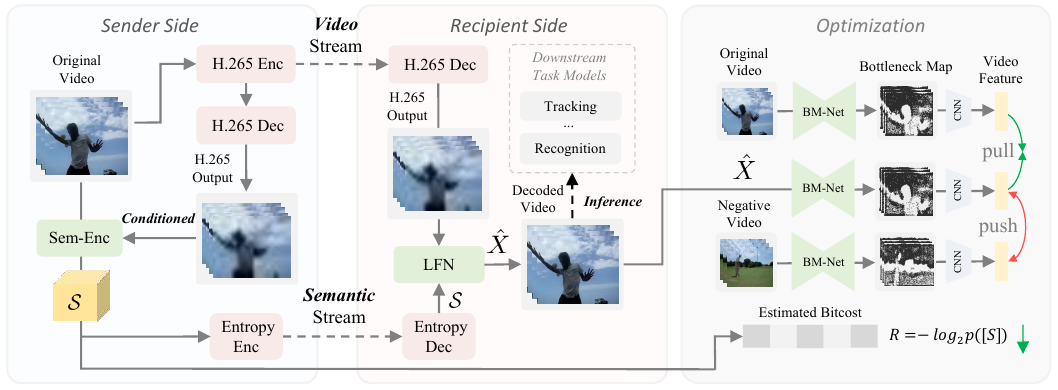}}
		\vspace{-1mm}
		\caption {
			{Overview of VCS framework.} It includes a \textit{semantic stream} $\mathcal{S}$ in addition to the video stream. $\mathcal{S}$ has no prior constraints (\textit{e.g.}, keypoint, edge, segmentation map) and is completely data-emerged.
			On the recipient side, the information within two streams is fused into the ultimately decoded video $\hat{X}$.
			{$\hat{X}$ can be directly deployed to the downstream task models, which are not involved in the training procedure.}
			The semantic completeness of $\hat{X}$ is ensured through a bottleneck map-based contrastive learning objective, wherein $\hat{X}$ is pulled closer to the original video, but pushed far away from the negative video.
			The bitcost of the semantic stream is also estimated and optimized to be minimal. Here, [$\cdot$] and $p$ represent the $round$ function and a learnable prior probability model, respectively. The H.265 codec is used as an example illustration.
		}
		\vspace{-3mm}
		\label{fig_framework}
	\end{figure*}

	\textbf{Video analysis in the compressed domain.}
	The early works~\cite{pilu1998using}\cite{smolic2000low}\cite{niu2009moving}\cite{chen2011joint}\cite{chen2011moving} mainly exploit the MV field to perform global motion estimation and moving object segmentation.
	\cite{mezaris2004real} extracts meaningful foreground spatio-temporal objects from the MPEG-2 compressed stream, assembling the object descriptors to a vocabulary for video indexing and retrieval.
	\cite{khatoonabadi2012video} presents a compressed-domain object tracking method that uses MVs and block coding mode information to perform fast tracking.
	In \cite{liu2022real}, an efficient video object tracking framework is proposed, where a heavy CNN is utilized for object detection on keyframes, and a lightweight CNN is applied to the MV and residual information of non-key frames for propagating the detected objects by heavy CNN.
	Besides, there are some works~\cite{wang2019fast}\cite{li2022end}\cite{wang2021real} exploiting the MV and residual information to perform video object detection and generic event boundary detection tasks.
	As for high-level {action recognition} task, the pioneering work~\cite{zhang2016real} uses the MV field to replace the computationally extensive optical flow modality in two-stream networks~\cite{feichtenhofer2016convolutional}.
	Later works further improve this work by exploiting the residual frames~\cite{wu2018compressed} or refined MV~\cite{shou2019dmc}.

	These methods focus on proposing better or more efficient video understanding models in the moderately compressed domain, whereas our work aims to propose a semantic coding framework, in which the decoded video at low bitrate levels enables superior downstream analysis performance.

	\textbf{Deep Video understanding.}
	\textit{Action recognition} is the most basic task for video semantics understanding.
	Early, two-stream CNNs~\cite{simonyan2014two}\cite{feichtenhofer2016convolutional} leverage the optical flow input as the motion representation.
	Later, the works mainly model the temporal cues in feature space for efficiency.
	For example, TSN~\cite{wang2018temporal} adopts a simple averaging function to aggregate features of each frame.
	The subsequent works (TRN~\cite{zhou2018temporal}, TSM~\cite{lin2019tsm}, TEA~\cite{li2020tea}, TDN~\cite{wang2021tdn}, EAN~\cite{tian2022ean}) improves TSN by introducing more sophisticated temporal modeling modules. To simultaneously learn the temporal evolution along with the spatial information, 3D networks~\cite{tran2015learning}\cite{tian2019video}\cite{carreira2017quo}\cite{feichtenhofer2019slowfast}\cite{tian2020self} as well as their computationally-efficient variants such as X3D networks~\cite{feichtenhofer2020x3d} are continuously proposed.
	Very recently, Gedas \etal~\cite{bertasius2021space} proposed the pure-Transformer architecture TimeSformer for video understanding, which involves self-attention designs over space-time dimensions.
	Please refer to~\cite{sun2022human} for the review of other recent methods.
	We adopt TSN, TSM, SlowFast, and TimeSformer as the downstream models, which range from 2D CNN to Transformer.
	\textit{Multiple object tracking (MOT)} includes two subtasks, \textit{i.e.}, object detection and object association,
	which are fundamental techniques of various video analysis applications.
	We adopt the state-of-the-art ByteTrack~\cite{zhang2021bytetrack} as the baseline method.
	\textit{Video object segmentation (VOS)} predicts the pixel-level labels for the objects in videos, which is used to validate our method can also preserve the fine-grained semantic information of the videos.
	We adopt the state-of-the-art XMem~\cite{cheng2022xmem} as the baseline method.
	We also mention that, in addition to the above video understanding tasks, there are also many famous computer vision tasks, such as object detection~\cite{ren2016faster}\cite{zou2023object}, adversarial machine learning~\cite{yan2021dehib}\cite{yan2023dhbe}, image/video quality assessment~\cite{li2023agiqa}\cite{chen2024gaia}\cite{li2024aigiqa}\cite{yi2021attention}, image enhancement~\cite{duan2022develop}, medical image analysis~\cite{chen2024cross}\cite{he2023transformers}\cite{jiang2024zept}, etc.
	These tasks are also worth re-evaluating in low-bitrate settings.

	\section{Approach}
	
	We propose the Video Coding for Semantics framework VCS, as shown in Fig.~\ref{fig_framework}, which exploits the superior content-coding capabilities of traditional codecs and the remarkable semantic coding capabilities of neural networks.
	The goal of VCS is to ensure that the ultimate decoded video preserves semantics and can be directly consumed by various video understanding models.
	We first give a brief introduction to the framework.

	\textbf{Video stream.}
	The original video $X$ is first compressed to a lossy video $\tilde{X}$ by a traditional video codec such as H.265.
	
	\textbf{Semantic stream.}
	A semantic stream $\mathcal{S}$ is introduced to efficiently transport the semantic part of $X$, which is produced by a learnable NN-based encoder Sem-Enc.
	The encoding procedure is conditioned on the video stream to reduce the overall bit cost.
	Moreover, adaptive and dynamic modeling schemes are introduced for extracting accurate semantic representation.
	
	\textbf{Recipient side.}
	After the video and semantic streams are received, a latent fusion network (LFN) is employed to fuse the information of two streams in a deep latent space and produce the video $\hat{X} := \operatorname{LFN}(\tilde{X}, \mathcal{S})$, which will be consumed by machine models.
	{The decoded result $\hat{X}$ comprises RGB video frames that adhere to the natural image prior, rather than free-form features, forming a task-decoupled input applicable to a wide range of video tasks and models.
		This task-decoupled philosophy also reduces the effort required in designing task-specific adapters between the coding procedure and downstream model.}

	\subsection{Semantic Stream Encoder (Sem-Enc)}\label{sec:method_enc_net}
	The goal of Sem-Enc is to extract semantics from the input video and encode it into a compact bitstream.
	Specifically, we first adaptively extract the semantic feature from each single frame by conditioning the video stream, and then utilize temporal fusion operations to aggregate these frame-level features into a video-level feature, which are then encoded into the semantic stream.

\begin{figure}[!tbp]
	\centering
	\includegraphics[width=8cm]{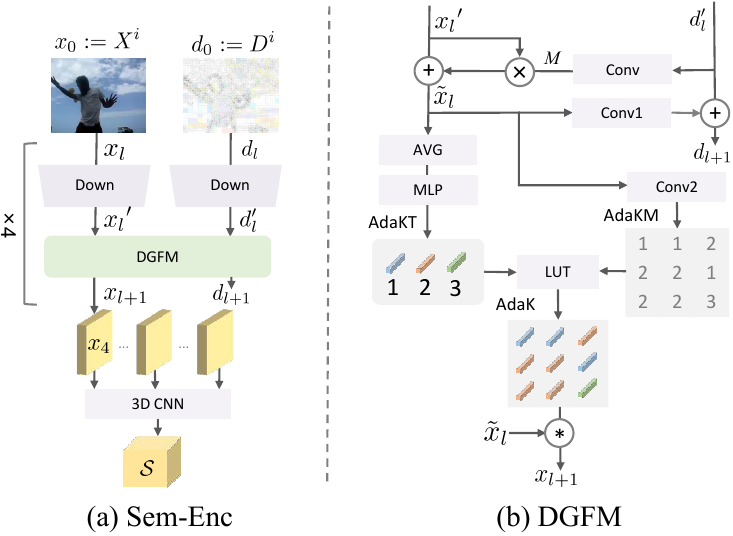}
	\vspace{-1mm}
	\caption{(a) {Sem-Enc} hierarchically transforms the input video $X$ and the difference map $D$ from pixel space to semantic space, producing the semantic feature $\mathcal{S}$.
		(b) {DGFM} is introduced to fuse the intermediate features from the two pathways of Sem-Enc.
		$\oplus$ and $\otimes$ denote the element-wise summation and multiplication, respectively.
		$\circledast$ denotes the pixel-adaptive convolution operation.
	}
	\label{fig_csgfm}
	\vspace{-6mm}
\end{figure}

	\textbf{Single frame feature extraction.}
	We exploit the video compression difference map $D := X - \tilde{X}$ to guide the feature extraction procedure, which aims to eliminate the redundancies between the video stream and the semantic stream.
	As shown in Fig.~\ref{fig_csgfm} (a), the network architecture follows a dual-pathway scheme, \textit{i.e.}, a feature encoding pathway, and a difference encoding pathway.
	Both two pathways include four stages.
	After each stage, the features from the two pathways are inter-fused via a difference-guided fusion module (DGFM).
	The procedure can be formulated as follows,
	\begin{equation}
	\begin{split}
	& x_{l+1}, d_{l+1} = \mathsf{DGFM}({Down}(x_l), {Down}(d_l) ), \\
	& x_0 = X^i, d_0 = D^i,
	\end{split}
	\end{equation}
	where $i$ indicates the frame index, $l \in [0,3]$ denotes the stage index, and ${Down}$ represents the three-layer CNNs with spatially downscaling ratio two.
	The difference pathway is four times lightweight than the feature pathway in terms of the channel number of convolutions, due to the sparsity of difference signal.
	The parameters of the two pathways are not shared.

	\textbf{Temporal fusion.}
	The ultimate-stage features $x_4$ (of all frames) are further processed by a lightweight 3D CNN to remove the inter-frame redundancy, producing the video-level semantic feature $\mathcal{S}\in \mathbb{R}^{T \times 256 \times \frac{H}{32} \times \frac{H}{32}}$, where $T$ and $H \times W$ are the temporal length and the spatial scale of the input video.
	The 3D CNN consists of two stacking 3D group convolutions with spatial-temporal kernel size five.
	We use causal convolution~\cite{carreira2018massively} for the temporal axis, so that information from future frames does not leak into the processing of the current frame, which is necessary for uni-directional compression modes such as low-delay mode with P frames (LDP). The temporally causal convolution-based aggregation network operates as an auto-regressive model, so that the semantic feature $\mathcal{S}^i = \mathcal{S}[i]$ of the $i$-th frame is aggregated from the information of all preceding frames.

	\textbf{Entropy coding.}
	First, a fully factorized parametric density model~\cite{balle2018variational} is adopted to fit the real distribution of $\mathcal{S}$. Then, adaptive arithmetic coding algorithm~\cite{langdon1984introduction} is applied to encode $\mathcal{S}$ into the semantic stream.

	\textbf{Difference-guided fusion module (DGFM).}
	As shown in Fig.~\ref{fig_csgfm} (b),
	DGFM inter-fuses the features from the two pathways of Sem-Enc by leveraging the attention mechanism.
	Additionally, we incorporate the region-adaptive dynamic convolution to extract the region-specific semantic representation with high flexibility.
	DGFM consists of three sub-modules:
	
	\textit{(1) Feature enhancement.}
	Given the frame feature $x_l'$ with channel number $C$ and the difference feature $d_l'$, we adopt two sequential convolutions denoted as ${Conv}$ to produce the attention map ${M}$, which reflects the poor-quality compressed regions and selectively enhances $x_l'$ by the element-wise multiplication operation, producing the enhanced feature $\tilde{x}_l$. \revisetext{Since we mainly emphasize spatial quality statistics rather than inter-channel dependencies, ${Conv}$ is instantiated as a group-wise convolution of group number eight to achieve a balanced trade-off between spatial modeling capability and computational cost.}
	
	Besides, the difference feature is also modulated by the enhanced frame feature, 
	$
	d_{l+1} = d_l' + {Conv1}(\tilde{x}_l),
	$
	aiming to provide more precise guidance information for the next stage.
	\revisetext{${Conv1}$ is a two-layer convolution network with ReLU non-linearity, aligning the channel number of the features between the two pathways.
	The kernel size of the convolution operators in ${Conv1}$ is set to one for simplicity, and we empirically find that using larger kernel size does not improve the performance.}

	\textit{(2) Adaptive kernel table generation.}
	Then, we leverage the global statistics of $\tilde{x}_l$ to estimate an adaptive kernel table {\footnotesize $\mathsf{AdaKT}$} $\in \mathbb{R}^{N \cdot C \cdot 5 \cdot 5  }$, which can be viewed as $N$ depth-wise convolutions with kernel size five and channel number $C$.
	$N$ is set to 10 for achieving a good trade-off between the modeling capability and the computational consumption. 
	Compared to the plain static CNN architectures, the kernels within {\footnotesize ${\mathsf{AdaKT}}$} are specified for the current frame $X^i$, thus of better representative capability.
	The global statistics are computed by an average pooling operation followed by a two-layer multi-layer perceptron (MLP)~\cite{hu2019squeeze}.
	
	\textit{(3) Region-adaptive feature extraction.}
	The kernels in {\footnotesize ${\sf{AdaKT}}$} are further allocated to the appropriated regions \revisetext{by considering the local contexts}, which upgrades the instance-level adaptiveness  to region-level adaptiveness.
	\revisetext{For simplicity,} we adopt a convolution ${Conv2}$ with kernel size three to dynamically generate a region-adaptive kernel index map {\footnotesize ${\mathsf{AdaKM}}$} $\in \mathbb{R}^{W \times H }$,
	which is then inflated into a spatial-adaptive kernel {\footnotesize $\mathsf{AdaK}$} $\in \mathbb{R}^{C \times W \times H \times 5 \times 5} $ by replacing each index with the kernel from {\footnotesize $\mathsf{AdaKT}$} using look-up-table ({\footnotesize $\mathsf{ LUT}$}) operation.
	Given {\footnotesize $ \mathsf{AdaK}$} and $\tilde{x}_l$, the frame feature $x_{l+1}$ is extracted by the depth-wise pixel-adaptive convolution operation~\cite{su2019pixel}, and then fed to the next stage.
	The {\footnotesize $\mathsf{LUT}$} operation can be approximated by a differential soft-argmax during training.

		\begin{figure}[!tp]
		\centering
		\includegraphics[width=8cm]{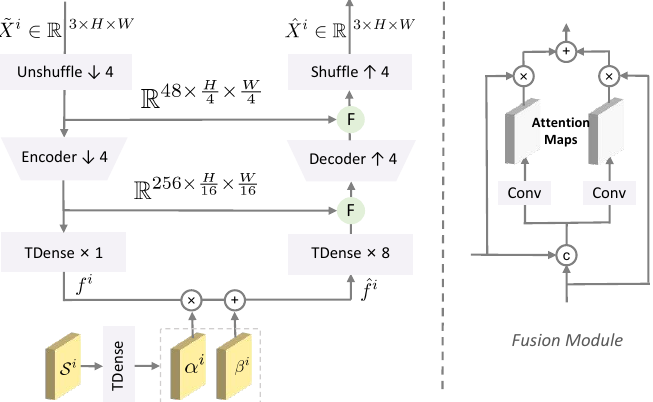}
		\caption {{Architecture of LFN}. $\downarrow 4$ and $\uparrow 4$ denote the operations that downscale and upscale the spatial scale of the input by four times, respectively. An attention-based feature fusion module ({F}) is adopted for adaptively fusing the information in the down and up pathways.
			$\oplus$, $\otimes$ and \textcopyright~denote the element-wise summation, multiplication, and channel-wise concatenation, respectively.
		}
		\label{fig_mgenet}
		\vspace{-5mm}
	\end{figure}

	\subsection{Latent Fusion Network (LFN)}

	 We adopt a UNet-style~\cite{ronneberger2015u} neural network to fuse the raw decoded video $\tilde{X}^i$ (from the video stream) and the semantic feature $\mathcal{S}$ (from the semantic stream) in the latent space.
	 
	\textbf{Architecture.}
	As shown in Fig.~\ref{fig_mgenet}, a pixel unshuffle operator~\cite{choi2020channel} first losslessly downscales the input video frame $\tilde{X}^i$ by a factor of four.
	Then, an encoder CNN is leveraged to further downscale the spatial resolution of the feature by a factor of four. The transformed features are then processed by a temporal modeling-enhanced dense block (TDense), producing the latent feature $f_i$. In this deep latent space (with $\frac{1}{16}$ smaller spatial resolution than the original input image), we exploit the received feature $\mathcal{S}$ to compensate the video semantics.
	\revisetext{Typically, we adopt the affine transformation to fuse the information in $\mathcal{S}$ and $f^i$, producing
	$
	\hat{f}^i = f^i \cdot \alpha^i + \beta^i,
	$
	where $\alpha^i$ and $\beta^i$ are the first and second half parts of the tensor transformed from the frame-wise semantic information $\mathcal{S}^i$ by a TDense block, $\alpha^i \circ \beta^i = {TDense}(\mathcal{S}^i)$ and $\mathcal{S}^i = \mathcal{S}[i]$.}
	Then, $\hat{f}^i$ is decoded into the video frame $\hat{X}^i$ by eight TDense blocks, which are followed by a lightweight decoder network and the pixel shuffle operation~\cite{shi2016real}.

	\textbf{Down-up pathway fusion (F).} Instead of utilizing the simple channel-wise concatenation operation for fusing the features of down and up pathways, we introduce the attention mechanism to adaptively preserve the information beneficial to downstream tasks. The motivation is that the original features may contain compression artifacts, and should be masked by an attention map.
	The attention maps are produced by a lightweight CNN ${Conv}$, which is a convolution of group size eight and kernel size three followed by a pointwise convolution~\cite{sandler2018mobilenetv2}.

	\textbf{TDense block.} The TDense block is modified from the vanilla Dense2D block~\cite{huang2017densely}\cite{zhang2020residual} by replacing the last spatial convolution layer with the multi-scale temporal convolutions, in which the first and the second half number channels are fed into the temporal causal convolutions~\cite{carreira2018massively} of kernel sizes three and five, respectively.
	\revisetext{As TDense blocks need to access the features of previous frames for temporally causal modeling, we incorporate a feature buffer memory within each block to store historical information.}

	\subsection{Optimization of Framework}\label{method_learning}
	The overall learning target of the proposed framework is to enforce the ultimately fused video $\hat{X}$ of both rich semantics and good visual quality, while minimizing the bitrate of the semantic stream, which can be formulated as follows,
	\begin{align}\label{eq:loss_rd}
		\mathcal{L} =    \underbrace{\alpha \mathcal{L}_{sem} }_{semantics}+ 
		\underbrace{\mathcal{L}_{lpips}  +  \mathcal{L}_{GAN}}_{visual~quality}
		+ \underbrace{R(\mathcal{S})}_{bitcost},
	\end{align}
	where $\mathcal{L}_{sem}$ denotes the proposed bottleneck-based semantic consistency loss, which will be elaborated in the next, $\alpha$ is the balancing weight.
	Following VQGAN~\cite{esser2021taming}, we introduce the combined $\mathcal{L}_{lpips}  +  \mathcal{L}_{GAN}$ item to keep good video visual quality at high compression rate, where $\mathcal{L}_{lpips}$ denotes the learned image perceptual loss~\cite{zhang2018unreasonable}, and $\mathcal{L}_{GAN}$ denotes the GAN loss.
	$R(\mathcal{S})$ represents the estimated bitcost of the semantic stream.
	\revisetext{In our loss function, neither task-specific designs nor video labels are involved, ensuring the optimization procedure of our framework is task-decoupled and label-free.
	}

	\textbf{Bitcost estimation}.
	The semantic feature $\mathcal{S}$ will be quantized and transformed into the bitstream for transmission by performing entropy coding.
	The bitcost of the semantic stream is estimated as $R(\mathcal{S}) = -log_2p(\mathrm{[}\mathcal{S}\mathrm{]})$, where $p$ is the fully factorized probability model proposed in~\cite{minnen2018joint}, and $\mathrm{[} \cdot \mathrm{]}$ denotes the rounding function.
	During the training procedure, we approximate the rounding operation by adding the uniform noise to $\mathcal{S}$, which follows~\cite{balle2018variational}.
	
	\textbf{Details of GAN}.
	The discriminator network architecture is same as that of PatchGAN~\cite{isola2017image}. LSGAN loss~\cite{mao2017least} is adopted as the training loss for its better stability than vanilla GAN~\cite{goodfellow2020generative}.

	\subsection{BottleNeck-based Semantic Consistency Loss}\label{method_ged}
	We employ the contrastive learning objective~\cite{he2020momentum}\cite{wang2022contrastive} to regularize the semantics of the decoded $\hat{X}$, which has been widely verified to be effective in previous unsupervised semantic learning methods.
	During the learning procedure, the feature of $\hat{X}$ is \textit{pulled} close to the original video, and \textit{pushed} far away from other videos in the semantic space, so that the most distinguishable and meaningful information is preserved in $\hat{X}$.
	Moreover, we adopt a one-bit map as the intermediate bottleneck representation, which suppresses the leakage of low-level texture information into the feature, and focuses on encoding the high-level semantic information such as object shapes.

\begin{figure}[!tp]
	\centering
	\includegraphics[width=8cm]{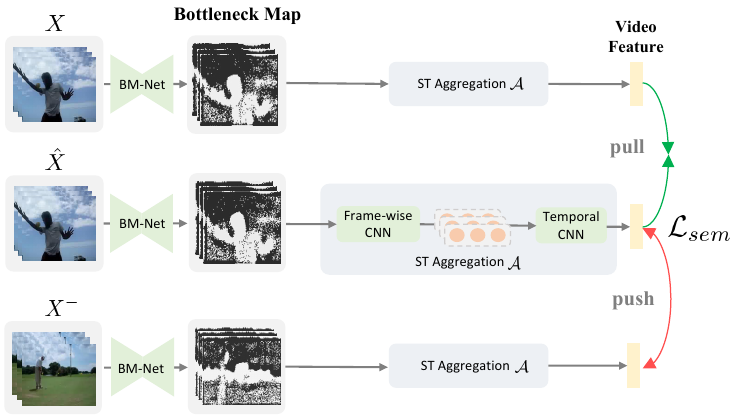}
	\caption {
	The contrastive learning procedure is applied to the decoded video $\hat{X}$, employing a one-bit bottleneck map (BM) representation. The original video $X$ serves as the positive sample, while $X^-$ represents a set of videos randomly sampled from the training dataset, utilized as negative samples for contrastive learning. For clarity, we illustrate with one negative video sample.
	}
	\vspace{-5mm}
	\label{fig_uvc_edge_optimize}
\end{figure}

	\textbf{Contrastive learning scheme.}
	The learning objective is to enforce the distance $d$ between the decoded video $\hat{X}$ and the original video $X$ is much smaller than that between $\hat{X}$ and any other video instance in the feature space, as shown in Fig~\ref{fig_uvc_edge_optimize}. This objective can be formulated as follows,
	\begin{equation}\label{eq_con_frame}
		\setlength{\jot}{7pt}
		\begin{split}
			\mathcal{L}_{sem} =&\log \frac{\exp(d( \mathcal{F} (\hat{X}) ,\mathcal{F}(X)) / \tau  )}
			{\sum_{i = 1}^{N} \exp(  d( \mathcal{F} (\hat{X}) ,\mathcal{F}(X^-_i)) / \tau)}, \\
			& ~~~~~~~~~\mathcal{F}  = \mathcal{A} \circ \operatorname{BM-Net}, \\
			& ~~~~~d(a,b) = \cos(\phi( a), \rho(b)),
		\end{split}
	\end{equation}
	where $\phi$ and $\rho$ denote two-layer MLPs with hidden number 512, $\cos$ denotes the cosine similarity, and $\tau$ is a temperature hyper-parameter adjusting the scale of cosine similarities.
	$X^-$ denotes the set of videos randomly sampled from the training dataset, serving as the negative samples for contrastive learning.
	$N$ denotes the negative sample number, which is enlarged to a large value such as 65536 by using the MoCo-style feature queue~\cite{he2020momentum}.
	Different from previous contrastive learning methods that reduces the feature dimension in the last stage, \textit{e.g.}, the last averaging pooling layer in ResNet, our method leverage the BM-Net to eliminate the redundant information in the 	very early stage, and then perform feature extraction with the spatial-temporal (ST) aggregation network $\mathcal{A}$.
	\revisetext{Compared to the semantic information extracted by a a pre-trained classification~\cite{yang2020discernible} or segmentation~\cite{duan2022jpd} network in previous works, our semantic representation extracted by the composite function $\mathcal{F}$ is purely emerged from unlabeled videos, jointly optimized with other components of the framework.}

	\textbf{Bottleneck representation with BM-Net.}
	We propose to use an information bottleneck map (BM) as an intermediate representation to enforce the discarding of hard-to-compress texture-related information.
	This not only saves the bit cost, but also suppresses the semantics-unrelated information being encoded, which may be noise to the downstream tasks.
	
	Specifically, we adopt a shallow network BM-Net to condense the original 24bit RGB video into a one-bit bottleneck map.
	The architecture of BM-Net follows an encoder-decoder scheme, where the encoder is implemented as two stacking residual blocks with a downsampling ratio of four, and the decoder part is symmetrical to that of the encoder. The tail of the BM-Net is a Sigmoid activation function, followed by a gumble-softmax~\cite{jang2016categorical} layer, so that the generated map is strictly 0/1 binarized.

	\textbf{Spatial-Temporal (ST) aggregation network $\mathcal{A}$.}
	Given the bottleneck map $M_i$ of the frame $X^i$, we adopt a two-layer MLP to compress each 16$\times$16 patch of $M^i$ as a 128-dimensional vector. These vectors form a feature map of size $\mathbb{R}^{ 128 \times \frac{H}{16} \times \frac{W}{16}}$.
	Then, a frame-wise CNN containing two residual blocks followed by an average pooling operation produces the frame-level feature.
	Finally, the frame-level descriptors are aggregated into a video-level representation by a lightweight temporal convolution network.

\section{Experiments}

\subsection{Video Datasets}
For the \textit{action recognition} task, we evaluate it on 5 large-scale video datasets, {UCF101}~\cite{soomro2012ucf101}, {HMDB51}~\cite{kuehne2011hmdb}, {Kinetics}~\cite{carreira2017quo}, {Something}~\cite{goyal2017something}, and {Diving48}~\cite{li2018resound}.
For the \textit{multiple object tracking (MOT)} task, we use the {MOT17}~\cite{milan2016mot16} and {SFU-HW-Tracks-v1}~\cite{tanaka2021sfu} datasets.
For the \textit{video object segmentation (VOS)} task, we use the {DAVIS2017}~\cite{pont20172017} dataset.

\textit{-Kinetics400.}
Kinetics400~\cite{carreira2017quo} is a very large-scale human action recognition dataset, which contains about 23K videos of 400 human action classes.
60K high-resolution videos (720p or 1080p) of the training set is used for the training of the framework.
Note that we do not use any annotation of these videos.
The validation set is adopted for the evaluation of our framework. 

\textit{-UCF101.}
UCF101~\cite{soomro2012ucf101} dataset contains 13,320 video clips that are divided into 101 action classes.
The first split of the official testing protocol is adopted for evaluation.

\textit{-HMDB51.}
HMDB51~\cite{kuehne2011hmdb}
dataset is a large collection of videos from various sources, which is composed of 6,766 video clips from 51 action categories.
The first split of the official testing protocol is adopted for evaluation.

\textit{-Something.}
SomethingV1 dataset~\cite{goyal2017something}
contains about 110K videos covering 174 fine-grained action categories.
This dataset is different from Kinetics/UCF101/HMDB51 in that it is motion-biased rather than scene-biased, which is suitable for evaluating the motion fidelity of the videos from our framework.
The validation set is adopted for evaluation.

\textit{-Diving48.}
Diving48~\cite{li2018resound} is with more than 18K video clips for 48 unambiguous diving classes.
The videos in this dataset are discriminated by the long-term temporal dynamics.
The validation set is adopted for evaluation.

\textit{-MOT17.}
MOT17 contains video sequences in unconstrained environments filmed with both static and moving cameras. There are 42 sequences (21 training, 21 test) with 33,705 frames.
We adopt the first half of each video in the training set of MOT17 for training the MOT tracker and the last half for validation.

\textit{-SFU-HW-Tracks-v1.}
SFU-HW-Tracks-v1~\cite{tanaka2021sfu} is an object tracking dataset extended from SFU-HW-Objects-v1 dataset~\cite{choi2021dataset}.
The raw frames available in this dataset allow more rigorous comparison of different video coding methods on MOT.

\textit{-DAVIS2017.}
The validation split of DAVIS 2017~\cite{pont20172017} consists of 30 videos with 59 objects, and the training split contains 60 videos with 138 objects.
The validation set is adopted for the evaluation of our framework on VOS task.

\textit{-HEVC Test Sequences.}
We also evaluate the video coding performance of our framework on HEVC Test Sequences~\cite{sullivan2012overview}, using the traditional visual quality metrics, \textit{i.e.}, Peak Signal-to-Noise Ratio (PSNR) and MS-SSIM~\cite{wang2003multiscale}.
\revisetext{We adopt the videos in HEVC Class B, C, D and
E with diverse resolutions from 416$\times$240 to 1920$\times$1080.}

\textbf{Dataset processing.}
During the \textit{training} procedure, we downsample the shortest side of the 60K training videos from Kinetics dataset to 256px for removing compression artifacts introduced by prior codecs on YouTube, which follows~\cite{wu2018video}.
For the evaluations of videos of \textit{action recognition} datasets, we also pre-downsample the shortest side of them to 256px and crop the video to the size 224$\times$224 before the coding procedure.
For the evaluation on \textit{MOT} task, we adopt the original MOT17 dataset of resolution 1920$\times$1080 because many tracking methods require high-resolution inputs. Since the videos in MOT17 are already compressed and may have slight impact on the evaluation, we also evaluate the MOT task on the SFU-HW-Tracks dataset, which contains uncompressed raw videos.
For \textit{VOS} task, we download the high-resolution version of DAVIS2017, which contains videos from 720p to 1080p, and then downsample them to 480p (854$\times$480), which is the input resolution of most VOS methods.

\subsection{Downstream Task Models}
For the \textit{action recognition} task, we adopt the following popular models as our baselines, \textit{i.e.}, {TSM}~\cite{lin2019tsm}, {TSN}~\cite{wang2018temporal}, {SlowFast}~\cite{feichtenhofer2019slowfast}, and {TimeSformer}~\cite{bertasius2021space}, including 2D CNN, 3D CNN and the recent Transformer architectures.
Both the pre-trained models and the sampling strategies are adopted from the MMAction2 framework~\cite{2020mmaction2}, except that we always use the simple-clip\&center-crop setting during evaluation. The reason is that the ``multiscale evaluation+online compression'' scheme costs in-affordable evaluation time.
For the \textit{MOT} task, we adopt {ByteTrack}~\cite{zhang2021bytetrack}, of which the pre-trained model is provided by the MMTracking framework~\cite{mmtrack2020}.
For the \textit{VOS} task, we adopt {XMem}~\cite{cheng2022xmem}, of which the pre-trained model is officially released by the authors.

\subsection{Evaluation Metrics}
We use bpp (bit per pixel) to measure the average number of bits used for one pixel in each frame.
For the \textit{action recognition} task, we adopt the Top1 accuracy as the performance indicator.
For the \textit{MOT} task, the metrics are more complicated, since the task should be evaluated from multiple aspects.
MOTA\greenuparrow (multiple object tracking accuracy)~\cite{kasturi2008framework} combines three metrics including FP\greendownarrow (false positives), FN\greendownarrow (false negatives) and IDs\greendownarrow (identity switches). MOTP\greenuparrow (multiple object tracking precision) denotes the precision of the output trajectories against ground truth. IDF1\greenuparrow~\cite{ristani2016performance} is the ratio of correctly identified detections over the average number of ground truth and computed detections. 
The indicators \greenuparrow/\greendownarrow mean the higher/lower the better.
For the \textit{VOS} task, the standard metrics Jaccard index $\mathcal{J}$, contour accuracy $\mathcal{F}$ and the average of $\mathcal{J}$ and $\mathcal{F}$ ($\mathcal{J\&F}$) are adopted.
We also report the contour recall $\mathcal{F}$-$Rec$.
We use Bjøntegaard Delta (BD)~\cite{bjontegaard2001calculation} algorithm to calculate the averagely improved performance or the saved bit cost percentage for all tasks.

\subsection{Implementation Details}

The Constant Rate Factor (CRF) of the video codec is randomly selected from \{51,47,43,39,35\}.
The encoder is set to low-delay mode with P frames (LDP).
$\alpha$ and $\lambda$ are set to 0.1 and 1, respectively.
For the contrastive learning loss item, the temperature $\tau$ is set to 0.2, the momentum update from one encoder to another is set to 0.999 and the negative feature queue size is set to 65536.
We use the Adam optimizer~\cite{kingma2014adam} by setting the learning rate as 0.0001, $\beta_1$ as 0.9 and $\beta_2$ as 0.999, respectively.
The resolution of training videos is 256 $\times$ 256 and the video clip length is 16.
The trained model can be applied to videos of any spatial resolution and temporal length, due to its fully convolutional characteristic.
The training iteration number is 500k.
\revisetext{When using PSNR and MS-SSIM for performance evaluation, we further fine-tune the model by 100k iterations with the rate-distortion loss, which is widely adopted by the recent learnable codecs~\cite{hu2021fvc}\cite{li2021deep}\cite{hu2022coarse}.}
The mini-batch size is 24.
The whole system is implemented with Pytorch~\cite{paszke2019pytorch} and takes about five days to train the model using eight Nvidia 2080Ti GPUs.
Since video standard reference software (such as VTM~\cite{VTM_code}) is extremely slow, taking approximately 1000 seconds (for VTM) to encode a single 1080p frame, using them to encode large-scale video understanding datasets is quite time-consuming. Therefore, we employ the fast VVC implementation, VVenC 1.5.0~\cite{VVenC}, as the traditional codec layer within our framework for AI task evaluation.
\revisetext{When our framework is evaluated using the PSNR and MS-SSIM metrics on the HEVC Class B/C/D/E datasets, which comprise about twenty videos, we utilize the video coding standard reference software, namely VTM 20.0~\cite{VTM_code} (for VVC) and HM 16.25~\cite{HM_code} (for H.265), as the traditional codec layer to compare with the performance of recent state-of-the-art learnable codecs.}



	\begin{figure*}[!hthp] 
		\centering
		\renewcommand\arraystretch{0.01}
		\newcommand{\widthscalefive}{0.235}
		\tabcolsep = 1mm
		\scalebox{1.0}{
			\hspace{-6mm}\begin{tabular}{cccc}
				\includegraphics[width=0.237 \textwidth]{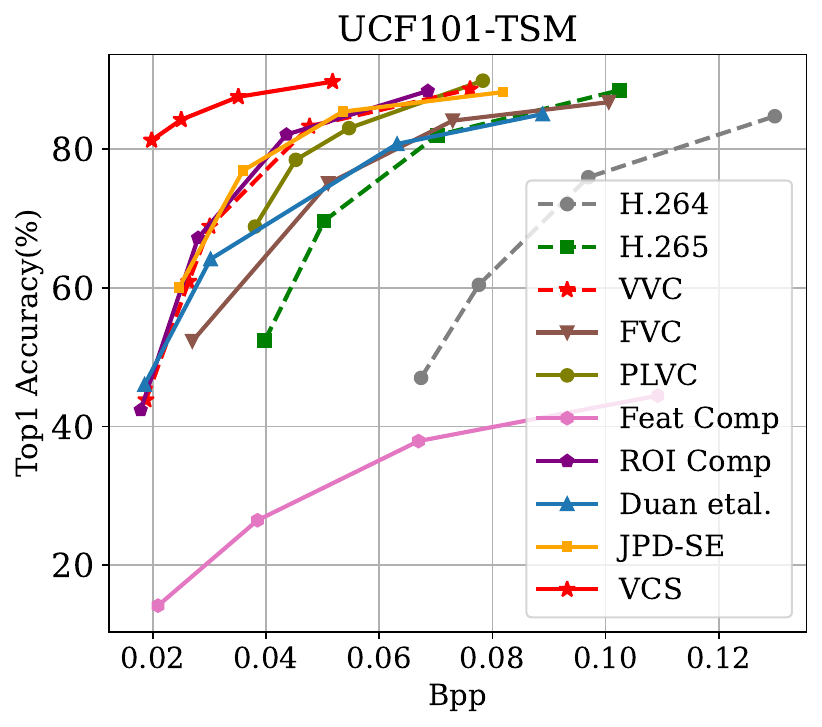} &
				\includegraphics[width=0.238 \textwidth]{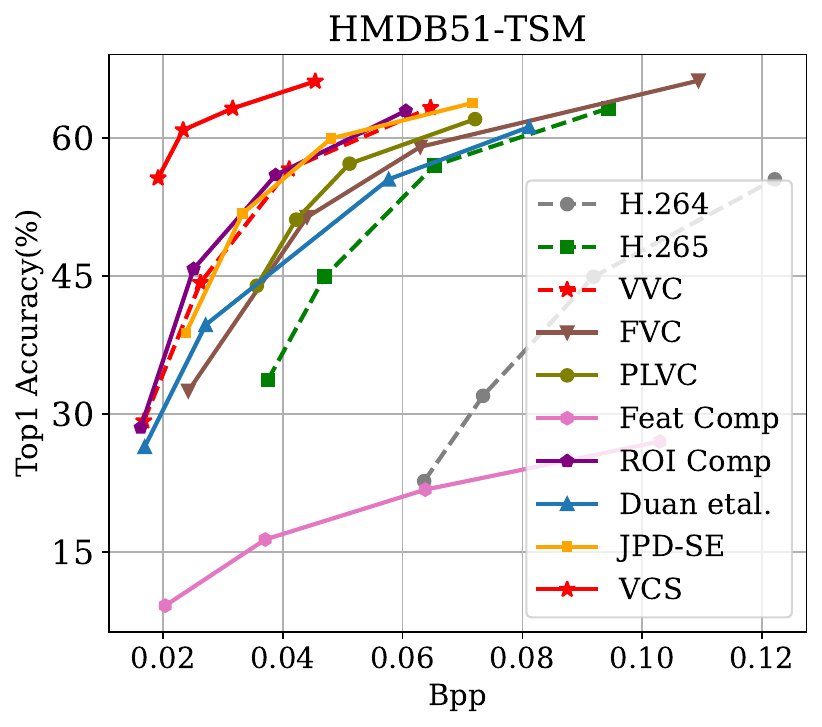} &
				\includegraphics[width=\widthscalefive \textwidth]{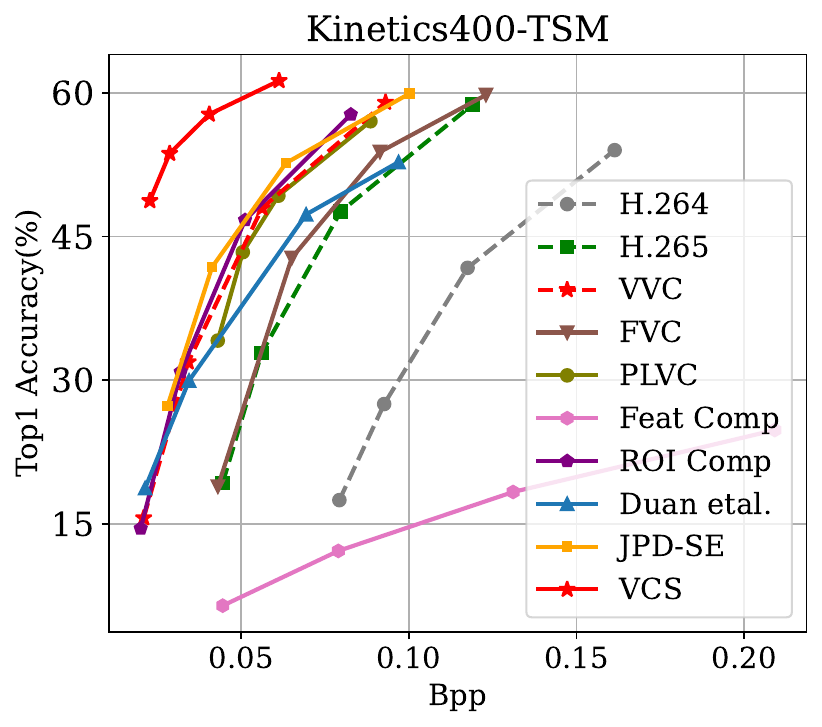} &
				\includegraphics[width=\widthscalefive \textwidth]{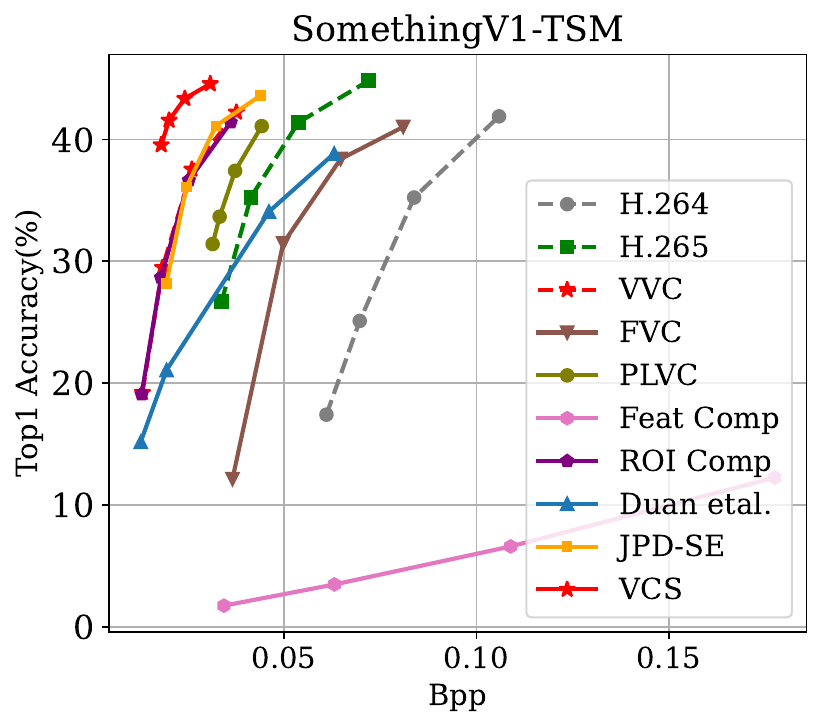} \\
				\includegraphics[width=\widthscalefive \textwidth]{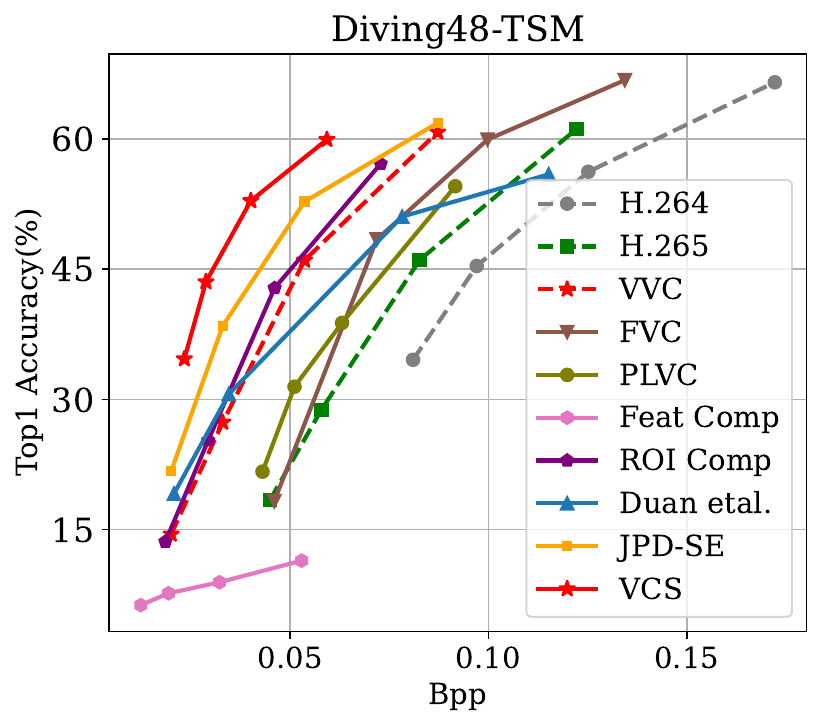}&
					\includegraphics[width=0.240 \textwidth]{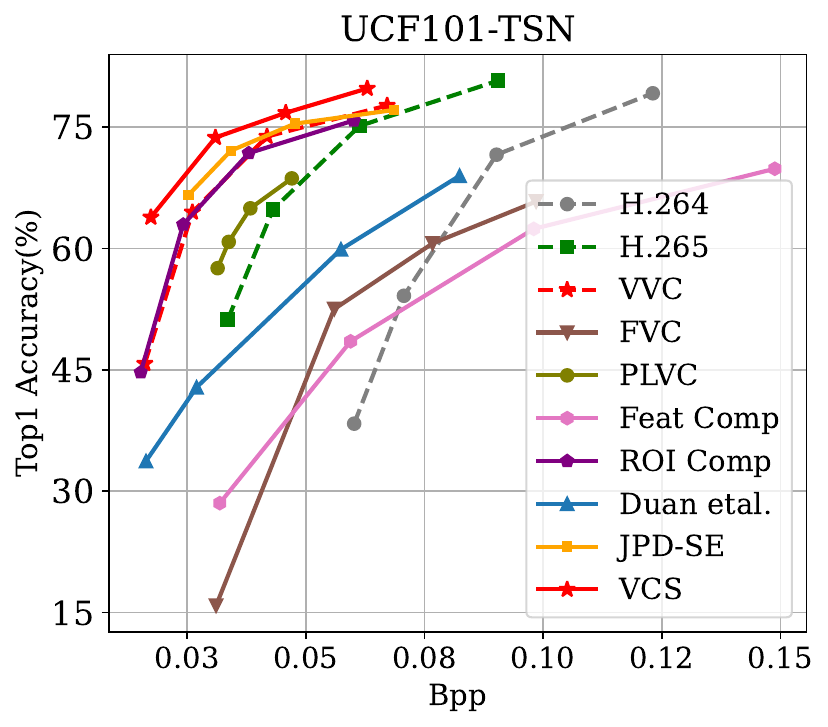} & 
				\includegraphics[width=\widthscalefive \textwidth]{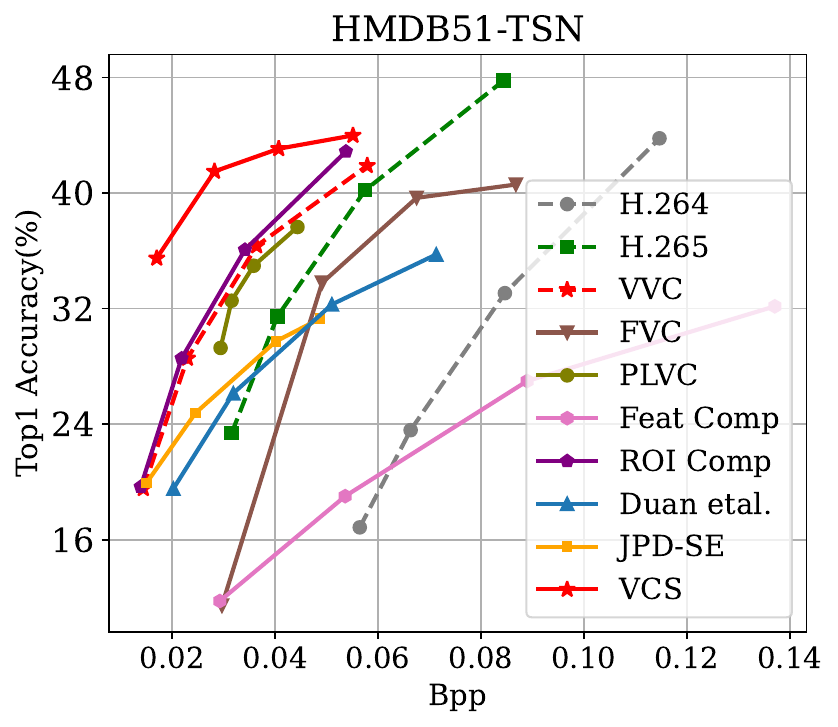} &
				\includegraphics[width=\widthscalefive \textwidth]{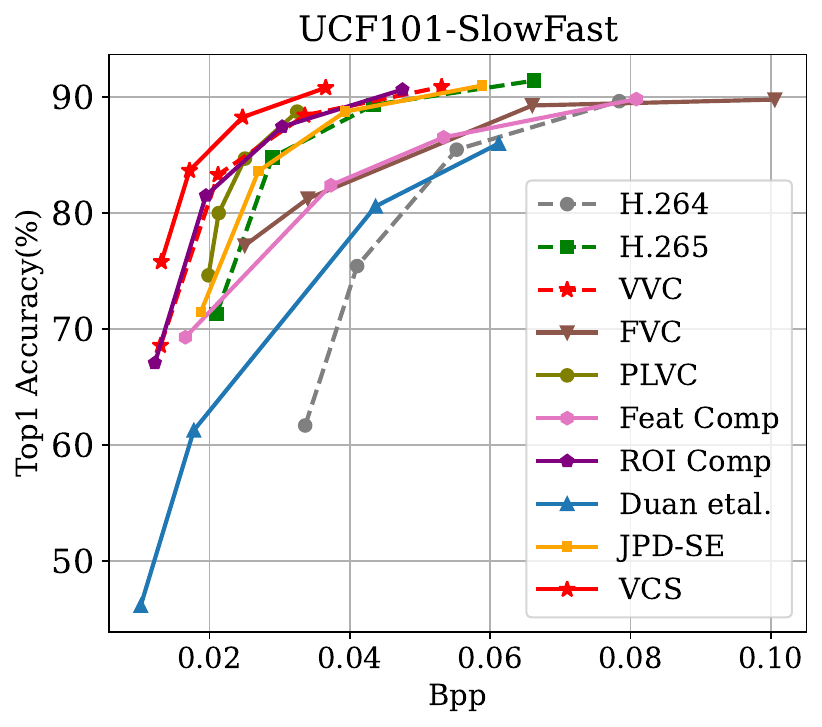} \\
				\includegraphics[width=\widthscalefive \textwidth]{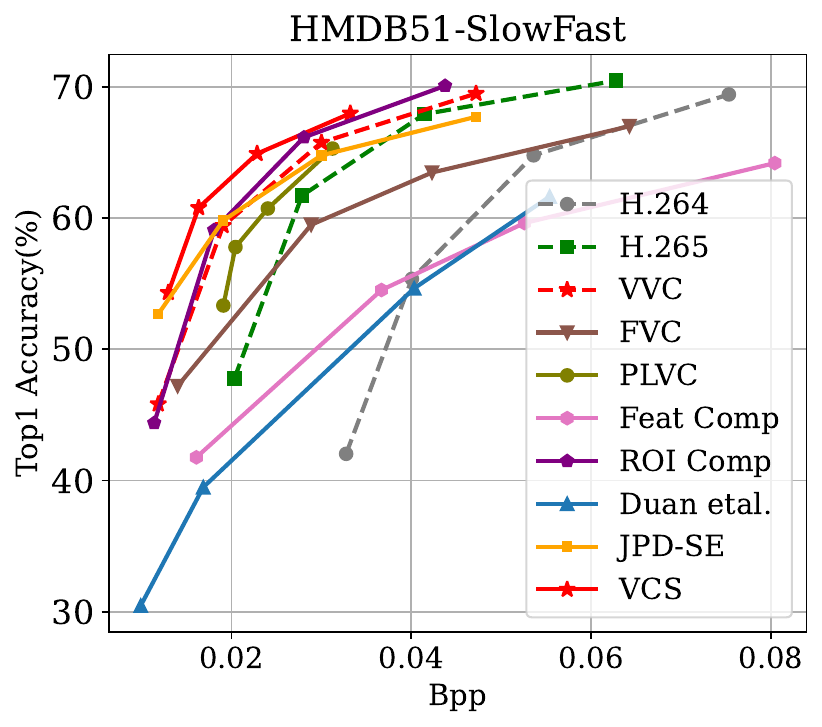} &
				\includegraphics[width=\widthscalefive \textwidth]{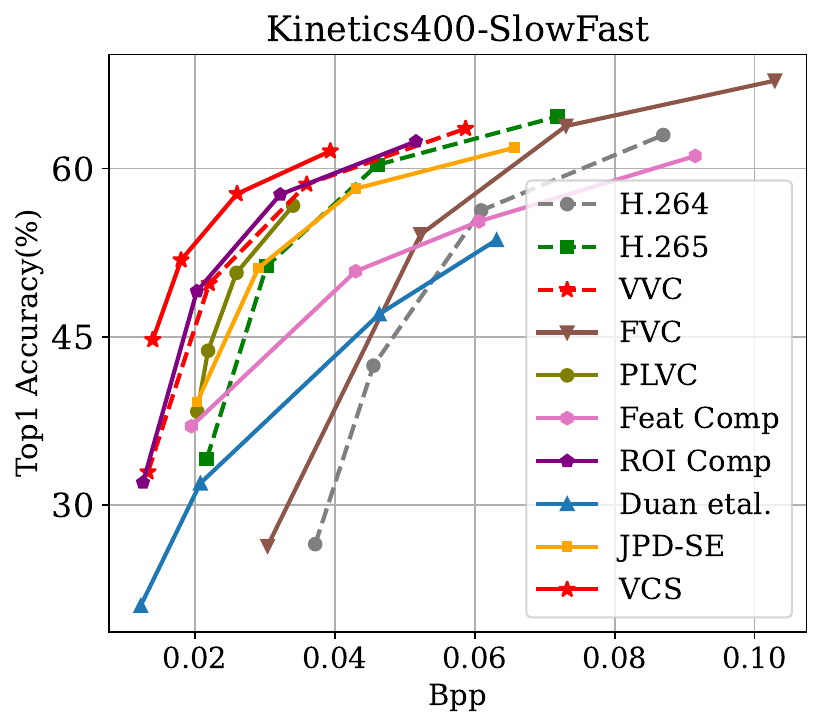} &
				\includegraphics[width=\widthscalefive \textwidth]{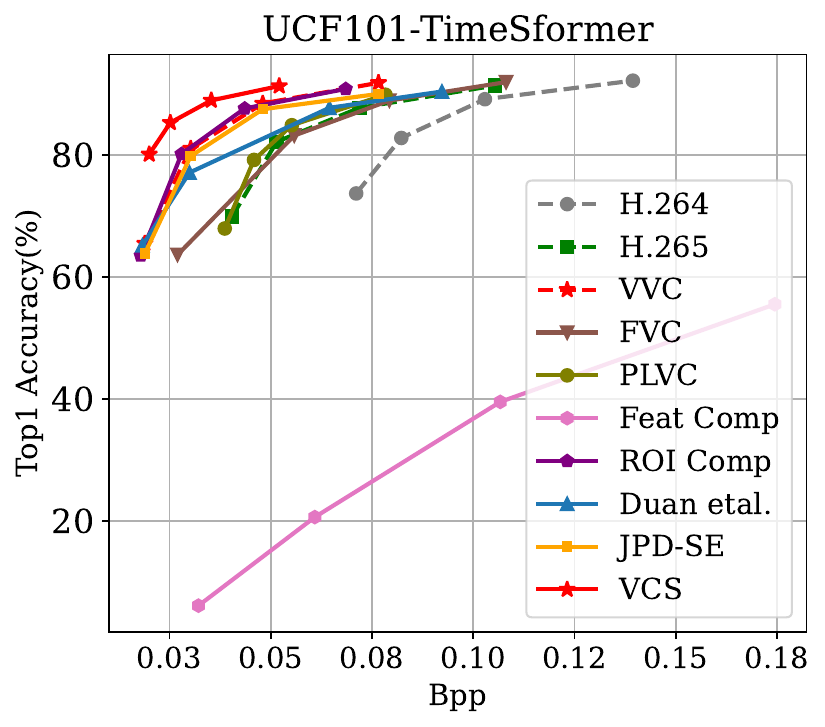}& 
				\includegraphics[width=\widthscalefive \textwidth]{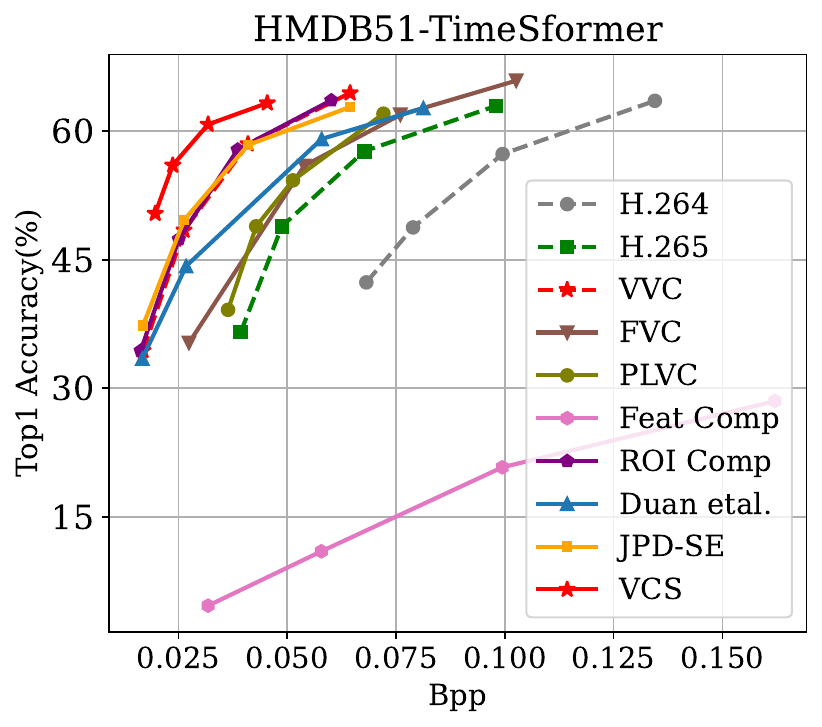}
				 \\
			\end{tabular}
		}
	\vspace{-2mm}
		\caption{
			RP curves on action recognition task.
			The plot titles are in the format \{\textit{Dataset}\}-\{\textit{Action model}\}.
		}
		\vspace{-1mm}
		\label{fig:lbuv_ar_blind}
	\end{figure*}

\begin{table*}[!thbp]
	
	\centering
	\renewcommand\arraystretch{1.05}
	\tabcolsep=0.8mm

	\begin{tabular}{|cc c||Hcc|cc|ccccc|c|}
	\hline
	\multirow{3}{*}{\tabincell{c}{\textit{Model}} } & \multirow{3}{*}{\tabincell{c}{\textit{Dataset}} }& \multirow{3}{*}{\tabincell{c}{\textit{@Bpp}} }&\multicolumn{11}{c|}{\textit{Top1 accuracy (\%)}}\\
	\cline{4-14} 
	& &  &{H.264}&  {H.265~\cite{sullivan2012overview}}&{VVC~\cite{bross2021overview}}&{FVC~\cite{hu2021fvc}}&{PLVC~\cite{yang2021perceptual}}&\tabincell{c}{Feat\\ Comp~\cite{chen2019toward}}&\tabincell{c}{ROI\\ Comp~\cite{cai2021novel}}&\tabincell{c}{Duan~\\\etal~\cite{duan2020video}}
	&{JPD-SE~\cite{duan2022jpd}} &{VCS}  &{\textcolor{gray}{Original}}\\
	\hline 
	\multirow{4}{*}{TSM}	& UCF101 & 0.04&- & 52.70& 76.97 &64.61&71.40&14.14 & 78.63& 69.06&78.85 &\textbf{88.19} & \textcolor{gray}{93.97} \\ 
	
	& HMDB51 & 0.04&22.68  & 36.64&55.72&47.54&48.67&9.15&56.33&46.34&55.48&  \textbf{64.98} & \textcolor{gray}{72.81}\\
	
	& Kinetics400 & 0.05 & 17.47& 25.68& 43.41 &26.39&42.75&6.44&45.76&37.67&46.08&\textbf{59.38} & \textcolor{gray}{70.73} \\
	
	& Something & 0.02 & 17.41&33.66&42.23&17.06&38.84&1.73&41.40&46.34&42.75&  \textbf{44.65} & \textcolor{gray}{47.85}\\
	
	& Diving48& 0.05 &-&22.48&42.75&22.94&30.13&6.29&44.94&51.52&50.27& \textbf{56.54}& \textcolor{gray}{75.99}\\
	
	\arrayrulecolor{gray}\cdashline{1-14}[5pt/3pt]
	\multirow{2}{*}{TSN}& UCF101  & 0.04 &38.33&60.39&72.74&28.96&65.66&34.40&72.14&50.12&73.51& \textbf{75.56} &\textcolor{gray}{87.31} \\ 
	
	& HMDB51 &  0.04& 16.86&31.06&37.26&23.31&36.28&14.96&38.11&46.34&29.68 &\textbf{42.98} &\textcolor{gray}{55.69}  \\ 
	
	\arrayrulecolor{gray}\cdashline{1-14}[5pt/3pt]
	\multirow{3}{*}{SlowFast}&UCF101&  0.03 &-&85.13&86.93&79.43&87.38&69.28&87.25&70.38&84.86 & \textbf{89.39}  &\textcolor{gray}{94.92}\\ 
	
	& HMDB51 & 0.02&- & 47.78&59.95&52.15&56.33&41.76&60.44&41.51&60.22 & \textbf{63.09} &\textcolor{gray}{72.03} \\ 
	
	& Kinetics400 & 0.03 &-&50.80&54.79&26.35&53.70&37.03&56.08&37.38&51.59& \textbf{58.87} &\textcolor{gray}{68.87}\\ 
	
	\arrayrulecolor{gray}\cdashline{1-14}[5pt/3pt]
	\multirow{2}{*}{TimeSformer}& UCF101 & 0.05 & -&80.53&88.68&79.25&81.71&6.16&88.45&83.27&87.67&\textbf{91.00} & \textcolor{gray}{95.43}\\ 
	
	& HMDB51 & 0.04& -&37.45&57.83&45.11&44.59&4.64&58.21&50.55&57.79& \textbf{62.27} & \textcolor{gray}{71.44}\\
	\boldhline
\end{tabular}
\vspace{-1mm}
\caption{
	Comparison of different video coding methods on action recognition task.
	``Original'' denotes the performance with original videos, which can be viewed as the upper bound on the low-bitrate settings.
}
\vspace{-4mm}
\label{tab:bdbr_ar}
\end{table*}

\subsection{Benchmark Video Coding Methods}
\revisetext{
Since our benchmark encompasses eight large-scale video datasets, three video analysis tasks, and several network backbones, evaluating supervised video coding for machine (VCM) methods \cite{choi2022scalable11}\cite{liu2021semantics} for this benchmark is quite challenging due to the significant time consumption of the dataset/task/backbone-specific training procedures. Therefore, we primarily assess traditional codecs, including \textit{H.264}, \textit{H.265}, and \textit{VVC}, as well as learnable codecs such as \textit{FVC}\cite{hu2021fvc} and \textit{PLVC}\cite{yang2021perceptual}, along with video-adapted unsupervised VCM methods, namely feature compression (\textit{Feat Comp})\cite{chen2019toward}, region-of-interest-based video compression (\textit{ROI Comp})\cite{cai2021novel}, {\textit{Duan \etal}}\cite{duan2020video}, and \textit{JPD-SE}\cite{duan2022jpd}. The details of these methods are provided below.
}

\revisetext{
-\textbf{\textit{H.264}}\cite{schwarz2007overview}/\textbf{\textit{H.265}}\cite{sullivan2012overview}. We utilize the FFmpeg software~\cite{tomar2006converting} along with the x264~\cite{x264_web}/x265~\cite{x265_web} library, where the preset is set to \textit{veryfast}, the tune is \textit{zerolantency}, and GOP is configured to 10, following the approach in~\cite{lu2020end}\cite{hu2021fvc}. The CRF value is chosen from the set \{47, 43, 39, 35\} to maintain a compressed video bitrate below 0.1bpp.
}

\revisetext{
-\textbf{\textit{VVC}}\cite{bross2021overview}. We utilize the VVenC 1.5.0 software\cite{VVenC} due to its rapid coding speed. We use the low-delay P frame mode (LDP) with the same GOP size and CRF value as H.264/H.265 to ensure a fair comparison. The settings for the VVC codec in our framework also adhere to these configurations.
}

\revisetext{
- \textbf{\textit{FVC}}\cite{hu2021fvc}. This is an exemplar deep video compression method that performs motion compensation and residual coding in the feature space. We train the low-bitrate FVC models using the officially released training codes\cite{fvc_code}.
}

\revisetext{
- \textbf{\textit{PLVC}}\cite{yang2021perceptual}. This learnable deep video compression method is focused on perceptual quality. It is a GAN-based approach that emphasizes object structure and video temporal smoothness. We train the low-bitrate PLVC models using the officially released training codes\cite{plvc_code}.
}

\revisetext{
- \textbf{\textit{Feat Comp}}\cite{chen2019toward}. This is a feature-based compression baseline method adapted from Chen~\etal~\cite{chen2019toward}. We encode the video features extracted by the task-specific backbone network and then utilize the VVC codec to compress these features. To ensure a fair comparison, we employ the same settings for the VVC codec as in the \textit{VVC} baseline. The feature layers for compression are meticulously tuned for various backbone network architectures to achieve optimal performance.
}

\revisetext{
-  \textbf{\textit{ROI Comp}}\cite{cai2021novel}. This is a region-of-interest (ROI)-based video coding baseline method adapted from Cai~\etal~\cite{cai2021novel}. We employ an object detection model~\cite{lin2017focal} to predict object regions and guide the VVC codec to allocate more bits to these regions.
The settings for the VVC codec are also the same as in the \textit{VVC} baseline for a fair comparison.
}

\revisetext{
- \textbf{\textit{Duan~\etal}}\cite{duan2020video}. This method, introduced by Duan~\etal, enhances video compression through keypoints. It utilizes sparse motion patterns~\cite{siarohin2019animating} to model motion between consecutive frames and employs a traditional HEVC codec to encode residual details. In our re-implementation, we substitute the HEVC codec with the more advanced VVC codec, for a fair comparison with our approach.
}

\revisetext{
- \textbf{\textit{JPD-SE}}~\cite{duan2022jpd}. This method, known as Joint Pixel-Domain Semantic Enhancement, enhances Image Coding for Machine (ICM) using segmentation maps. It involves extracting the segmentation map of the image on the encoder side, transmitting the map after compressing it with hand-crafted methods, and leveraging the map to enhance image structure on the decoder side. In our re-implementation, we adapt this method to the video domain by introducing temporal modeling. Specifically, we employ a small video compression model to compress the frame-wise segmentation maps, effectively eliminating inter-frame redundancies.
}

\vspace{-2mm}
\subsection{Results}

\textbf{Action recognition.}
As shown in Tab.~\ref{tab:bdbr_ar}, our framework substantially outperforms other benchmarks by a large margin.
For example, our framework saves the bit cost of the advanced traditional VVC codec by over 50\% ($\triangle$Bit\%\greendownarrow~ in Tab.~\ref{tab:bdbr_ar}) on UCF101, HDMB51, and Kinetics400 datasets, when using the TSM network to perform recognition.
The improvements with SlowFast and TimeSformer models are smaller than TSM,  because 3D CNN and Transformer are more robust to compression artifacts than 2D CNNs, \textit{i.e.}, less performance degradation when inputting compressed videos by traditional codecs.
Taking the HMDB51 dataset compressed by VVC as an example, the Top1 accuracy of SlowFast, TimeSformer, and TSM is decreased by 13\%, 32\%, and 40\% at 0.02bpp level, respectively.
We further give detailed analysis to the performance of other coding methods.

\revisetext{
As for learnable deep video compression methods, the performance of FVC is superior or comparable to the H.265 codec for most action model-dataset pair settings. For instance, at the 0.04bpp bitrate level, FVC achieves 52.15\% and H.265 achieves 47.78\% in terms of Top1 accuracy on the HMDB51-SlowFast setting. Furthermore, PLVC consistently outperforms FVC across all settings due to its additional constraint on the spatial-temporal structure of video, \textit{i.e.}, its discriminator is conditioned on the temporal and spatial features of the compressed video.
}

\revisetext{
Regarding the unsupervised VCM methods, we observe that the feature coding method Feat Comp exhibits poor performance. Despite the advanced VVC codec used for compressing the features, its performance is notably worse than that of the older H.264 codec. For instance, at the 0.075bpp bitrate level, it only achieves a Top1 accuracy of 16.12\% on the HMDB51-TimeSformer setting, while H.264 achieves 47.38\%.
The performance degradation phenomenon is also observed in the keypoint-based VCM method Duan~\etal, which might be attributed to the fact that the simple sparse motion modeling cannot effectively handle the complex motions present in action videos from diverse environments.
Furthermore, the ROI coding paradigm (ROI Comp) consistently enhances the performance of the baseline VVC codec across most settings. For example, it results in performance improvements of 1.66\%, 1.29\%, and 0.38\% on UCF101-TSM@0.04bpp, Kinetics400-SlowFast@0.03bpp, and HMDB51-TimeSformer@0.04bpp, respectively. This demonstrates that semantic priors such as object distribution can be advantageous for video analysis.
}

\revisetext{
Furthermore, we notice that JPD-SE achieves comparable performance to ROI Comp on most settings. However, it demonstrates notably stronger performance on the Diving48 dataset, achieving a Top1 accuracy of 51.52\% compared to 44.89\% at the 0.05bpp bitrate level.
This performance difference can be attributed to the segmentation map enhancement strategy employed by JPD-SE, which refines human body shapes. This enhancement is particularly effective for fine-grained diving sport actions in the Diving48 videos, where these actions are accurately discerned through dynamic changes in the human body.
In summary, incorporating semantic priors like object location and shape into the video coding process effectively enhances action recognition performance. Nonetheless, these hand-crafted or supervised semantic priors yield only marginal improvements. Furthermore, they exhibit unstable performance across various target datasets. In contrast, our approach's data-emerged semantic prior, learned from large-scale unlabeled videos through self-supervised learning, excels in terms of generalization. This results in our method consistently achieving remarkable performance across all datasets.
}

We also illustrate the rate-performance (RP) curves in Fig.~\ref{fig:lbuv_ar_blind}, in which Top1 accuracy and bit per pixel (bpp) are adopted as the performance and bit cost metrics.
The performance of our framework is consistently superior to all other benchmark dataset on different action recognition models and datasets.

Although our framework has achieved impressive performance by being optimized with task-agnostic self-supervised learning objective, we expect better results after jointly fine-tuning it with downstream models and video labels.
As shown in Tab.~\ref{tab:bdbr_customized}, the recognition accuracy is further improved by 2.10\% and 1.33\% at the 0.02bpp level on UCF101 and HMDB51, respectively.
\begin{table}[!t]
	\centering
	\renewcommand\arraystretch{1.05}
	\tabcolsep=1.2mm
	\begin{tabular}{|c||cc|cc|}
		\hline
		\textit{\multirow{2}{*}{Method}} &\multicolumn{2}{c|}{\textit{\textit{Top1(\%) on UCF101@}}}	&\multicolumn{2}{c|}{\textit{\textit{Top1(\%) on HMDB51@}}} \\
		& 0.02bpp  & 0.03bpp & 0.02bpp  & 0.03bpp   \\
		\hline
		{VCS} 		& 85.04& 89.35 & 63.62 & 66.98  \\ 
		\hline
		{VCS+Label} & {87.14}(+2.10) & {90.61}(+1.26) & {64.95}(+1.33) & {67.73}(+0.75) \\ 
		\hline
	\end{tabular}
	\vspace{-1mm}
	\caption{
		Improving the VCS framework by fine-tuning it with the downstream task dataset labels.
		The action model is SlowFast.
	}
	\label{tab:bdbr_customized}
\end{table}
\begin{table}[!t]
	
	\centering
	\renewcommand\arraystretch{1.05}
	\tabcolsep=0.6mm
	
	\begin{tabular}{|cc ||cc|cc|}
		\hline
		\multirow{2}{*}{\tabincell{c}{\textit{Model}} } & \multirow{2}{*}{\tabincell{c}{\textit{Dataset}} }
		& \multicolumn{2}{c|}{\tabincell{c}{\textit{ VCS$_\mathrm{H.264}$ \textit{v.s.} H.264}}} & \multicolumn{2}{c|}{\tabincell{c}{\textit{ VCS$_\mathrm{H.265}$ \textit{v.s.} H.265}}}\\
		&&\textit{$\triangle$Top1(\%)\greenuparrow}&\textit{$\triangle$Bit(\%)\greendownarrow}&\textit{$\triangle$Top1(\%)\greenuparrow}&\textit{$\triangle$Bit(\%)\greendownarrow}\\
		\hline 
		\multirow{4}{*}{TSM}	& UCF101 & 13.61 &-37.92& 7.38 &-31.24\\ 
		
		& HMDB51 &  20.08& -56.40& 3.83 &-13.67\\
		
		& Kinetics400 &21.15& -45.51& 14.70& -40.76 \\
		
		& Something & 6.34& -20.64& 2.40 &-15.62\\
		
		& Diving48&  6.02 &-17.48& 10.29 &-22.35\\
		
		\arrayrulecolor{gray}\cdashline{1-6}[5pt/3pt]
		\multirow{2}{*}{TSN}& UCF101  & 9.05& -17.61& 3.54 &-11.61 \\ 
		
		& HMDB51 &  11.85 &-34.76& 5.09 &-20.8  \\ 
		
		\arrayrulecolor{gray}\cdashline{1-6}[5pt/3pt]
		\multirow{3}{*}{SlowFast}&UCF101&   7.24& -19.21& 1.90 &-16.43\\ 
		
		& HMDB51 &  4.72 &-11.50& 1.20 &-8.63\\ 
		
		& Kinetics400 &  8.52 &-14.98& 4.15 &-15.77\\ 
		
		\arrayrulecolor{gray}\cdashline{1-6}[5pt/3pt]
		\multirow{2}{*}{TimeSformer}& UCF101& 2.13& -12.89& 5.35 &-26.97\\ 
		
		& HMDB51 &  6.22& -23.67& 7.50& -24.88\\
		\boldhline
	\end{tabular}
\vspace{-1mm}
	\caption{
		Effectiveness of VCS framework on H.264 and H.265 codecs in terms of action recognition task.
	}
	\vspace{-5mm}
	\label{tab:perf_improve_action}
\end{table}
Furthermore, to demonstrate that our method is orthogonal to the adopted traditional codec, we apply it to the two widely-used commercial codecs, H.264 and H.265.
It is also worthwhile to mention that we directly deploy the VCS model trained for VVC codec to these two codecs without any codec-specific adaption procedure.
We calculate the averagely improved recognition accuracy (\textit{$\triangle$Top1(\%)\greenuparrow}) and the averaged saved bitcost percentage (\textit{$\triangle$Bit(\%)\greendownarrow}) by the Bjøntegaard delta algorithm~\cite{bjontegaard2001calculation}.
As shown in Tab.~\ref{tab:perf_improve_action}, our framework also substantially improves H.264 and H.265 on all settings.

\begin{figure*}[!thbp]
	\centering
	\scriptsize
	\tabcolsep=1mm
	\newcommand{\widthscalefive}{17.0cm}
	\newcommand{\myclipl}{1cm}
	\includegraphics[width=\widthscalefive]{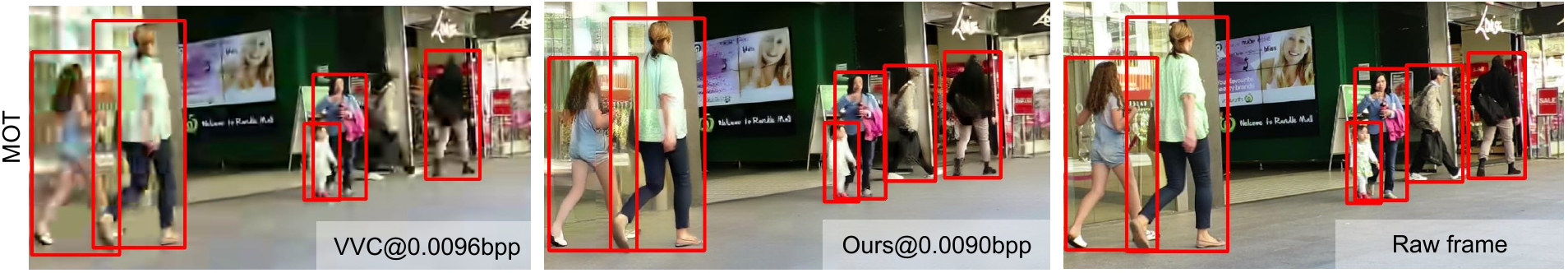}
		\vspace{-2mm}
\end{figure*}

	\begin{figure*}[!thbp]
	\centering
	\scriptsize
	\tabcolsep=1mm
	\newcommand{\widthscalefive}{17.0cm}
	\newcommand{\myclipl}{1cm}
	\includegraphics[width=\widthscalefive]{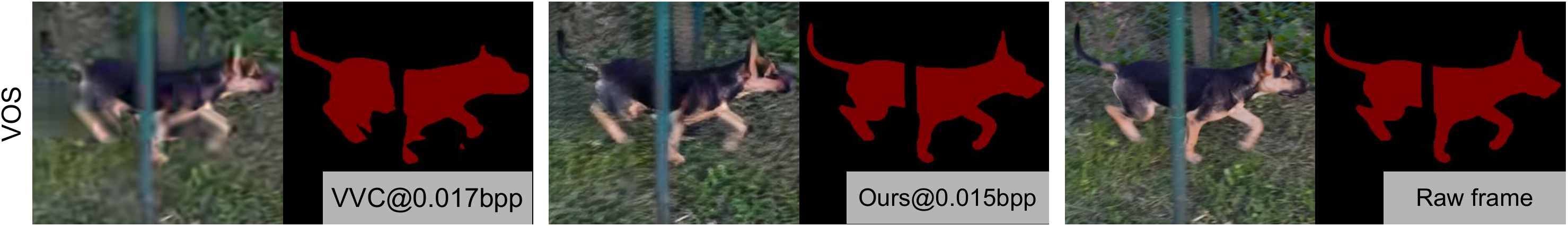}
	\vspace{-1mm}
		\caption{
		Visualization of the MOT and VOS task results.
		The \textit{MOT17-09} 207\textit{th} frame and \textit{DAVIS2017-libby} 2\textit{th} frame are illustrated.
	}
\label{fig:qua_mot_vos_blind}
	\vspace{-2mm}
\end{figure*}

\begin{figure*}[!thhp] 
		\centering
		\newcommand{\widthscalefive}{0.235}
		\tabcolsep = 1.2mm
		\scalebox{1}{
			\hspace{-2mm}\begin{tabular}{cccc}
				\includegraphics[width=\widthscalefive \textwidth]{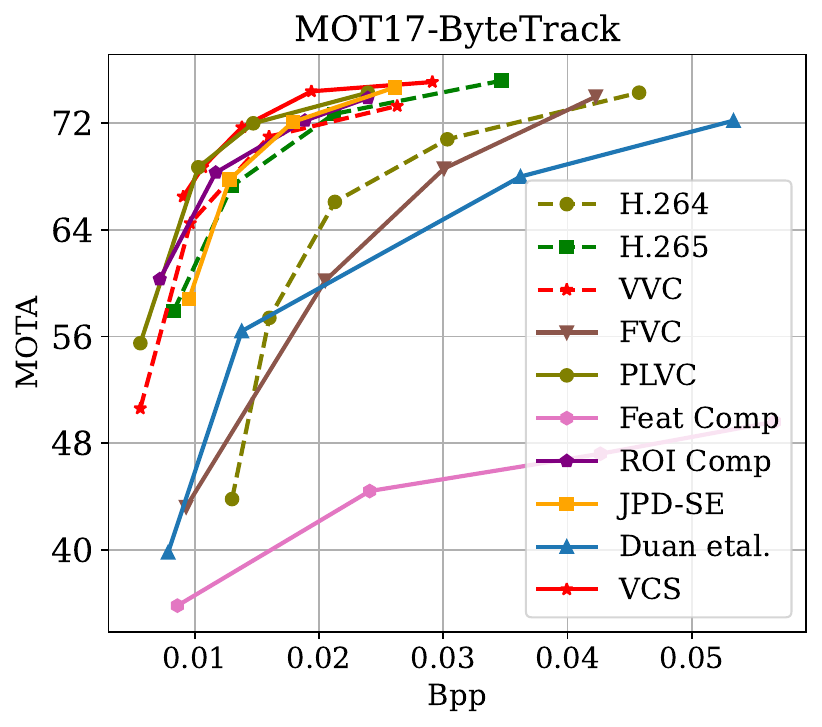} &
				\includegraphics[width=\widthscalefive \textwidth]{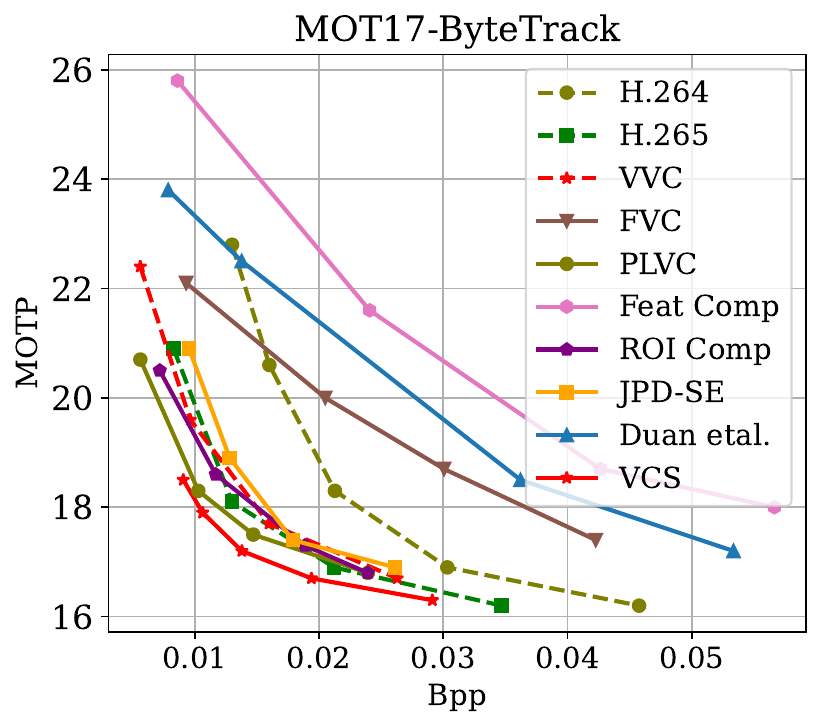} &
				\includegraphics[width=\widthscalefive \textwidth]{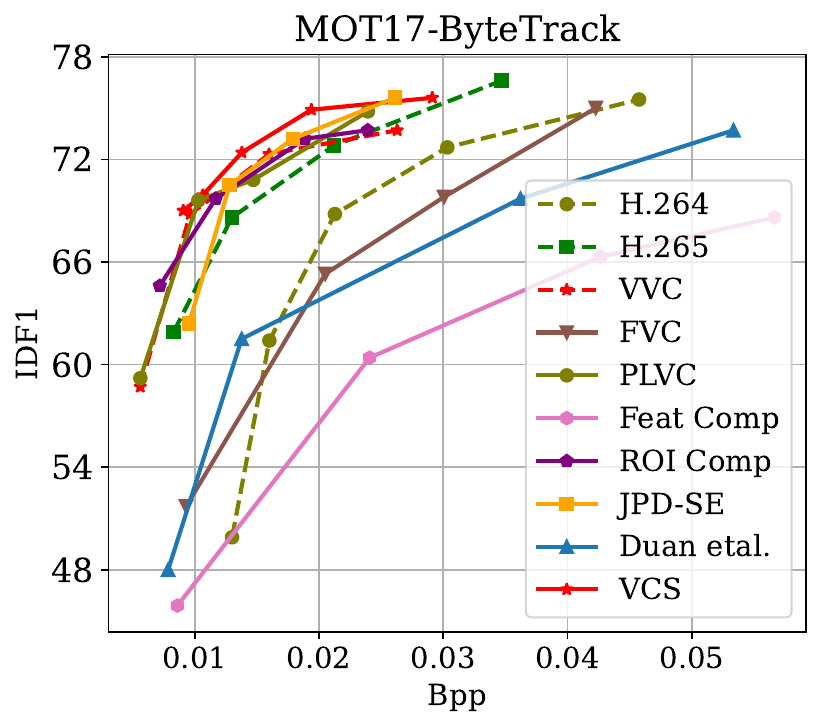} &
				\includegraphics[width=\widthscalefive \textwidth]{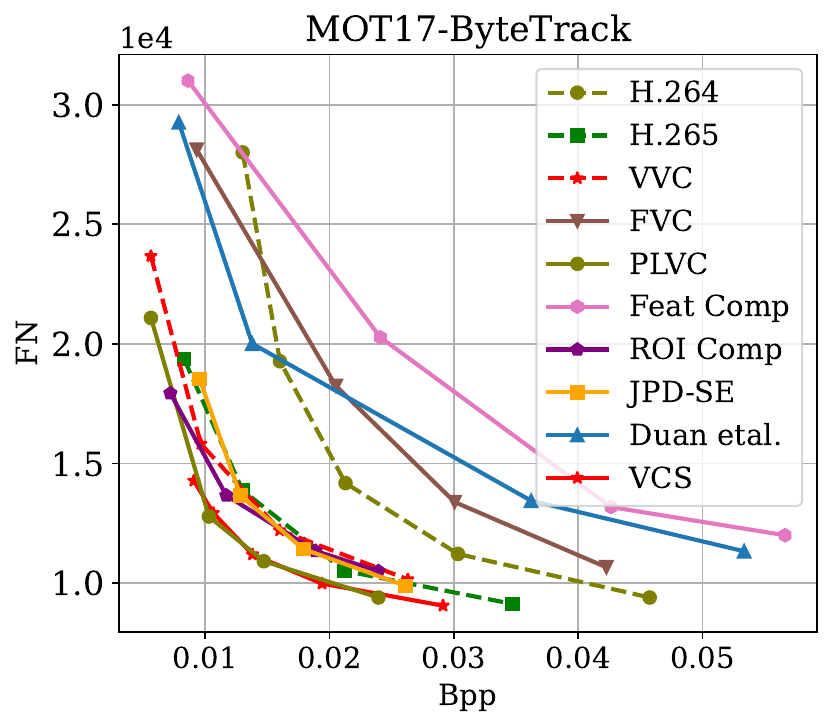} 
			\end{tabular}
		}
	\vspace{-6mm}
	\end{figure*}

\begin{figure*}[!hthp] 
	\centering
	\newcommand{\widthscalefive}{0.232}
	\tabcolsep = 1.2mm
	\scalebox{1}{
		\hspace{-2mm}\begin{tabular}{cccc}
			\includegraphics[width=\widthscalefive \textwidth]{\detokenize{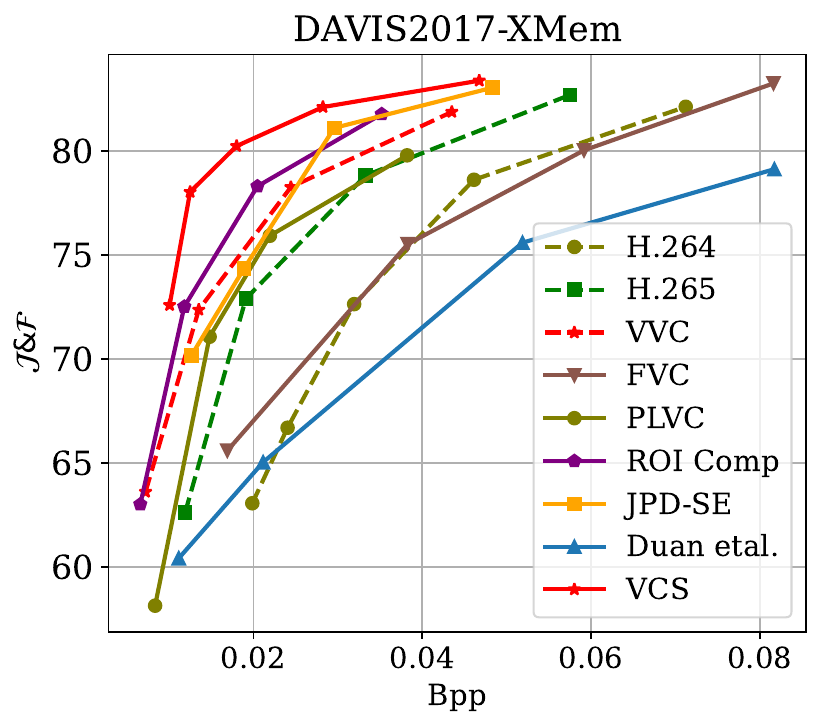}} &
			\includegraphics[width=0.24 \textwidth]{\detokenize{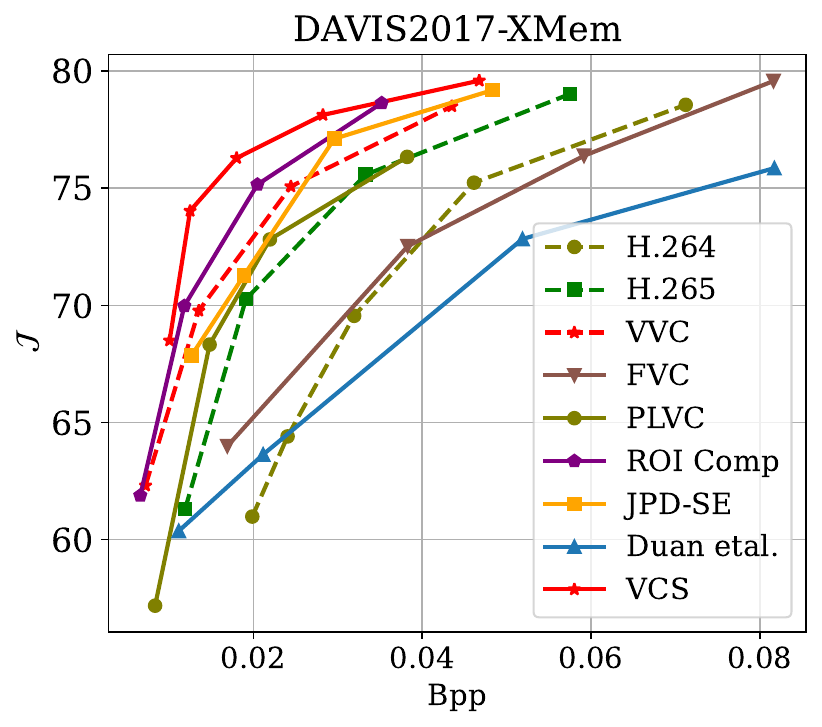}} &
			 \includegraphics[width=\widthscalefive \textwidth]			   {\detokenize{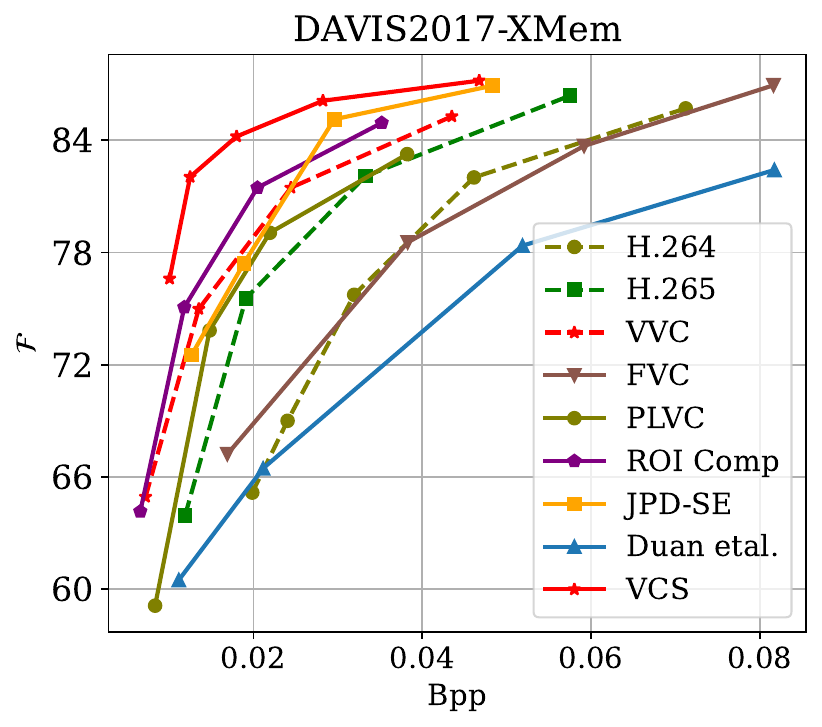}} &
			\includegraphics[width=\widthscalefive \textwidth]{\detokenize{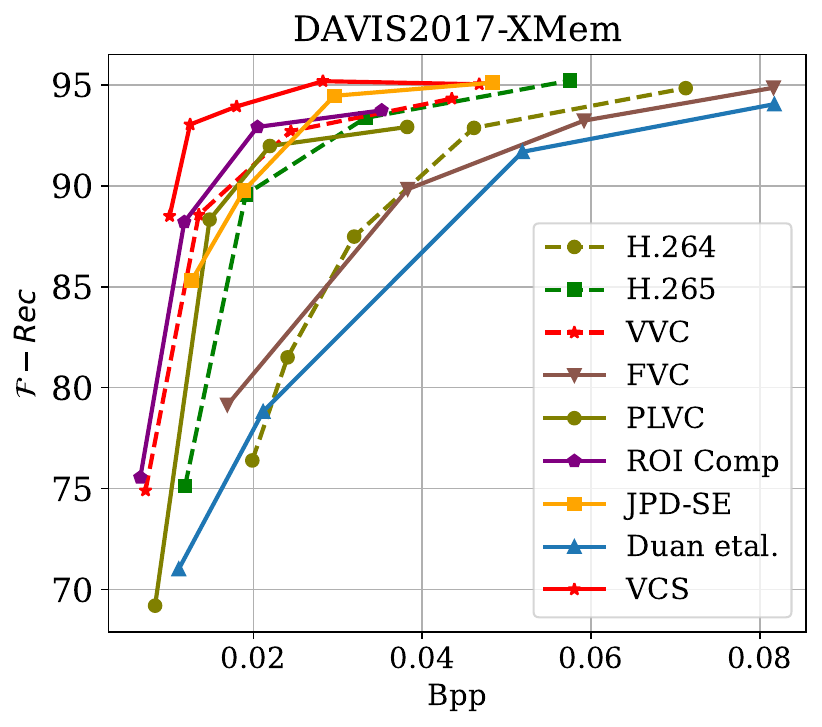}} 
		\end{tabular}
	}
	\vspace{-3mm}
	\caption{
		RP curves of different video coding methods on MOT (first row) and VOS (second row) tasks.
	}
			\vspace{-4mm}
	\label{fig:lbuv_mot_vos_blind}
\end{figure*}

\textbf{Multiple object tracking (MOT).}
Apart from evaluating the coding methods on the action recognition task, which is intricately linked to global semantic cues, we also assess these methods on a dense prediction task, namely MOT. This task entails not only object location detection but also entails capturing semantic connections among objects across different frames.
As shown in Tab.~\ref{tab:MOT_SOTA}, our method achieves the best performance in terms of all metrics when compared to other traditional, learnable and VCM method.
Specifically, our method outperforms VVC codec by 2.93\% (67.79\% \textit{vs.} 64.86\%) in terms of MOTA at the 0.01bpp bitrate level.
\revisetext{
Regarding the learnable codecs, it is notable that FVC exhibits subpar performance on the MOT task, even worse than the old H.264 codec. This contrasts with the performance trend observed in the action recognition task, where FVC outperforms H.265 on most datasets. We conjecture the performance decline may be attributed to the domain shift between I-frames and P-frames in FVC. Specifically, I-frames are compressed using the hand-crafted BPG codec, while P-frames are compressed using a neural codec. This domain shift negatively impacts the cross-frame object association process.
This phenomenon also elucidates why PLVC outperforms all other coding methods except our VCS approach, as it mitigates frame-wise domain inconsistencies through GAN loss.
In terms of the VCM methods, the performance trend aligns with that observed in action recognition task, wherein ROI Comp yields the best results and Feat Comp yields the worst.
}

\begin{table}[!t]
	\centering
		\renewcommand\arraystretch{1.05}
	\tabcolsep=0.35mm
	\begin{tabular}{|c||cccc|cccc|}
		
		\hline
		\multirow{2}{*}{{\textit{\makecell{Method}}}}&\multicolumn{4}{c|}{\textit{Performance@0.01bpp}}&\multicolumn{4}{c|}{\textit{Performance@0.02bpp}}\\
		&{\makecell{MOTA\greenuparrow}} & {\makecell{MOTP\greendownarrow}}
		& {\makecell{IDF1\greenuparrow}}& {\makecell{FN\greendownarrow}}
		&{\makecell{MOTA\greenuparrow}} & {\makecell{MOTP\greendownarrow}}
		& {\makecell{IDF1\greenuparrow}}& {\makecell{FN\greendownarrow}}   \\
		\hline
		{{H.265}}&61.30  &19.88 &64.32 &17377    	& 71.90 &17.07&72.18 & 11005\\
		{{VVC}}  &64.86  &19.49 &68.99 &15621  		& 71.89 &17.31&72.84 &11408\\
		\hline
		{{FVC}}  &44.24  &21.97 &52.53 &27508 		& 59.43 &20.09&64.68&18688\\
		{{PLVC}}  &67.87 &18.44 &68.95 &13299  		& 73.31 &17.09&73.09&10042\\
		\hline
		{{Feat Comp}}  &36.57  &25.42  &47.20 &30040 	& 42.13 &22.70&56.57 &23103\\
		{{ROI Coding}}  &65.29  &19.31 &67.78 &15270 & 72.58  &17.18&73.31&11154\\
		{{Duan~\etal}}  &45.79  &23.33 &52.87 &25920 & 59.61 &21.39&63.77&18187\\
		{{JPD-SE}}  &60.05  	&20.62     &63.52 &17847  	&72.76 &17.27&73.81&11029\\
		{{VCS}} &\textbf{67.79}  &\textbf{18.14} &\textbf{69.53} &\textbf{13484} & \textbf{74.44} &\textbf{16.67}&\textbf{74.94}&\textbf{9922}\\
		\hline
		{\textcolor{gray}{Original}} &\textcolor{gray}{78.60}  &\textcolor{gray}{15.80} &\textcolor{gray}{79.00} &\textcolor{gray}{7000}&\textcolor{gray}{78.60} & \textcolor{gray}{15.80} &\textcolor{gray}{79.00} &\textcolor{gray}{7000} \\
		\hline
	\end{tabular}
\vspace{-1mm}
	\caption{
		Performance comparison of different coding methods on MOT17.
		``Original'' denotes the performance with original videos, which can be viewed as the upper bound on the low-bitrate settings.
	}
	\label{tab:MOT_SOTA}
\end{table}

\begin{table}[!t]
	\centering
	\vspace{-2mm}
	\tabcolsep=0.25mm
	\centering
	\begin{tabular}{|c||cccc|cccc|}
		\hline
		&\multicolumn{4}{c|}{\textit{$\mathbf{\triangle}$Performance}} & \multicolumn{4}{c|}{\textit{$\triangle$Bit(\%)\greendownarrow}} \\
		& {MOTA\greenuparrow} & {MOTP\greendownarrow} &{IDF1\greenuparrow} & {FN\greendownarrow}   &   {MOTA} & {MOTP} &{IDF1} & {FN} \\
		\hline
		\makecell{\scriptsize  \textbf{\tiny VCS$_\mathrm{H.264}$}  {\textit{v.s.} \scriptsize H.264}}&3.05\%  &-0.48\% &0.84\% &-1567 &-22.75 &-11.02 &-9.09 &-18.58 \\
		\hline
		\makecell{\scriptsize \textbf{\tiny VCS$_\mathrm{H.265}$}  {\textit{v.s.} \scriptsize H.265}}&2.06\%  &-0.46\% &2.02\% &-1336 &-17.14 &-17.83 &-22.42 &-17.54 \\
		\hline
	\end{tabular}
	\vspace{-1mm}
	\caption{
		Effectiveness of VCS for H.264 and H.265 on MOT task.
	}
			\vspace{-4mm}
	\label{tab:bdbr_MOT}
\end{table}

Then, we illustrate the RP curves on MOT task in the first row of Fig.~\ref{fig:lbuv_mot_vos_blind}, where our framework consistently outperforms all other methods.
In the top row of Fig.~\ref{fig:qua_mot_vos_blind}, we present the tracking results. Notably, our method successfully detects the fifth person, who is wearing clothing similar in color to the background, whereas VVC fails to do so.
Finally, We report the quantitative gains of our framework when applying it to two widely-deployed industrial codecs (H.264/H.265).
As shown in Tab.~\ref{tab:bdbr_MOT}, our framework surpasses the popular H.264 codec by 3.05\% in terms of MOTA, but also saves the bit cost by 22.75\%.

\begin{figure}[!thbp]
	\centering
	\footnotesize
	\tabcolsep=0mm
	\newcommand{\widthscalefive}{4.4cm}
	\hspace{-5mm}
	\begin{tabular}{cc}
		\includegraphics[width=0.237 \textwidth]{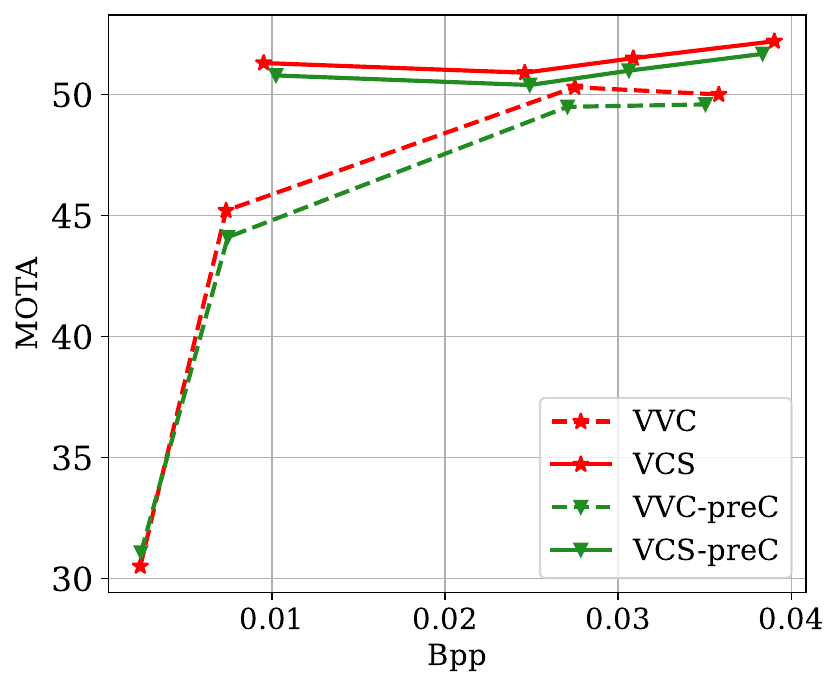}&
		\includegraphics[width=0.237 \textwidth]{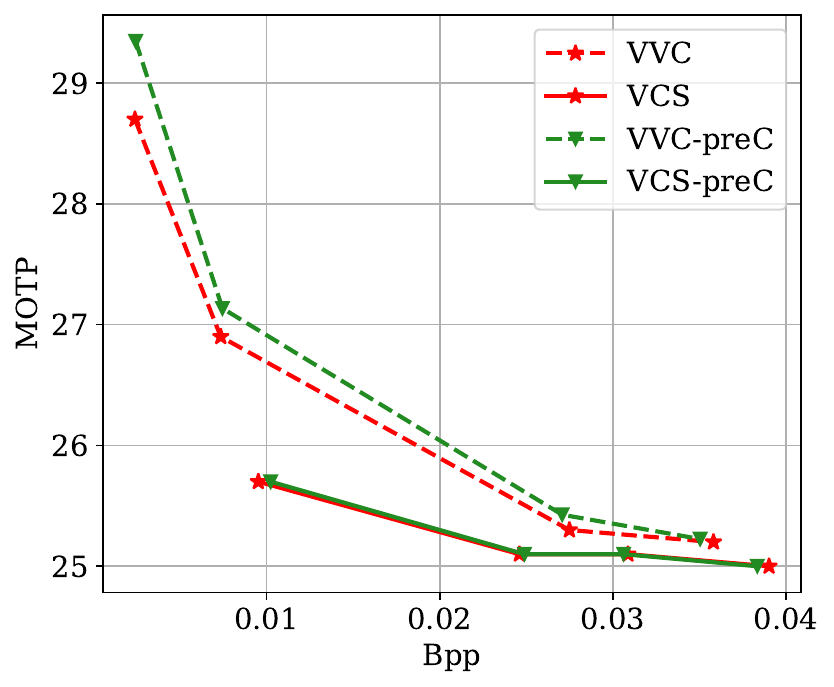}
	\end{tabular}
	\vspace{-3mm}
	\caption {
		MOT results on SFU-HW-Tracks.
		``preC'' denotes the source videos are already compressed before the evaluation.
	}
	\vspace{-6mm}
	\label{fig:result_sfu_hw}
\end{figure}

We further employ the SFU-HW-Tracks dataset to validate if our framework is still effective on raw videos.
As shown in the red curves of Fig.~\ref{fig:result_sfu_hw}, our method also consistently improves the tracking results of the VVC codec on raw videos in this dataset. For example, the performance gain of our framework over the vanilla VVC is 4.9\% at the 0.01bpp level in terms of MOTA. We further explore the impact of the pre-coding (preC) procedure, where the videos are already compressed for one time before our evaluation procedure. We set the CRF of the pre-coding codec to 32, \textit{i.e.}, the default value in VTM configuration file.

	\begin{figure*}[!thhp] 
	\centering
	\renewcommand\arraystretch{0.2}
	\newcommand{\widthscalefive}{0.235}
	\tabcolsep = 1.2mm
	\scalebox{1}{
		\begin{tabular}{cccc}
			\includegraphics[width=0.237 \textwidth]{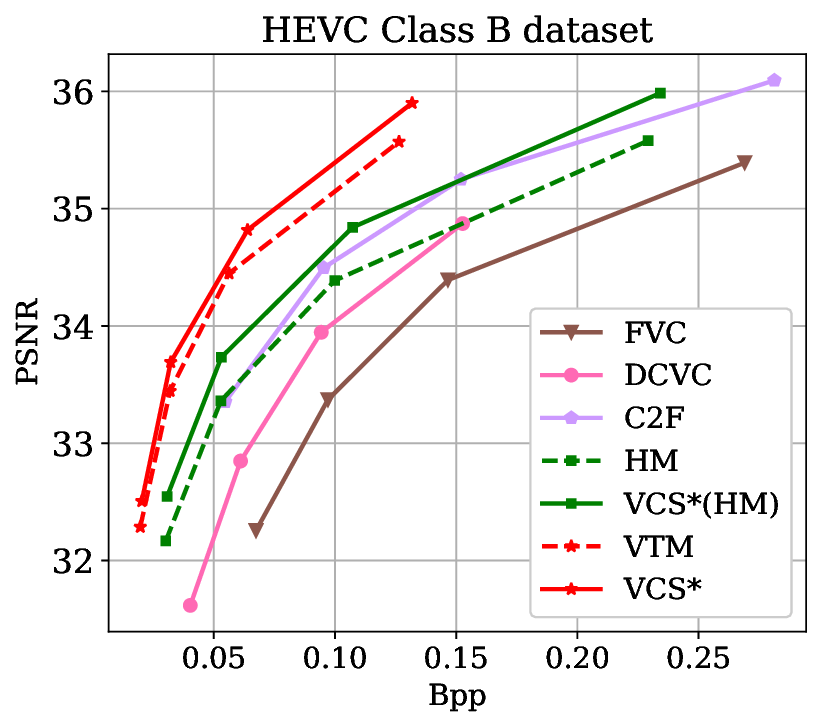} &
			\includegraphics[width=0.238 \textwidth]{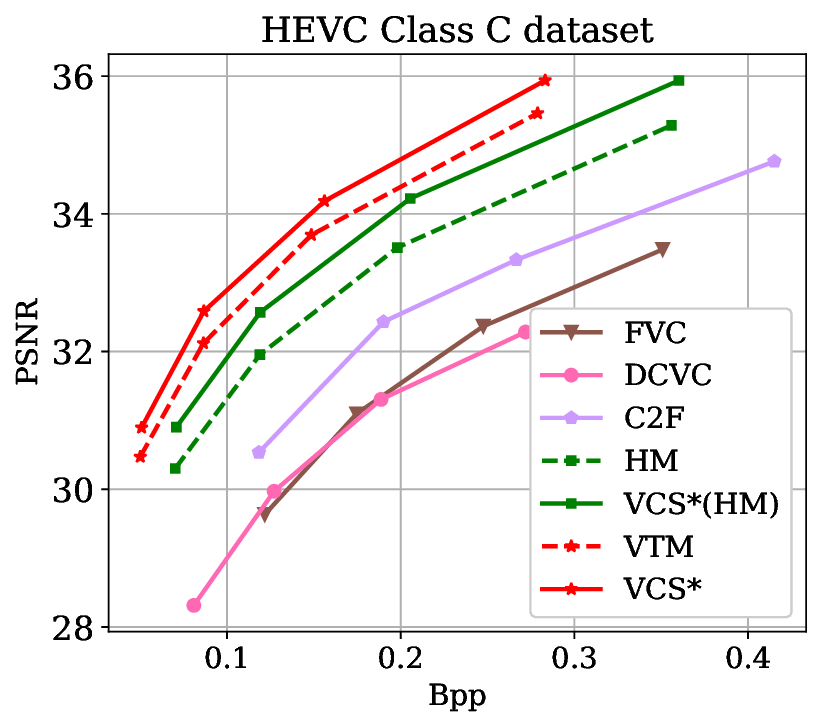} &
			\includegraphics[width=0.238 \textwidth]{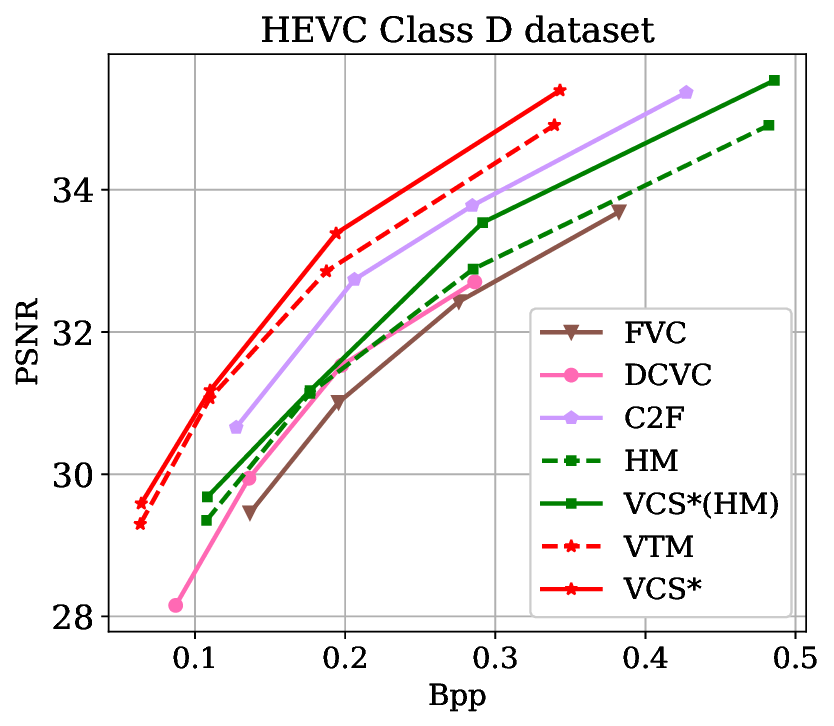} &
			\includegraphics[width=\widthscalefive \textwidth]{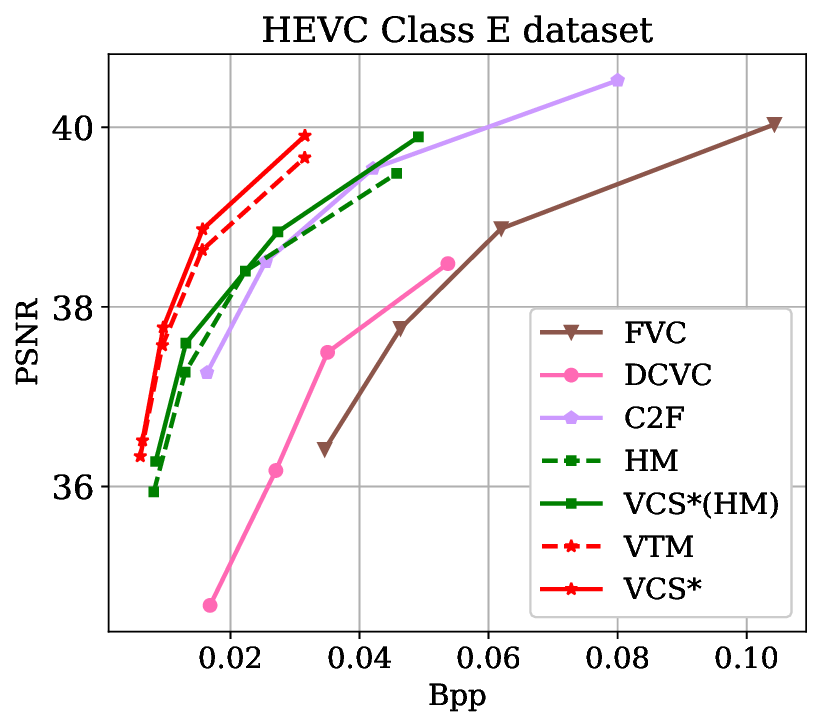} \vspace{-1mm}\\
			\includegraphics[width=\widthscalefive \textwidth]{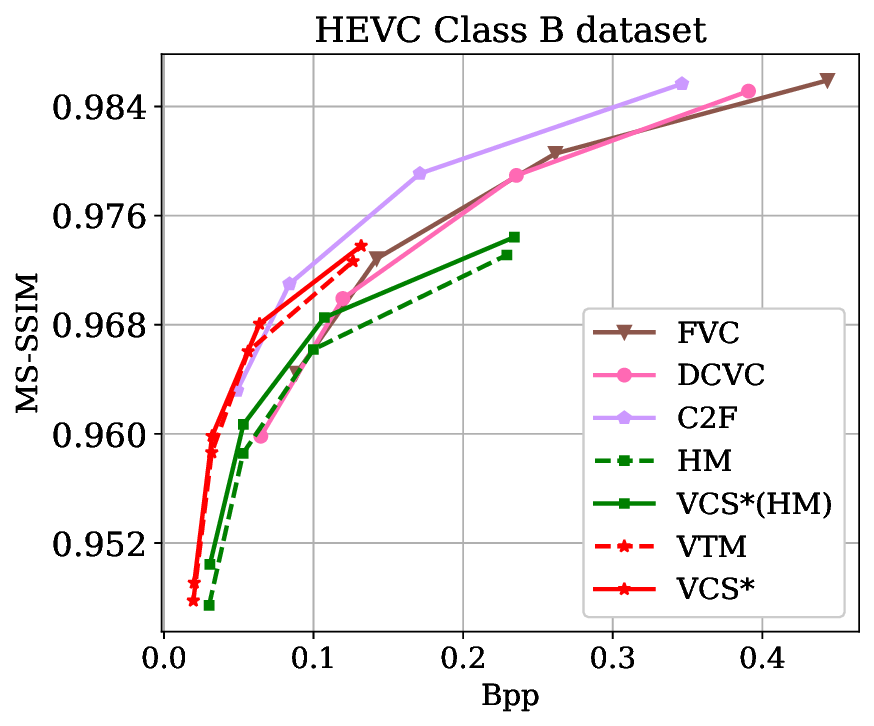}&
			\includegraphics[width=0.238 \textwidth]{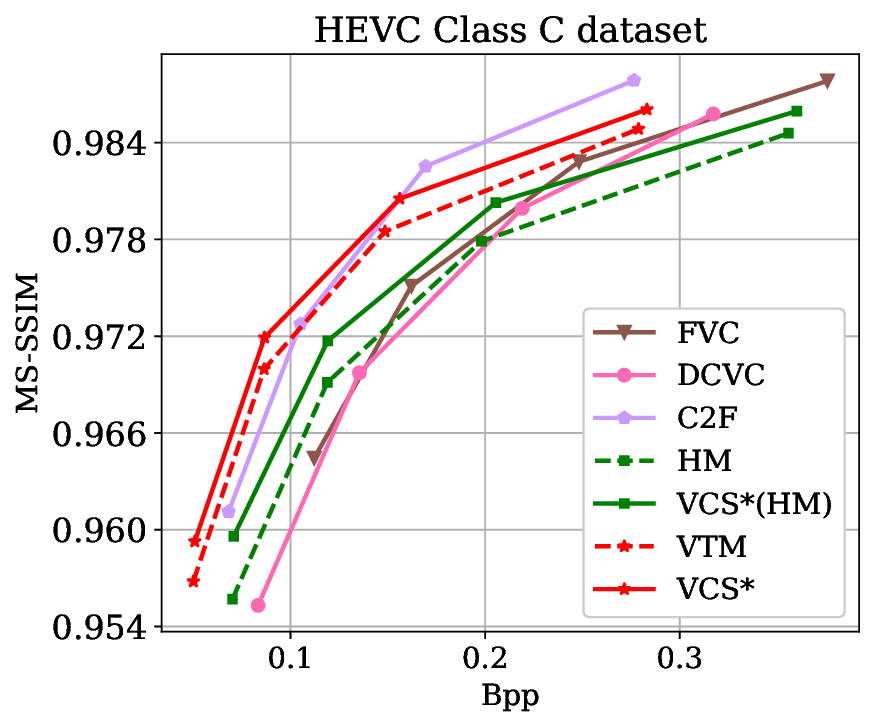} & 
			\includegraphics[width=\widthscalefive \textwidth]{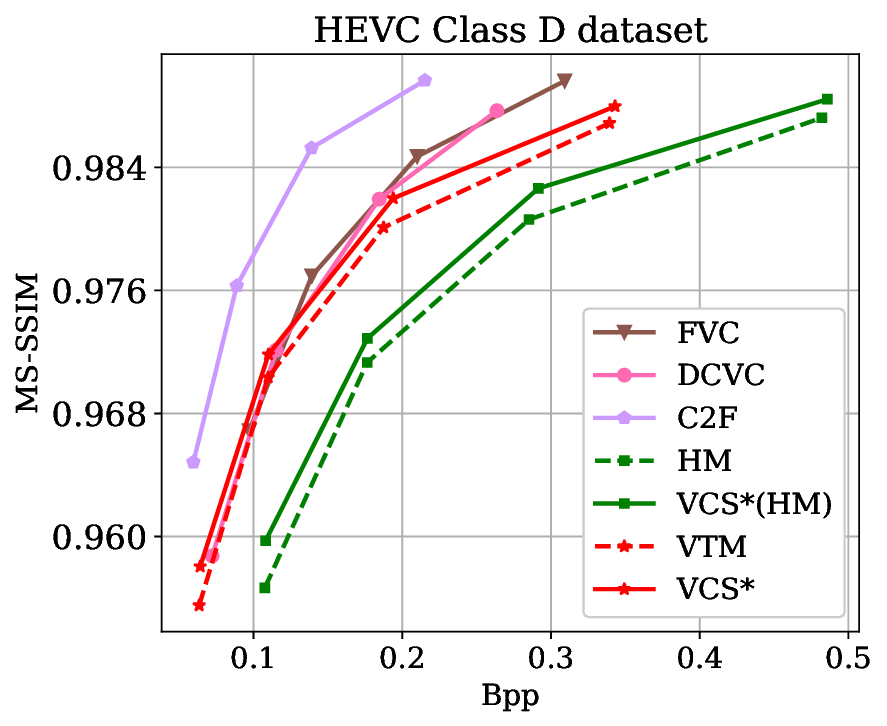} &
			\includegraphics[width=\widthscalefive \textwidth]{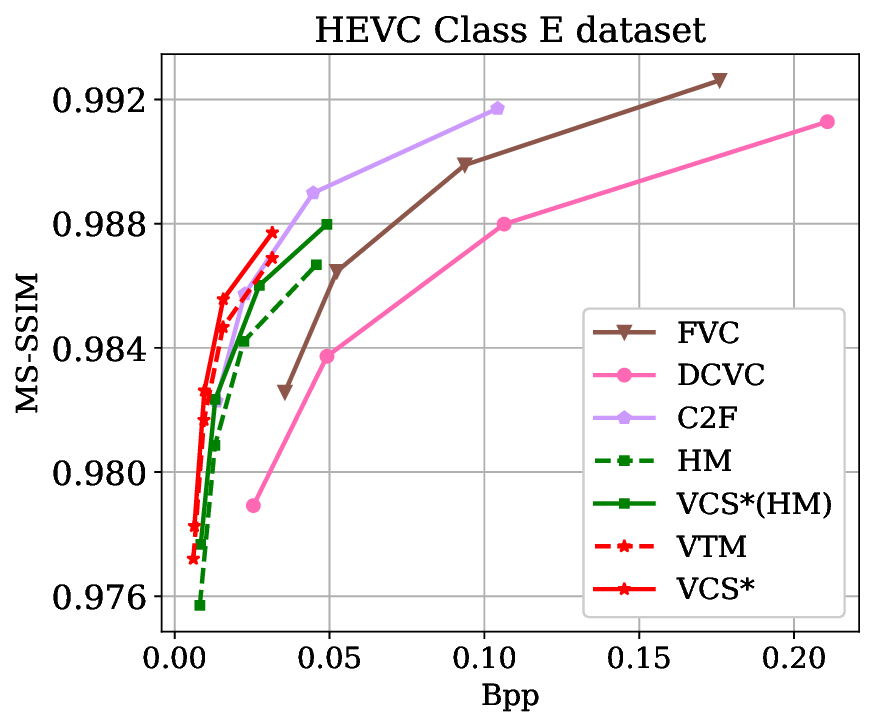} 
			\\
		\end{tabular}
	}
	\vspace{-3mm}
	\caption{
	Comparison of different coding methods in terms of PSNR and MS-SSIM. VCS* model is fine-tuned with the RD loss~\cite{lu2019dvc}.
	}
	\vspace{-4mm}
	\label{fig:video_compression_sota}
\end{figure*}

As shown in the green curves of Fig.~\ref{fig:result_sfu_hw}, the relative improvements of our framework to the baseline VVC codec are consistent no matter the pre-coding procedure is introduced or not.
We think the reason is that the medium CRF values (23 and 32 for H.264 and VVC codecs, respectively) have little degradation to the semantics within images/videos, as empirically studied by the previous works~\cite{tanaka2022does}\cite{zhang2021just}.
The similar pre-coding procedure is also adopted by UGC video websites such as YouTube for saving the storage space, where most video understanding datasets are collected.
Therefore, although the videos from current large-scale video understanding benchmarks are pre-encoded for more efficient storage, they are semantically near-lossless and still effective for evaluating the semantic coding capability of different video compression methods.

\begin{table}[!t]
	\centering
	\renewcommand\arraystretch{1.05}
	\tabcolsep=0.9mm
	\begin{tabular}{|c||cccc|cccc|}
		
		\hline
		\multirow{2}{*}{\textit{\makecell{Method}}}&\multicolumn{4}{c|}{\textit{Performance@0.01bpp}}&\multicolumn{4}{c|}{\textit{Performance@0.02bpp}}\\
		&{$\mathcal{J}$\&$\mathcal{F}$} & \textbf{$\mathcal{J}$} &\textbf{$\mathcal{F}$} & {$\mathcal{F}$-$Rec$}  &{$\mathcal{J}$\&$\mathcal{F}$} & \textbf{$\mathcal{J}$} &\textbf{$\mathcal{F}$} & {$\mathcal{F}$-$Rec$} \\
		\hline
		{{H.264}}&-  &- &- &- 							&63.18  &61.09 &65.28 &76.54  \\
		{{H.265}}&57.68  &56.84 &58.51 &67.44 			&73.28  &70.59 &75.96 &89.81  \\
		{{VVC}}  &67.47  &65.59 &69.36 &80.92 			&75.87  &72.91 &78.82 &91.02  \\
		\hline
		{{FVC}}  &-  &- &- &-			&67.03  &65.22 &68.85 &80.68  \\
		{{PLVC}}  &61.45  &60.02 &62.87 &74.07 			&74.60  &71.58 &77.62 &90.97  \\
		\hline
		{{ROI Coding}}  &69.22  &67.16 &71.28 &83.80 	&77.99  &74.87 &81.10 &92.65  \\
		{{Duan~\etal}}  &59.89  &59.99 &59.79 &70.10 	&64.51  &63.25 &65.77 &77.91  \\
		{{JPD-SE}}  &61.48  &59.86 &63.09 &73.31 		&75.10  &71.89 &78.31 &90.57  \\
		{{VCS}} &\textbf{72.44}  &\textbf{68.34} &\textbf{76.44} &\textbf{88.36}       &\textbf{80.61}  &\textbf{76.64} &\textbf{84.57} &\textbf{94.17} \\
		\hline
		\textbf{\textcolor{gray}{Original}} &\textcolor{gray}{87.70}  &\textcolor{gray}{84.06} &\textcolor{gray}{91.33} &\textcolor{gray}{97.02} &\textcolor{gray}{87.70}  &\textcolor{gray}{84.06} &\textcolor{gray}{91.33} &\textcolor{gray}{97.02}\\
		\hline
	\end{tabular}
\vspace{-1mm}
	\caption{
		Performance comparison of different coding methods on DAVIS2017.
		``Original'' denotes the performance with original videos, which can be viewed as the upper bound on the low-bitrate settings.
	}
	\label{tab:VOS_SOTA}
\end{table}

\begin{table}[!t]
	\vspace{-2mm}
	\centering
	\renewcommand\arraystretch{1.05}
	\tabcolsep=0.6mm
	\centering
	\begin{tabular}{|c||cccc|cccc|}
		\hline
		&\multicolumn{4}{c|}{\textit{$\mathbf{\triangle}$Performance(\%)\greenuparrow}} & \multicolumn{4}{c|}{\textit{$\triangle$Bit(\%)\greendownarrow}} \\
		& {$\mathcal{J}$\&$\mathcal{F}$} & \textbf{$\mathcal{J}$} &\textbf{$\mathcal{F}$} & {$\mathcal{F}$-$Rec$}   &   {$\mathcal{J}$\&$\mathcal{F}$} & \textbf{$\mathcal{J}$} &\textbf{$\mathcal{F}$} & {$\mathcal{F}$-$Rec$} \\
		\hline
		\makecell{\scriptsize \textbf{\tiny VCS$_\mathrm{H.264}$}  {\scriptsize \textit{v.s.} H.264}}&6.02  &4.95 &7.08 &5.40 &-36.51 &-31.03 &-41.64 &-45.27 \\
		\hline
		\makecell{\scriptsize \textbf{\tiny VCS$_\mathrm{H.265}$}  {\scriptsize \textit{v.s.} H.265}}&4.35  &3.08 &5.62 &2.91 &-36.50 &-29.33 &-42.23 &-39.90 \\
		\hline
	\end{tabular}
	\vspace{-1mm}
	\caption{
		Effectiveness of VCS for H.264 and H.265 on VOS task.
	}
		\vspace{-5mm}
	\label{tab:bdbr_VOS}
\end{table}

\textbf{Video object segmentation (VOS).}
We further benchmark different coding methods on VOS task, which is more fine-grained than the MOT task, also requires both accurate pixel-level details and high-level region semantics to obtain the ultimate object segmentation results.
\revisetext{As depicted in Tab.~\ref{tab:VOS_SOTA}, our approach exhibits superior performance across all metrics when compared to other traditional, learnable, and VCM methods. Notably, our method significantly surpasses the VVC codec by 4.97\% (72.44\% \textit{v.s.} 67.47\%) in terms of $\mathcal{J}\&\mathcal{F}$ at the 0.01bpp bitrate level.
}
\revisetext{
The performance trends for learnable codecs and VCM methods align with those observed in the MOT task. We have omitted reporting the feature compression method for this task due to challenges in harmonizing the feature compression process with intricate feature buffering and reusing strategies present in state-of-the-art VOS models like XMEM.
This observation underscores that task-coupled methods, such as feature compression, which necessitates integrating the compression process into the task model, face difficulties in being adopted for intricate downstream tasks and models, such as video object segmentation. In contrast, our task-decoupled scheme is free from this.}

We illustrate the RP curves on VOS task in the bottom row of Fig.~\ref{fig:lbuv_mot_vos_blind}, which clearly illustrate the advantages of our method over others.
Also, we demonstrate the segmentation results by adopting VVC and our framework, as shown in the bottom row of Fig.~\ref{fig:qua_mot_vos_blind}.
Our decoded frame demonstrates clearer object structures and sharper edges, which contributes to better segmentation results.
Finally, we evaluate the quantitative gains of the VCS framework when deploying it to H.264 and H.265 codecs.
As shown in Tab.~\ref{tab:bdbr_VOS}, our framework largely improves the H.264 codec by 6.02\% in terms of $\mathcal{J}\&\mathcal{F}$, while saving the bitcost by 36.51\%.



\revisetext{
\textbf{Video compression.}
We further explore that if the learned semantics by our self-supervised learning objective can benefit the video compression task, where conventional image signal fidelity metrics such as PSNR and MS-SSIM are adopted for evaluation.
Following other recent neural network-based video compression methods~\cite{hu2021fvc}\cite{hu2022coarse}, we adopt the high-performance VVC reference software VTM as the baseline codec and fine-tune our framework with the rate-distortion (RD) loss, producing the VCS* model, for a fair comparison with other learnable deep video codecs~\cite{hu2021fvc}\cite{hu2022coarse}.
As shown in Fig.~\ref{fig:video_compression_sota}, our approach outperforms all traditional codecs including the strong VVC standard reference software VTM and recent learnable codecs such as DCVC~\cite{li2021deep}, FVC~\cite{hu2021fvc} and C2F~\cite{hu2022coarse} by a large margin in terms of PSNR. When measured by the MS-SSIM metric, our method also performs favorably against VTM and comparably to the state-of-the-art learnable codec C2F on most datasets.
To prove that our method is still effective for enhancing other traditional codecs on video compression task, we further apply it to the H.265 standard reference software HM.
As compared by green curves of Fig.~\ref{fig:video_compression_sota}, our method also substantially improves the performance of HM software.
When our framework is applied to an older H.264 reference software JM~\cite{JM_code}, the compression performance of JM is largely improved by 0.42db, 0.79dB, 0.46dB and 0.31dB on HEVC Class B, C, D and E datasets, respectively.
}

\revisetext{
To investigate the source of the performance improvement, we additionally trained various variant models and reported their compression performance.
(1) VCS-TuneDec, which shares the same encoder (as well as the semantic bitstream) as the vanilla VCS model, but its decoder part is tuned with $\ell1$ loss.
(2) LFN-Only, where the semantic encoder is removed and the semantic stream $\mathcal{S}$ is replaced with an all-zero tensor. Since no extra bits are transported, LFN-Only is a video enhancement model. This baseline is for rigorously analyzing the performance gain between our \textit{codec+semantic stream} and \textit{codec+post enhancement} schemes without being confused by the network architecture gap.
(3) The state-of-the-art video compression enhancement method BasicVSR++~\cite{chan2022basicvsr++}, where the bi-directional feature propagation is replaced with the forward-only propagation scheme, to adhere to the frame-dependency rule of the P frame coding mode.
All variant models above are equipped with the VTM software as the traditional video codec for fair comparison.
}

\begin{table}[!t]
	\centering
	\renewcommand\arraystretch{1.05}
	\tabcolsep=2.0mm
	\begin{tabular}{|c||ccccc|}
		\hline
		\textit{\multirow{2}{*}{Method}}&\multicolumn{5}{c|}{\textit{PSNR (dB) on HEVC Class C@}} \\ 
		& \textit{0.05bpp}  & \textit{0.10bpp} &\textit{0.15bpp} & \textit{0.20bpp} &\textit{0.25bpp}   \\
		\hline
		{VTM} 		&30.47		& 32.46 &33.71 & 34.39 & 35.06\\ 
		
		\hline
		{VCS} 		&30.22		& 32.38 &33.61 & 34.27  & 34.92\\ 
		{VCS-TuneDec} 		&30.80		& 32.83 &33.94 & 34.62  & 35.33 \\ 
		{VCS*} 		& \textbf{30.85}		& \textbf{32.89} &\textbf{34.04} & \textbf{34.79}& \textbf{35.48}\\ 
		\hline
		{LFN-Only} 		&30.68		& 32.64 &33.82	 & 34.51&35.20 \\ 
		{ BasicVSR++} 		&30.71		& 32.68 &33.88	 & 34.59&35.31 \\ 
		\hline
	\end{tabular}
	\vspace{-2mm}
	\caption{
		Comparison of different video compression methods.
		The semantic stream is included in the bitcost for VCS-series methods.
		VCS* model is fine-tuned with the RD loss~\cite{lu2019dvc}.
	}
	\vspace{-7mm}
	\label{tab:invest_psnr_source}
\end{table}

\revisetext{
As shown in Tab.~\ref{tab:invest_psnr_source}, our vanilla VCS model is inferior to the original VTM software.
This is reasonable from the rate-perceptual-distortion theory~\cite{blau2019rethinking}, \textit{i.e.}, optimizing the coding system with GAN loss, will degrade the PSNR performance.
Furthermore, when using a PSNR-oriented decoder to decode the bitstreams of the vanilla VCS model, the resulted VCS-TuneDec model substantially outperforms VTM, especially in the low bitrate levels. For example, the performance gain is 0.37db at the 0.1bpp bitrate level.
When optimizing both the encoder and the decoder of VCS, the resulted VCS* model achieves further better video compression performance.
}

\revisetext{
The video enhancement model LFN-Only indeed improves the performance, but is noticeably inferior to VCS* model.
The performance disparities, \textit{e.g.}, 0.25dB at the 0.10bpp bitrate level, arise from the introduction of additional semantic information by our method, which can considerably improve video construction quality.
Employing a more advanced video enhancement method BasicVSR++ leads to a moderate improvement in video quality, but still falls notably short of our \textit{codec+semantic stream} scheme, \textit{e.g.}, 0.21dB performance gap at the 0.10bpp bitrate level.
}

Additionally, we compare our approach with the previous perceptual coding method PLVC in terms of two perceptual image quality metrics, \textit{i.e.}, LPIPS~\cite{zhang2018unreasonable} and FID~\cite{heusel2017gans}.
As shown in Tab.~\ref{tab:comp_percep}, our method consistently performs better.


\begin{table}[!t]
	\centering
	\renewcommand\arraystretch{1.05}
	\tabcolsep=1.5mm
	\begin{tabular}{|c||ccc|ccc|}
		\hline
		\textit{\multirow{2}{*}{Method}} &\multicolumn{3}{c|}{{\textit{LPIPS$\greendownarrow$@}}}	&\multicolumn{3}{c|}{\textit{{FID$\greendownarrow$@}}} \\
		& 0.02bpp  & 0.03bpp &0.04bpp & 0.02bpp  & 0.03bpp  &0.04bpp  \\
		\hline
		{PLVC} 		&0.4195		& 0.3732 &0.3337 & 50.25& 44.72 & 40.86 \\ 
		\hline
		{VCS} 		&\textbf{0.3159} & \textbf{0.2898} & \textbf{0.2726} &\textbf{48.23}& \textbf{42.49} & \textbf{38.74} \\ 
		\hline
	\end{tabular}
\vspace{-1mm}
	\caption{
		Comparison of different methods on perceptual quality.
	}
	\vspace{-2mm}
	\label{tab:comp_percep}
\end{table}
\begin{table}[!t]
	\centering
	\renewcommand\arraystretch{1.05}
	\tabcolsep=1.4mm
	\begin{tabular}{|c||ccc|ccc|}
		\hline
		\textit{\multirow{2}{*}{Method}} &\multicolumn{3}{c|}{{\textit{SIFT Distance$\greendownarrow$@}}}	&\multicolumn{3}{c|}{\textit{{SURF Distance$\greendownarrow$@}}} \\
		& 0.02bpp  & 0.03bpp &0.04bpp & 0.02bpp  & 0.03bpp  &0.04bpp  \\
		\hline
		{VVC} 		&265		& 252 &243 & 0.262& 0.248 & 0.237 \\ 
		\hline
		{VCS} 		&\textbf{251} & \textbf{243} & \textbf{236} &\textbf{0.257}& \textbf{0.240} & \textbf{0.231} \\ 
		\hline
	\end{tabular}
	\vspace{-1mm}
	\caption{
		Comparison of different methods on detected traditional interest point quality.
	}
		\vspace{-6mm}
	\label{tab:improve_sift}
\end{table}
\textbf{Traditional interest point quality.}
In addition to AI models, there are also many applications relying on traditional interest points detection algorithms such as SIFT~\cite{lowe1999object} and SURF~\cite{bay2006surf}.
As shown in Tab.~\ref{tab:improve_sift}, our framework consistently improves the VVC codec in terms of the quality of the detected interest points.
The quality is measured by the point matching distance between the compressed frames and the raw frames.
The raw \textit{RaceHorse} sequence in HEVC test sequences is adopted for evaluation.

\section{Framework Analysis}\label{sec:method_ana}




 \begin{figure}[!tbp]
	\centering
	\newcommand{\widthscalefive}{0.239}
	\tabcolsep=-0mm
	\begin{tabular}{cc}
		\includegraphics[width=\widthscalefive \textwidth]{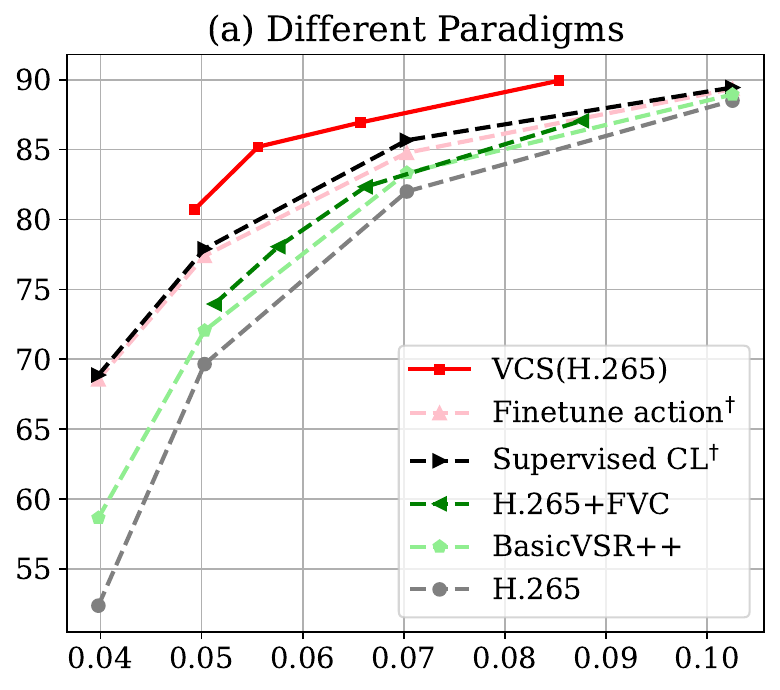}&
		\includegraphics[width=\widthscalefive \textwidth]{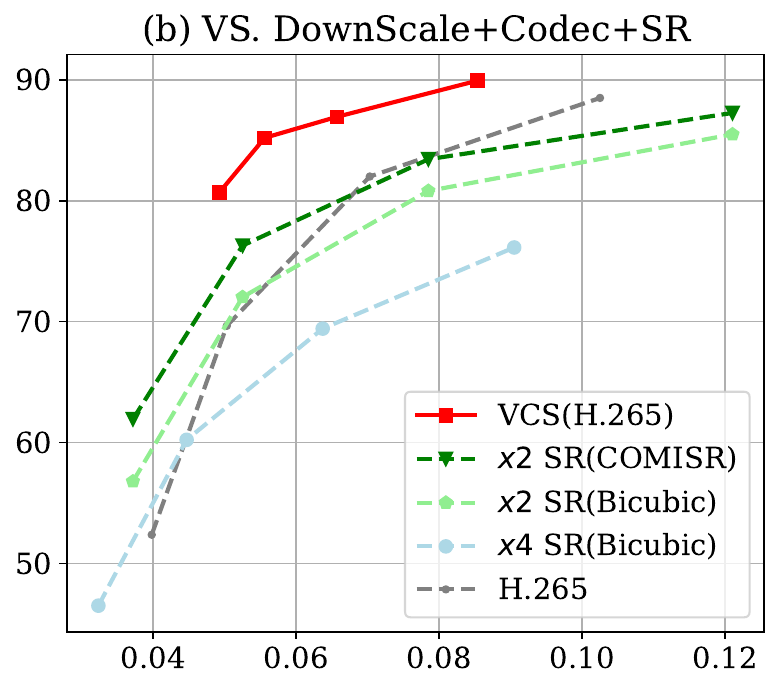}
	\end{tabular}
	\vspace{-3mm}
	\caption {
		{Ablation studies on the framework.}
		$\dag$ denotes the method is dataset-specific due to using dataset labels.
		The horizontal and vertical axes represent the bitcost (Bpp) and the UCF101-TSM Top1 accuracy (\%).
		Experiments are conducted with H.265 codec for a fair comparison with previous video restoration methods.
	}
	\label{fig:ablation}
	\vspace{-6mm}
\end{figure}

\subsection{Comparison of Different Paradigms}
Since we aim to build a unified and practical coding framework that can smoothly transition from currently widely used traditional codecs and be readily deployed to various downstream tasks and applications, we choose the ``traditional codec + unsupervised semantic stream'' paradigm.
However, without considering the constraints above, there are also several potential solutions, including the following two task-unspecific methods.
(1) BasicVSR++, where a recent state-of-the-art (SOTA) video restoration method {BasicVSR++}~\cite{chan2022basicvsr++} is adopted to enhance the lossy video by codec and then fed into the downstream task.
\revisetext{
(2) H.265+FVC, where a recent learnable codec FVC~\cite{hu2021fvc} is delicately trained as the texture detail enhancement layer to encode the residule part between the original video and the lossy video compressed by the traditional codec.
Additionally, we also propose two following task-specific methods.}
(3) Supervised fine-tuning, where the downstream action model is fine-tuned on the compressed version of the dataset with the standard cross-entropy loss.
\revisetext{
(4) Supervised CL, where the downstream action model is fine-tuned with the supervised contrastive learning (CL) loss~\cite{khosla2020supervised}. Compared to the previous supervised fine-tuning paradigm, supervised CL also enforces the features of the compressed videos similar to the original video, learning better compression-agnostic video representations.
}

As shown in Fig.~\ref{fig:ablation} (a), our framework with semantic stream outperforms the \textit{traditional codec+post restoration} method BasicVSR++ by a large margin. For example, our method outperforms BasicVSR++ model by 8\% at the 0.06bpp bitrate level.
\revisetext{
Then, we introduce a textual detail enhancement layer implemented with FVC to the base codec. The resulting approach, referred to as \textit{traditional codec+learnable detail stream} method H.265+FVC, demonstrates improved performance compared to BasicVSR++. However, it remains significantly inferior to our approach \textit{traditional codec+semantic stream}, exhibiting a performance drop of over 5\% at the 0.06bpp level.
This observation emphasizes the necessity of an additional stream to transport and compensate for the distorted information in videos compressed by traditional codecs. Furthermore, it highlights that our learned semantic stream is notably more effective than the conventional residual stream for downstream AI tasks.
To summarize, enhancing continuously evolving traditional codecs with a data-emerged semantic prior is a promising way to achieve a high-performance low-bitrate video understanding systems.
}

Further, one may wonder that although the human-aware structure information in the video is distorted by the codecs, the left part is still possible to be recognized by machine models. Therefore, we explore how the dataset annotation can improve the performance. As shown in Fig.~\ref{fig:ablation} (a), simply fine-tuning the downstream model on the compressed training dataset, largely outperforms BasicVSR++ and H.265+FVC.
\revisetext{
After introducing the contrastive learning objective, the performance of the resulting Supervised CL model is further improved, but there is still a noticeable gap compared to our method. This indeed proves that, even with the guidance of downstream task dataset annotations, machine models struggle to uncover more hidden semantic cues using a simple fine-tuning strategy without the additionally transported semantic information.
In contrast, our semantic stream can effectively address this limitation without requiring any data annotations. Additionally, the task-specific training procedures mentioned above introduce significant time costs prior to deployment. Taking the lightweight action recognition model TSM as an example, fine-tuning TSM on compressed mainstream video datasets like UCF101, Kinetics400, and Something takes over four days, even with eight Nvidia 2080Ti GPUs. In comparison, our framework is task-decoupled, directly deployable across various datasets, and incurs zero time costs.
This showcases the practicality of the task-decoupling and label-free principles for video analysis-oriented video coding methods.
}

Finally, we investigate the impact of different video enhancement methods to the downstream task performance.
Concretely, we evaluate different video restoration methods, \textit{i.e.}, (1) {LFN} in our method by masking the semantic stream with zero values,
(2) a recent state-of-the-art (SOTA) video restoration method {BasicVSR++}~\cite{chan2022basicvsr++} 
and (3) a video restoration method {STDF}~\cite{deng2020spatio} that is specified for removing compression artifacts.
To enable a fair comparison, we train the BasicVSR++ and STDF on our training dataset after initializing them with the officially pre-trained weights.
The baseline codec is H.265, as most video restoration methods are designed for it.

\begin{table}[!thbp]
	\centering
	\renewcommand\arraystretch{1.05}
	\tabcolsep=1.9mm
	\begin{tabular}{|c||ccccc|}
		\hline
		\textit{\multirow{2}{*}{Method}}&\multicolumn{5}{c|}{\textit{Top1 Accuracy(\%) on UCF101-TSM @}} \\ 
		& \textit{0.04bpp}	&\textit{0.05bpp}& \textit{0.06bpp}& \textit{0.08bpp} & \textit{0.1bpp}   \\ 
		\hline
		H.265 & 52.70 &69.16& 75.64 &83.95 & 87.99  \\
		\hline
		STDF & 58.44 &69.93& 77.09 &84.65 & 88.38  \\
		BasicVSR++ & 58.91 & 71.67& 77.52 &85.02 & 88.50  \\
		BasicVSR++* & 68.40 & 77.01& 80.86 &86.15 & 89.20  \\
		\hline
		LFN(Ours) & 71.13 &78.15& 81.45 & 86.30 &89.55  \\
		\hline
	\end{tabular}
\vspace{-1mm}
	\caption{
		Comparison of different post video enhancement methods in terms of the action recognition task performance.
		The BasicVSR++* model is fine-tuned from the vanilla BasicVSR++ model by incorporating a combination of semantic and visual quality loss terms, similar to our loss function.
	}
	\vspace{-2mm}
	\label{tab:comp_post_enahce}
\end{table}

As shown in Tab.~\ref{tab:comp_post_enahce}, our framework with semantic stream outperforms all video enhancement methods by a large margin. Even after we retrain the BasicVSR++ model with our GAN+Semantics loss function, which is much more beneficial to semantic-related tasks than the original $\ell1$ loss, the produced BasicVSR++$^*$ model still lefts behind our method by about 5\%@0.06bpp.
The reason is that the semantic information in the video is already discarded by the traditional codecs at low bitrate levels and is impossible to be restored by even a SOTA video restoration network.

Furthermore, we observe that the performance of STDF is much inferior to BasicVSR++, due to its much fewer parameter number ({0.47M v.s. 6.88M}).
Moreover, our LFN achieves similar (sometimes even higher) performance to BasicVSR++$^*$, but consumes 16$\times$ fewer computational cost ({5.85 v.s. 93.20 GFLOPs}).
This proves that performing restoration in the deep latent space is more effective for the structure distortions at low-bitrate compression levels, compared to the explicit motion compensation scheme adopted by BasicVSR++.

\begin{figure}[!tbp]
	\centering
	\newcommand{\widthscalefive}{0.15}
	\tabcolsep=0.5mm
	\scriptsize
	\begin{tabular}{ccc}
		\includegraphics[width=\widthscalefive \textwidth]{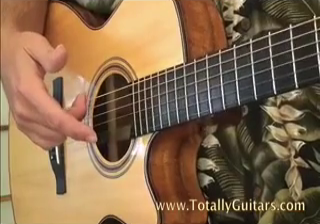}&
		\includegraphics[width=\widthscalefive \textwidth]{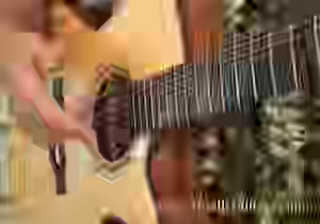}&
		\includegraphics[width=\widthscalefive \textwidth]{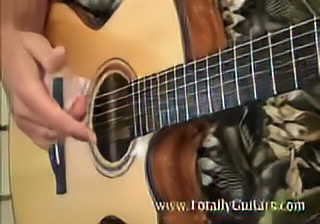}
		\\
		(a) Original & (b) H.265  &(c) H.265+Ours \\
		\includegraphics[width=\widthscalefive \textwidth]{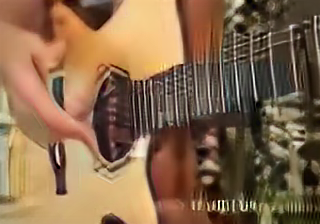}&
		\includegraphics[width=\widthscalefive \textwidth]{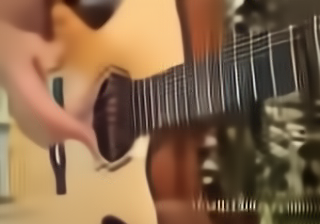}&
		\includegraphics[width=\widthscalefive \textwidth]{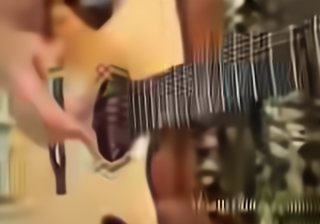} \\
		(d) H.265+BasicVSR++$^*$ & (e)  H.265+BasicVSR++ &(f) H.265+STDF \\
	\end{tabular}
	\vspace{-2mm}
	\caption {
		Visual comparison of different coding methods.
		The video bitrates of our method and other methods are 0.021bpp and 0.023bpp, respectively.
		The BasicVSR++* model is fine-tuned from the vanilla BasicVSR++ model by incorporating a combination of semantic and visual quality loss terms, similar to our loss function.
		\textit{Best to view by zooming in.}
	}
	\label{fig:ablation_vis}
	\vspace{-5mm}
\end{figure}

Then, we demonstrate the decoded videos from different coding schemes, as shown in Fig.~\ref{fig:ablation_vis}.
The video restoration methods such as BasicVSR++ and STDF can not repair the compression distortion of the human hand and guitar string, and the frames are also blurry.
Even trained with our GAN+Semantics loss, the BasicVSR++$^*$ model just introduces some meaningless textures, \textit{e.g.}, some unordered guitar string.
In contrast, the videos decoded from our method are much similar to the original video in terms of object structure, and also demonstrate better visual comfort.


We also investigate if the hand-crafted semantic preservation strategy is effective for the low-bitrate video understanding task.
Downsampling is well known to be effective in preserving the low-frequency structure information of images/videos, which is closely related to semantics.
Therefore, we also compare our method with the \textit{downsampling$\rightarrow$traditional codec$\rightarrow$upsampling} scheme.
As shown in Fig.~\ref{fig:ablation} (b), this simple semantic preservation strategy can bring marginal performance gains at very low bitrate levels (0.035$\sim$0.05bpp), but introduces large drops at other bitrate levels (0.05$\sim$0.12bpp).
Even when we adopt the SOTA compressed video super-resolution (SR) method COMISR~\cite{li2021comisr} to upsample the videos, the performance on the downstream task is still behind the original H.265 codec and much inferior to our method, at bitrate levels larger than 0.08bpp.
This proves that the hand-crafted semantic preservation strategy is inadequate for building a stable low-bitrate video understanding system.
\revisetext{Our data-emerged semantic preservation strategy is superior and performs steadily.}

\subsection{Comparison of Different Semantic Priors}
One key component of our method is employing a contrastive learning objective to learn data-emerged semantic priors from the web-scale unlabeled video data.
To verify the superiority of our data-emerged semantic prior, we train several variant models by leveraging different common semantic representations to regularize the semantic consistency, \textit{i.e.}, (1) features output from the pre-trained {VGG16} network~\cite{simonyan2014very}, (2) perceptual edge maps output the {HED} model~\cite{xie2015holistically}, and (3) semantic segmentation map output from the {DeepLabV3} model~\cite{chen2017rethinking}.

\begin{figure}[!t]
	\vspace{-2mm}
	\centering
	\newcommand{\widthscalefive}{0.239}
	\tabcolsep=-0mm
	\begin{tabular}{cc}
		\includegraphics[width=\widthscalefive \textwidth]{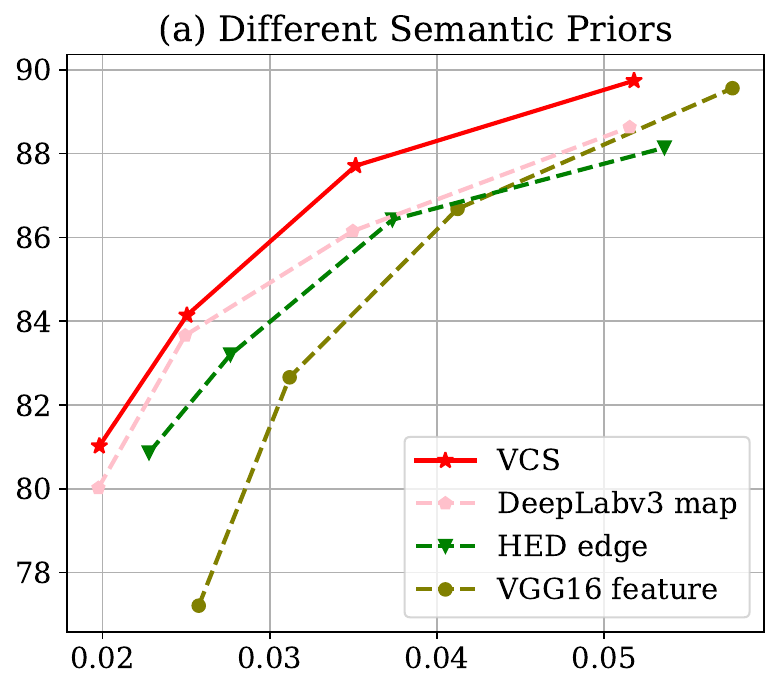}&
		\includegraphics[width=\widthscalefive \textwidth]{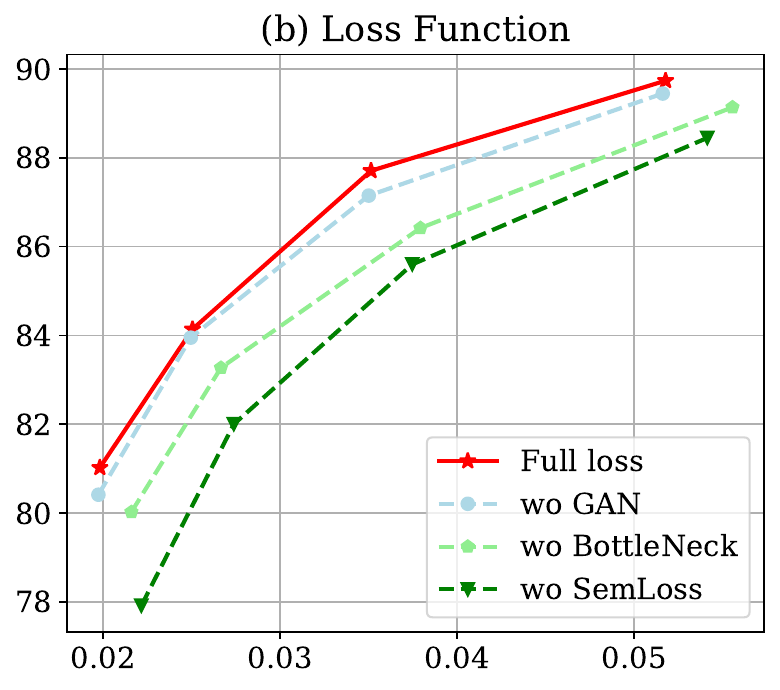}
	\end{tabular}
	\vspace{-3mm}
	\caption {
		{Ablation studies on different semantic priors and the loss function.}
		The horizontal and vertical axes represent the bitcost (Bpp) and the UCF101-TSM Top1 accuracy (\%).
		``wo'' denotes ``without''.
	}
	\label{fig:ablation_loss}
	\vspace{-3mm}
\end{figure}

As shown in Fig.~\ref{fig:ablation_loss} (a), the model trained with VGG16 feature consistency loss demonstrates much inferior performance to other losses,
while also making the semantic stream transport more redundant information.
Specifically, when the codec CRF is set to 47 (the second point on the curves), the bitrates of the semantic stream for VGG feature, HED, DeepLabV3, and our contrastive learning-based semantic consistency loss are 0.0123bpp, 0.0088bpp, 0.0062bpp, and 0.0062bpp, respectively.
The reason is that,
compared to high-dimensional feature maps from VGG16, both our bottleneck map, HED edge map and DeepLabV3 segmentation map represent the decoded videos with a sparse information bottleneck, which encourages discarding the semantic-less information.

It is also observed that our method outperforms HED and DeepLabV3 by a large margin, \textit{i.e.}, about 2\%@0.04bpp, which can be attributed to the fact that both the other two models are learned on human-labeled annotations and may contain the information unnecessary to machine intelligence.
\revisetext{In contrast, our model adheres to a completely label-free scheme, unburdened by the constraints of annotating intricate image edge or segmentation datasets.
}
We also notice that the model trained with DeepLabV3 performs better than the model trained with HED, especially in the low bitrate range (0.02$\sim$0.035bpp), because the segmentation map from DeepLabV3 facilitates the intra-object semantic consistency as well, compared to HED's edge-only regularization.
\begin{table}[!tp]
	\centering
	\renewcommand\arraystretch{1.05}
	\tabcolsep=2.4mm
	\begin{tabular}{|c||c|c|c|c|c|}
		\hline
		$\alpha$ 		& 0	& 0.001& 0.01 & 0.1 & 1  \\ 
		\hline
		UCF101-TSM Top1(\%) & 82.2 & 83.9& \textbf{88.11} & 87.79 & 81.8  \\
		\hline
	\end{tabular}
	\vspace{-1mm}
	\caption{
		Impact of $\alpha$.
		The target codec@bitrate is VVC@0.04bpp.
	}
		\vspace{-6mm}
	\label{tab:hyper_alpha}
\end{table}
\vspace{-2mm}
\subsection{Ablation Study on the Loss Function}

As shown in Fig.~\ref{fig:ablation_loss} (b), removing the GAN Loss only marginally lowers the performance, as it mainly facilitates the realness of the local texture patches instead of the structures. Despite the little impact on downstream tasks, removing this loss severely degrades the fidelity of the decoded videos, \textit{i.e.}, the FID metric is increased from 46 to 70 at 0.02bpp level.
Then, we remove the proposed semantic consistency loss $\mathcal{L}_{sem}$, which aims to learn a robust video semantic representation from large-scale unlabeled videos. The performance is drastically degraded by over 3\%@0.03bpp.

\begin{figure}[!bp]
	\vspace{-4mm}
	\centering
	\newcommand{\widthscalefive}{0.239}
	\tabcolsep=-0mm
	\begin{tabular}{cc}
		\includegraphics[width=\widthscalefive \textwidth]{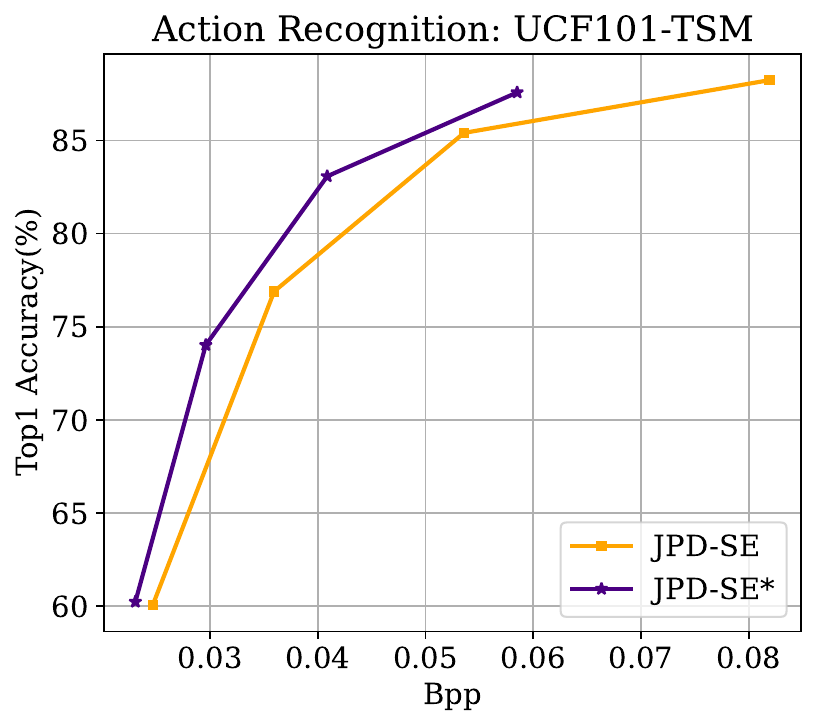}&
		\includegraphics[width=\widthscalefive \textwidth]{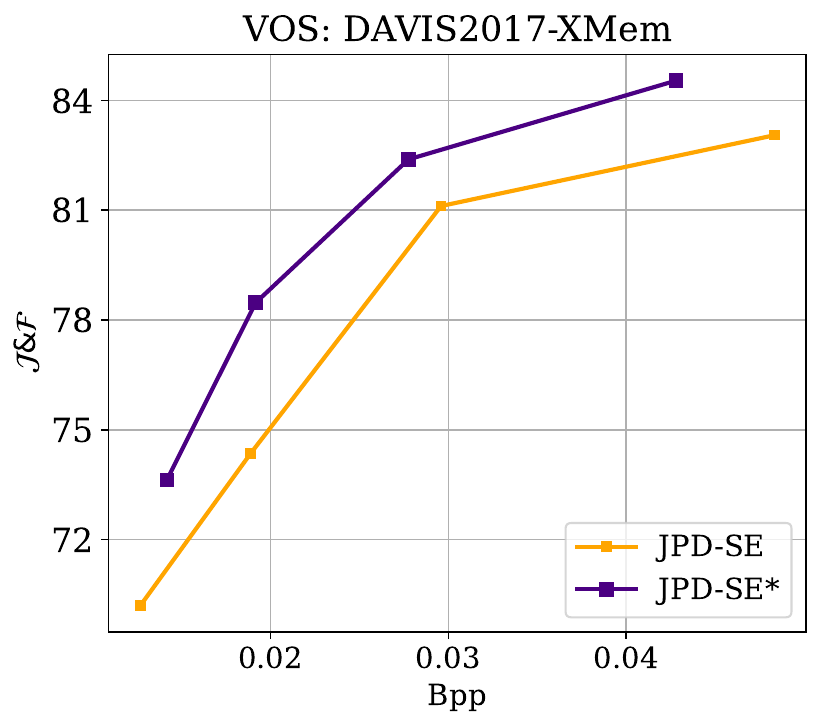}
	\end{tabular}
	\vspace{-3mm}
	\caption {
		Effectiveness of the contrastive objective $\mathcal{L}_{sem}$ on JPD-SE~\cite{duan2022jpd}. JPD-SE* denotes the JPD-SE model trained with $\mathcal{L}_{sem}$.
	}
	\label{fig:apply_to_jpdse}
\end{figure}

When we remove the hard binarized bottleneck map (BM) design and directly perform contrastive learning on the full-color videos, the performance is dropped obviously (2\%@0.05bpp). This proves that our early bottleneck-based contrastive learning scheme is effective and necessary for a transportation-efficient semantic coding system.
We also try to remove the $\mathcal{L}_{lpips}$ loss, but the model can not converge, as it is the only item for constraining the video color space.
Finally, we analyze the impact of the weight $\alpha$ for $\mathcal{L}_{sem}$, as shown in Tab.~\ref{tab:hyper_alpha}.
The performance increases consistently with the larger $\alpha$.
When $\alpha$ is set to a very large value, $\alpha=1$, the performance is decreased due to the training instability.

Furthermore, we investigate whether the proposed bottleneck map-based contrastive learning objective $\mathcal{L}_{sem}$ can enhance the performance of other video coding methods. Specifically, we integrate $\mathcal{L}_{sem}$ into the training objective of JPD-SE, resulting in the JPD-SE* model. As illustrated in Fig.~\ref{fig:apply_to_jpdse}, $\mathcal{L}_{sem}$ remarkably enhances the performance of JPD-SE on both action recognition and VOS tasks. For example, about 4\% improvement is observed for the action recognition task on the UCF101-TSM@0.04bpp setting.


\subsection{{Ablation Study on the Architecture of Sem-Enc}}

\textbf{Effectiveness of the dynamic convolution designs.}
As mentioned in Section~\ref{sec:method_enc_net}, Sem-Enc is adaptive to the current input videos by introducing the dynamic convolution (DyConv) design, \textit{i.e.}, composing the instance-adaptive kernel table ($AdaKT$) and region-adaptive kernel allocation map ($AdaKM$).
To verify the effectiveness of these designs, we change its network architecture and train several variant models.
The results are shown in Tab.~\ref{tab:ablation_enc_net}.
When removing the difference map pathway, the bit cost of the analytic stream is substantially increased by $+$68\%.
In this situation,
the information already encoded by the video stream may be also encoded into the semantic stream.
We also investigate the impact of the adaptive convolution design adopted by the feature fusion module DGFM.
Remember that we synthesize the dynamic convolution kernels by indexing $AdaKT$ with $AdaKM$.
When replacing the video instance-adaptive $AdaKT$ with the fixed kernel table $FixKT$ that is shared by all videos, the performances are degraded obviously, \textit{i.e.}, 17\% bit cost increase and 1.10\% accuracy drop.
Finally, we use plain convolutions to replace the dynamic convolution comprising both $AdaKT$ and $AdaKM$, the architecture of Sem-Enc degenerates into a completely static one. The performance is further substantially deteriorated in terms of the bit cost (+49\%) and action recognition accuracy (-3.02\%).
\begin{table}[!thbp]
	\centering
		\renewcommand\arraystretch{1.2}
	\tabcolsep=1.7mm
	\begin{tabular}{|c|c|c|c|}
		
		\hline
		\textit{Diff-Path}& \textit{Conv Type} & \textit{Bpp($\mathcal{S}$) \greendownarrow $\times10^{-3}$}& \textit{BD-Top1 (\%)} \greenuparrow  \\
		\hline
		
		\xmark & $AdaKT$+$AdaKM$ & 12.6 (+68\%) & 11.20 (-6.15)  \\
		\hline
		\cmark & $Plain$ & 11.2 (+49\%) & 13.46 (-3.89)  \\
		\hline
		\cmark & $FixKT$+$AdaKM$ & 8.8 (+17\%) & 16.14 (-1.21)  \\
		\hline
		\cmark & \textbf{$AdaKT$+$AdaKM$}  & \textbf{7.5} (+0\%) & \textbf{17.35} (-0)\\
		\hline
	\end{tabular}
	\vspace{-1mm}
	\caption{
		Ablation study on the dynamic convolution designs in Sem-Enc.
		The semantic stream bitcost \textit{Bpp($\mathcal{S}$)} is calculated when the base VVC codec CRF set to 51.
		\textit{BD-Top1} indicates the averagely improved Top1 accuracy over VVC on UCF101-TSM setting.
	}
	\label{tab:ablation_enc_net}
\end{table}

\revisetext{
\textbf{Effectiveness of difference map-based attention mechanism.}
We first mention that the entropy of the semantic stream can be reduced by being conditioned on the available information~\cite{ladune2020optical}\cite{li2021deep}, \textit{i.e.}, the lossy video $\tilde{X}$ decoded from the VVC video stream.
In our method, we leverage the difference map $D = X - \tilde{X}$ instead of $\tilde{X}$ as the condition variable due to the off-the-shelf guidance information embodied in $D$ for adaptive bit-allocation, \textit{e.g.}, patches corresponding to larger difference magnitude should be allocated with more bits.
The attention modules (masking the feature with the attention map produced from $D$) further explicitly apply the bit allocation strategy to the feature encoding procedure.
The enhanced feature focuses on encoding the corrupted regions, which avoids duplicated coding of the well compressed regions by the traditional codec.
}

\begin{table}[!thbp]
	\centering
	\renewcommand\arraystretch{1.2}
	\tabcolsep=0.8mm
	\begin{tabular}{|c|c|c|c|c|}
		
		\hline
		\textit{Model \#}& \textit{Cond; DFGM Arch} &\textit{Bpp($\mathcal{S}$) \greendownarrow $\times10^{-3}$}&\textit{BD-Top1 (\%)} \greenuparrow&Param \\
		\hline
		
		1&$\tilde{X}$; ResBlks+DyConv & {10.3}(+37\%) & 14.33(-3.02)& 12.91M  \\
		\hline
		2&$\tilde{X}$; AttBlks+DyConv  & {9.2}(+22\%) & 15.78(-1.57) &12.89M \\
		\hline
		3 (Ours)&$D $; AttBlks+DyConv  & \textbf{7.5}(+0\%)  & \textbf{17.35} (-0) &12.74M\\
		\hline
	\end{tabular}
		\vspace{-1mm}
	\caption{ Investigation of different condition variables and feature fusion strategies within the DFGM module.
	The calculation method for \textit{Bpp($\mathcal{S}$)} and \textit{BD-Top1} is same as that in Tab.~\ref{tab:ablation_enc_net}.
	}
	\label{tab:ablation_diff_attention}
\end{table}

To quantitatively verify the effectiveness of our approach in eliminating redundancy within the semantic stream, we trained several variant models and reported their performances in Tab.~\ref{tab:ablation_diff_attention}.
First, we replaced the difference map $D = X -\tilde{X} $ with the lossy video $\tilde{X}$ compressed by traditional codec (the video stream in Fig.~\ref{fig_framework}). The resulting model \#2 consumed 22\% more bitcost, indicating that the RGB difference signal better indicates the compression quality of the traditional video stream.
Furthermore, we replaced the attention blocks in model \#2 with simple residual blocks (ResBlks) as the guidance connection in the DFGM module. The performance of the produced model \#1 was dramatically reduced by 3.02\% in terms of Top1 accuracy.

\revisetext{
One might argue that the attention mechanism and dynamic convolution design could introduce significant additional inference time.
To investigate this, we conducted a comparison of model complexity and inference time on real hardware across different models.
We passed an eight-frame clip of spatial scale 256$\times$256 through the encoder 20 times and calculated the average inference time cost.
As shown in Tab.~\ref{tab:comp_dfgm_resblk}, the model implemented with only ResBlks is more hardware-friendly and slightly faster than our approach, with an inference time of 1.22ms compared to 1.31ms for our method. However, this gain in speed comes at the cost of noticeable performance degradation, specifically a 1.92\% Top1 accuracy drop.
Therefore, our method achieves a much better speed-performance trade-off.
}

	\begin{table}[!t]
			\vspace{-2mm}
		\renewcommand\arraystretch{1.2}
	\centering
	\tabcolsep=0.55mm
	\begin{tabular}{|c|c|Hc|c|c|}
		
		\hline
		\textit{DFGM  Arch}&	\textit{Param}& \textit{FLOPs} &\textit{Time}  &\textit{Bpp($\mathcal{S}$) \greendownarrow $\times10^{-3}$}&\textit{BD-Top1 (\%)} \greenuparrow\\
		\hline
	ResBlks+ResBlks &12.63M & 1.75G & \textbf{1.22ms} & 8.7& 15.43 (-1.92) \\
		\hline
		AttBlks+DyConv (Ours) & {12.74M}&  \textbf{1.81G} & 1.31ms& \textbf{7.5 }& \textbf{17.35}  \\
		\hline
	\end{tabular}
	\vspace{-1mm}
	\caption{
		Performance-speed tradeoff of  DFGM.
		The calculation method for \textit{Bpp($\mathcal{S}$)} and \textit{BD-Top1} is same as that in Tab.~\ref{tab:ablation_enc_net}.
	}
	\vspace{-3mm}
	\label{tab:comp_dfgm_resblk}
\end{table}

\subsection{{Ablation Study on the Architecture of LFN}}

\revisetext{
\textbf{Superiority of the light-down-heavy-up architecture.}
In Fig.~\ref{fig_mgenet}, the downscaling path is responsible for transforming the low-quality video compressed by the traditional codec into latent features. Since the video can be quite noisy, a lightweight network is sufficient for extracting useful information.
On the other hand, the upscaling path needs to generate high-quality and photorealistic videos, essentially functioning as a generative model. This process commonly relies on a larger number of parameters to generate plausible textures. As a result, we employ only one TDense block in the downscaling path, while utilizing eight blocks in the upscaling path. This arrangement forms an asymmetric architecture.
To further validate this design, we trained several variant models with similar numbers of parameters.
}

\begin{table}[!b]
	\renewcommand\arraystretch{1.2}
	\centering
		\vspace{-3mm}
	\tabcolsep=2.0mm
	\begin{tabular}{|c|c|c|c|c|}
		
		\hline
		\textit{Down Blocks}&	\textit{Up Blocks}&\textit{Param}&\textit{FLOPs}   &\textit{BD-Top1 (\%)} \greenuparrow\\
		\hline
		TDense$\times$5 &TDense$\times$5 &  9.31M & {6.28G} &  12.46 \\
		
		TDense$\times$8 &TDense$\times$1 &  8.65M & {5.85G} &  6.71 \\
		
		TDense$\times$1(Ours) &TDense$\times$8 & 8.65M & {5.85G} &  \textbf{17.35} \\
		\hline
		Dense$\times$1 &Dense$\times$8 &  8.21M & {5.29G}&  15.12 \\
		\hline
	\end{tabular}
\vspace{-1mm}
	\caption{
		Ablation study on LFN network architecture. The calculation method for \textit{BD-Top1} is same as that in Tab.~\ref{tab:ablation_enc_net}.
	}
	\label{tab:comp_LFN_arch}
\end{table}

\revisetext{
As shown in Tab.~\ref{tab:comp_LFN_arch}, our light-down-heavy-up scheme outperforms both the heavy-down-light-up scheme and the symmetric architecture by a large margin.
Moreover, we replace the TDense block with the vanilla Dense block, the resulted model is much inferior to our method due to the loss of the temporal modeling capability, as shown in the last row of Tab.~\ref{tab:comp_LFN_arch}.
}

\begin{figure}[!t]
	\centering
	\newcommand{\widthscalefive}{0.15}
	\tabcolsep=0.5mm
	\scriptsize
	\begin{tabular}{ccc}
		\includegraphics[width=0.150 \textwidth]{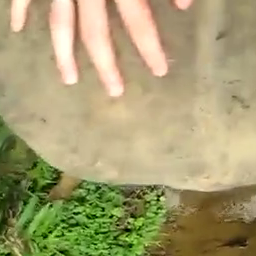}&
		\includegraphics[width=0.150 \textwidth]{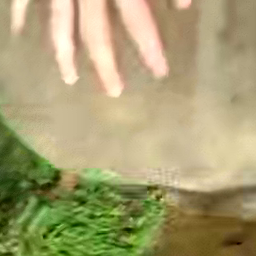}&
		\includegraphics[width=0.162 \textwidth]{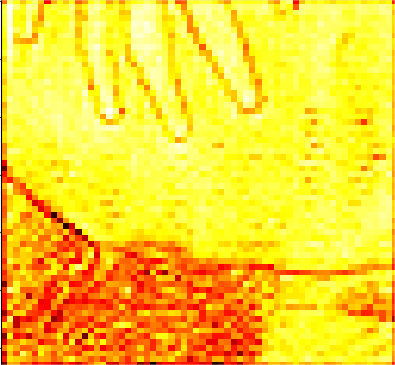}
		\\
		(a) Original frame & (b) Lossy frame  &(c) Attention map \\
	\end{tabular}
	\vspace{-2mm}
	\caption {
		Visualization of the attention map in LFN.
		Red color in attention map indicates the feature of the distorted region is suppressed.
	}
	\label{fig:vis_LFN_attention}
\end{figure}

\revisetext{
\textbf{Analysis of attention-based feature fusion scheme.}
LFN employs an attention-based fusion module instead of the simple concatenation plus convolution operation used in Unet~\cite{ronneberger2015u}. This choice arises from their fundamentally distinct motivations.
In Unet, the features are generally clean, so that concatenation plus convolution operations are sufficient for multi-scale modeling and for detecting object segmentation boundary cues from both local and global contexts.
In contrast, LFN capitalizes on a content-adaptive attention map to leverage the high-quality regions of the lossy video while concurrently sidestepping interference from distorted regions. This is exemplified in Fig.~\ref{fig:vis_LFN_attention}, where the attention map effectively detects and suppresses distorted areas, such as the ringing artifact surrounding the human finger and the structural distortions within the grassy region.
}

\revisetext{
Additionally, we perform a quantitative comparison of the two schemes. As shown in Tab.~\ref{tab:comp_LFN_fusion}, our explicit attention-based feature fusion scheme large outperforms the variant model equipped with residual blocks (ResBlks), \textit{i.e.}, 2.23\% improvement in terms of Top1 accuracy.
}
\begin{table}[!t]
	\renewcommand\arraystretch{1.2}
	\centering
	\tabcolsep=1.90mm
	\begin{tabular}{|c|c|c|c|}
		
		\hline
		\textit{Fusion Strategy}&\textit{Param}&\textit{FLOPs}   &\textit{BD-Top1 (\%)} \greenuparrow\\
		\hline
		Concatenation+ResBlks & 8.71M & {5.89G} &  {15.12}\\
		\hline
		Concatenation+Attention (Ours)  &  8.65M & {5.85G}&  \textbf{17.35} \\
		\hline
	\end{tabular}
	\vspace{-1mm}
	\caption{
		Ablation study on feature fusion strategy of LFN.
		The calculation method for \textit{BD-Top1} is same as that in Tab.~\ref{tab:ablation_enc_net}.
	}
	\vspace{-2mm}
	\label{tab:comp_LFN_fusion}
\end{table}
\subsection{Model Complexity}
We calculate the per-frame computational complexity of our framework with an input video clip of spatial scale $256\times256$, as shown in Tab.~\ref{tab:complexity}.
As for the \textit{encoder} side, the computational cost of our method is about 30$\times$ fewer than DVC due to not using the optical flow estimation network.
Our method is even 3$\times$ efficient than a simple image compression network Ball{\'e}~\etal~\cite{balle2018variational}.
Note that DVC and Ball{\'e}~\etal~\cite{balle2018variational} are not able to achieve the performance of H.265 codec, but our method can further improve and surpass the performance of the recent VVC codec.
As for the \textit{decoder} side, LFN in our method is about 16$\times$ efficient than the state-of-the-art video restoration method BasicVSR++.

 \begin{table}[!thbp]
 	\centering
 	\tabcolsep=4.99mm
 	\begin{tabular}{|c|c|c|c|}
 		\hline
 		& \textit{Model}  &\textit{Param} & \textit{FLOPs}   \\
 		\hline 
 		\multirow{3}{*}{\makecell[c]{Encoder\\Side}} & Ball{\'e}~\etal~\cite{balle2018variational} &3.39 & 5.35 \\
 		& DVC~\cite{lu2020end} &5.96M & 56.68G \\
 		& \textbf{Sem-Enc(Ours)} &12.74M &\textbf{1.81G} \\
 		\arrayrulecolor{gray}\cdashline{1-4}[5pt/3pt]
 		\multirow{3}{*}{\makecell[c]{Decoder\\Side}} & STDF &0.47M&24.06G \\
 		& BasicVSR++ &6.88M&93.20G \\
 		&\textbf{LFN(Ours)}  &8.65M &\textbf{5.85G} \\
 		\hline
 	\end{tabular}
 	\vspace{-1mm}
 	\caption{
 		Model complexity of video encoder and decoder.
 	}
 	\label{tab:complexity}
 \end{table}

	\section{Conclusion and Limitations}
	In this paper, we have proposed a semantic coding framework VCS for low-bitrate compressed video understanding.
	Our framework inherits the advantages of both the strong content-coding capability of traditional video codecs and the superior semantic coding capability of neural networks.
	Experimental results show that our approach improves the current traditional codecs by a large margin on three downstream video understanding tasks, \textit{i.e.}, action recognition, multiple object tracking and video object segmentation.
	Moreover, we have thoroughly built a benchmark for this novel problem over several large-scale video datasets.
	In our follow-up works, we have proposed more advanced semantic compression methods~\cite{tian2023non}\cite{tian2024smc++}\cite{tian2024free} by adapting the masked image modeling strategies and absorbing the semantics from large visual foundation models.
	
	One limitation is that our method processes the foreground and background objects equally, which may cause the learned semantic representations to be background-biased compared to human perception. We plan to address this problem by introducing the 3D geometry and physics priors to model the 3D object structure, which can better reason about video semantics.

	\noindent \textbf{Acknowledgment} This work was supported by the National Science Foundation of China (62225112), the Fundamental Research Funds for the Central Universities, National Key R\&D Program of China (2021YFE0206700), Shanghai Municipal Science and Technology Major Project (2021SHZDZX0102), STCSM under Grant 22DZ2229005, and Shanghai Artificial Intelligence Laboratory.

	\ifCLASSOPTIONcaptionsoff
	\newpage
	\fi

	
	
	%
	\bibliographystyle{IEEEtran}
	\bibliography{IEEEfull}
	
	\begin{IEEEbiography}
		[{\includegraphics[width=1in,height=1.25in,clip,keepaspectratio]{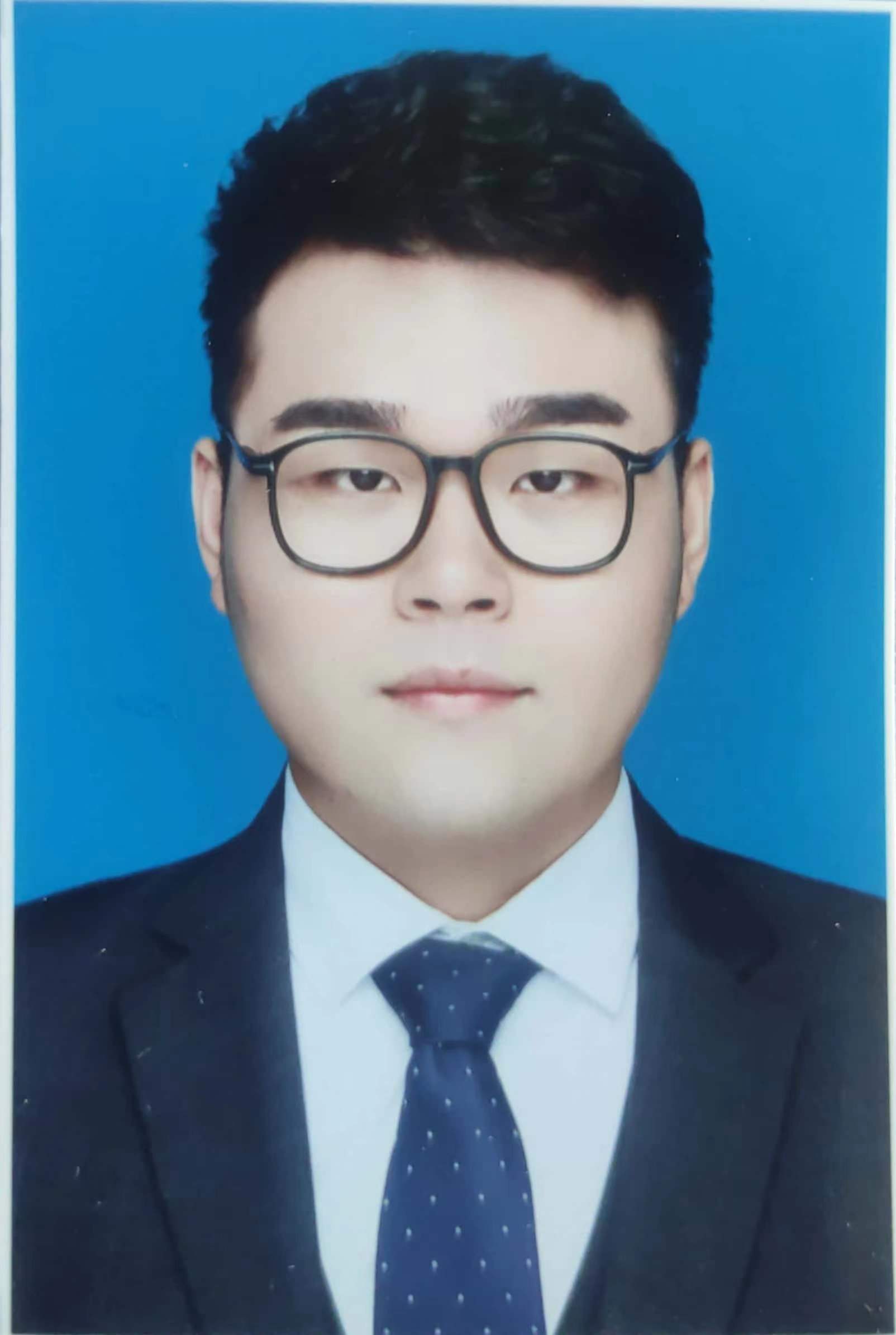}}]
		{Yuan Tian}
		received the B.Sc. degree in electronic engineering from Wuhan University, Wuhan, China, in 2017. He is currently
		pursuing the Ph.D. degree with the Department of Electronic Engineering, Shanghai Jiao Tong University, Shanghai, China.
		His works have been published in top-tier journals and conferences (e.g., IJCV, ICCV, and ECCV).
		His research interests include video understanding and video compression.
	\end{IEEEbiography}
	\begin{IEEEbiography}
			[{\includegraphics[width=1in,height=1.25in,clip,keepaspectratio]{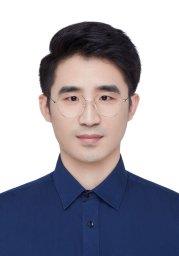}}]
		{Guo Lu}
		received the B.S. degree from the Ocean University of China in 2014 and the Ph.D. degree from Shanghai Jiao Tong University in 2020. Currently, he is an Assistant Professor with the Department of Electronic Engineering, Shanghai Jiao Tong University, Shanghai, China. His works have been published in top-tier journals and conferences (e.g., TPAMI, TIP, CVPR, and ECCV). His research
		interests include image and video processing, video compression, and computer vision.
	\end{IEEEbiography}
	\begin{IEEEbiography}
			[{\includegraphics[width=1in,height=1.25in,clip,keepaspectratio]{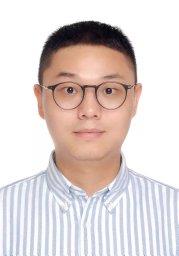}}]
		{Yichao Yan}
		received his B.E. and Ph.D degree in electrical engineering from Shanghai Jiao Tong University, in 2013 and 2019, respectively.
		Currently, he is an Assistant Professor with the AI Institute, Shanghai Jiao Tong University, Shanghai, China.
		He has authored/coauthored more than 10 peer-reviewed papers, including those in highly regarded journals and conferences such as TPAMI, CVPR, ECCV, ACMMM, IJCAI, TMM, etc. His research interests include object recognition, video analysis, and deep learning.
	\end{IEEEbiography}
	\begin{IEEEbiography}
			[{\includegraphics[width=1in,height=1.25in,clip,keepaspectratio]{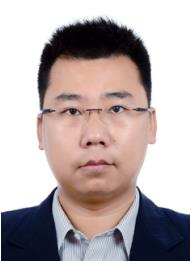}}]
		{Guangtao Zhai} received the B.E. and M.E. degrees from Shandong University, Shandong, China, in 2001 and 2004, respectively, and the Ph.D. degree from Shanghai Jiao Tong University, Shanghai, China, in 2009, where he is currently a Research Professor with the Institute of Image Communication and Information Processing. From 2008 to 2009, he was a Visiting Student with the Department of Electrical and Computer Engineering, McMaster University, Hamilton, ON, Canada, where he was a Post-Doctoral Fellow from 2010 to 2012. From 2012 to 2013, he was a Humboldt Research Fellow with the Institute of Multimedia Communication and Signal Processing, Friedrich Alexander University of Erlangen-Nuremberg, Germany. He received the Award of National Excellent Ph.D. Thesis from the Ministry of Education of China in 2012. His research interests include multimedia signal processing and perceptual signal processing.
	\end{IEEEbiography}
	\begin{IEEEbiography}
			[{\includegraphics[width=1in,height=1.25in,clip,keepaspectratio]{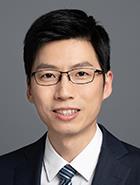}}]
		{Li Chen} received his B.S. and M.S. degrees, both from Northwestern Polytechnical University at Xian of China in 1998 and 2000, and the Ph.D. Degree from Shanghai Jiao Tong University, China, in 2006, all in electrical engineering. His current research interests include Image and Video Processing, DSP and VLSI for Image and video processing.
	\end{IEEEbiography}
	\begin{IEEEbiography}
			[{\includegraphics[width=1in,height=1.25in,clip,keepaspectratio]{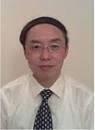}}]
		{Zhiyong Gao} received the B.S. and M.S. degrees in electrical engineering from the Changsha Institute of Technology (CIT), Changsha, China, in 1981 and 1984, respectively, and the Ph.D. degree from Tsinghua University, Beijing, China, in 1989. From 1994 to 2010, he took several senior technical positions in England, including a Principal Engineer with Snell \& Wilcox, Petersfield, U.K., from 1995 to 2000, a Video Architect with 3DLabs, Egham, U.K., from 2000 to 2001, a Consultant Engineer with Sony European Semiconductor Design Center, Basingstoke, U.K., from 2001 to 2004, and a Digital Video Architect with Imagination Technologies, Kings Langley, U.K., from 2004 to 2010. Since 2010, he has been a Professor with Shanghai Jiao Tong University. His research interests include video processing and its implementation, video coding, digital TV and broadcasting.
	\end{IEEEbiography}

\end{document}